\documentclass[pra]{revtex4}
\usepackage{graphicx}
\usepackage{amsfonts}

\begin{document}

\preprint{}
\title[Optical parallel computation similar to quantum computation]{Optical
parallel computation similar to quantum computation based on optical fields
modulated with pseudorandom phase sequences}
\author{Jian Fu}
\affiliation{College of Optical Science and Engineering, Zhejiang University, Hangzhou,
310027, China}

\begin{abstract}
We propose an optical parallel computation similar to quantum computation
that can be realized by introducing pseudorandom phase sequences into
classical optical fields with two orthogonal modes. Based on the
pseudorandom phase sequences, we first propose a theoretical framework of
\textquotedblleft phase ensemble model\textquotedblright\ referring from the
concept of quantum ensemble. Using the ensemble model, we further
demonstrate the inseparability of the fields similar to quantum
entanglement. It is interesting that a $N2^{N}$ dimensional Hilbert space
spanned by $N$ optical fields is larger than that spanned by $N$ quantum
particles. This leads a problem for our scheme that is not the lack of
resources but the redundancy of resources. In order to reduce the
redundancy, we propose a special sequence permutation mechanism to
efficiently imitate certain quantum states, including the product state,
Bell states, GHZ state and W state. For better fault tolerance, we further
devise each orthogonal mode of optical fields is measured to assign discrete
values. Finally, we propose a generalized gate array model to imitate some
quantum algorithms, such as Shor's algorithm, Grover's algorithm and quantum
Fourier algorithm. The research on the optical parallel computation might be
important, for it not only has the potential beyond quantum computation, but
also provides useful insights into fundamental concepts of quantum mechanics.
\end{abstract}

\date{today}
\startpage{1}
\email{jianfu@zju.edu.cn}
\maketitle

\section*{Introduction \label{Sec I}}

It has been widely known that quantum computation enormously promotes
computational efficiency by using several basic and purely physical features
of quantum mechanics, such as coherent superposition, quantum entanglement,
measurement collapse etc. \cite{Nielsen}. The accelerant ability of quantum
computation is related to quantum entanglement and tensor product structure,
which are essential to allow growing exponentially the computation resource
with the number of qubits \cite{Jozsa,Braunstein}. Yet the practical quantum
computations are difficult to be realized for restrictions of quantum system
controllability, decoherence property and measurement randomness \cite%
{Knill,Nielsen2,Browne,Kok}.

The classical simulation of quantum systems, especially of quantum
entanglement has been under investigation for a long time\ \cite%
{Cerf,Massar,Spreeuw}. In addition to easy implementations, the researches
on the classical simulations can help understand some fundamental concepts
in quantum mechanics. However, it has been pointed out by several
researchers that the classical simulation of quantum systems requires
exponentially scaling of physical resources with the number of quantum
particles \cite{Jozsa,Spreeuw}. In Ref. \cite{Spreeuw}, an optical analogy
to quantum systems was introduced, in which the number of light beams and
optical components required grows exponentially with the number of qubits.
In Ref. \cite{Vidal}, a classical protocol to efficiently simulate any
pure-state quantum computation is presented, yet the amount of entanglement
involved is restricted. In Ref. \cite{Jozsa}, it is elucidated that in
classical theory, the state space of a composite system is the direct
product of subsystems, whereas in quantum theory it is the tensor product.
It is generally accepted that the essential distinction between direct and
tensor products is precise the phenomenon of quantum entanglement, and
regarded as the origin of the limitation of any classical systems. Recently,
several researches have proposed a new concept of realization of classical
entanglement based on classical optical fields by introducing a new degree
of freedom, such as orbital angular momentum, to realize tensor product in
quantum entanglement \cite{Toppel}. However, the method cannot provide
enough orthogonal degrees of freedom, which the scalability might be
doubtful.

In this paper, we propose an optical parallel computation similar to quantum
computation that can be realized by introducing pseudorandom phase sequences
into optical fields with two orthogonal modes \cite{Fu1}. The two orthogonal
modes (polarization or transverse) of optical field are encoded as optical
analogies to quantum bits $\left\vert 0\right\rangle $\ and\ $\left\vert
1\right\rangle $ \cite{Fu,dragoman,Lee}. In wireless and optical
communications, orthogonal pseudorandom sequences have been widely applied
to Code Division Multiple Access (CDMA) communication technology as a way to
distinguish different users \cite{Golomb,Viterbi,Peterson}. A set of
pseudorandom sequences guided by a Galois field GF($p$) is generated by
using a linear feedback shift register method, which satisfies orthogonal,
closure and balance properties \cite{Golomb,Viterbi,Peterson}. In Phase
Shift Keying (PSK) communication technology \cite{PSK}, the information is
encoded in the phase of classical optical/electromagnetic fields, where the
phase values in $\left\{ 0,2\pi /p,\cdots ,2\pi \left( p-1\right) /p\right\} 
$ for the $p$-ary communication. Combining the two communication
technologies, we introduce the pseudorandom phase sequences in our scheme.
Guaranteed by the orthogonal property, the optical/electromagnetic fields
modulated with different pseudorandom phase sequences can transmit in one
communication channel simultaneously without crosstalk, and can be easily
distinguished by implementing a coherent demodulation \cite{Viterbi}.

Different from other schemes \cite{Spreeuw,Toppel}, the pseudorandom phase
sequences employed in our scheme are able to provide not only scalable
degrees of freedom to support arbitrary dimensional tensor product
structure, also a theoretical framework of \textquotedblleft phase ensemble
model\textquotedblright\ similar to the concept of quantum ensemble. Using
the ensemble model, we can demonstrate the inseparable correlation between
the optical fields with different pseudorandom phase sequences similar to
quantum entanglement. It is interesting that a $N2^{N}$ dimensional Hilbert
space can be spanned by $N$ optical fields which is larger than that spanned
by $N$ quantum particles. This leads a problem for our scheme that is not
the lack of resources but the redundancy of resources. In order to reduce
the redundancy, we have to introduce a sequential cycle permutation
mechanism based on coherent demodulation to realize the bijection imitation
of certain quantum states. Optical analogies to some typical quantum states
are also discussed, including Bell states, GHZ and W states. For better
fault tolerance, we devise each orthogonal mode of optical fields is
measured to assign discrete values. It means that a discrete computation
model is provided in our scheme. Furthermore, we propose a gate array model
to imitate quantum computation based on four kinds of mode control gates. As
some examples, we demonstrate the imitations of Shor's algorithm \cite{Shor}%
, Grover's algorithm \cite{Grover} and quantum Fourier algorithm \cite%
{Nielsen}. In order to verify the feasibility, we numerically simulate our
scheme using the widely used optical communication simulation software
OPTISYSTEM.

The paper is organized as follows: In Section \ref{Sec II}, we introduce
some preparing knowledges required later in this paper. In Section \ref{Sec
III}, a theoretical framework of \textquotedblleft phase ensemble
model\textquotedblright\ and the optical analogies of several typical
quantum states are then discussed. In Section \ref{Sec IV}, a gate array
model to imitate quantum computation is proposed. Finally, we summarize our
conclusions in Section \ref{Sec V}.

\section{Preparing knowledges \label{Sec II}}

In this section, we introduce some notations and basic results required
later in this paper. We first introduce pseudorandom phase sequences (PPSs)
and their properties. Then we introduce the scheme of modulation and
demodulation on optical fields with PPSs. Finally, we discuss the
similarities between an optical field and a single-particle quantum state.

\subsection{Pseudorandom phase sequences and their properties \label{Sec
II.A}}

As far as we know, orthogonal pseudorandom sequences have been widely
applied to CDMA communication technology as a way to distinguish different
users \cite{Golomb,Viterbi,Peterson}. A set of pseudorandom sequences is
generated from a shift register guided by a Galois field GF($p$), which
satisfies orthogonal, closure and balance properties. The orthogonal
property ensures that sequences of the set are independent and distinguished
each other with an excellent correlation property. The closure property
ensures that any linear combination of the sequences remains in the same
set. The balance property ensures that the occurrence rate of all
non-zero-element in every sequence is equal, and the the number of
zero-elements is exactly one less than the other elements. One famous
generator of pseudorandom sequences is linear feedback shift register
(LFSR), which can produce a maximal period sequence, called m-sequence \cite%
{Golomb}. We consider an m-sequence of period $N-1$ ($N=p^{s}$) generated by
a primitive polynomial of degree $s$ over GF($p$). Since the correlation
between different shifts of an m-sequence is almost zero, they can be used
as different codes with their excellent correlation property. In this
regard, the set of $N-1$ m-sequences of length $N-1$ can be obtained by
cyclic shifting of a single m-sequence.

In this paper, we employ PPSs with $2$-ary $\{0,\pi /2\}$ phase shift
modulation. Although the phases should uniformly distribute in $\left[
0,2\pi \right] $, the $2$-ary PSK is the most frequently used modulation due
to the phase symmetry in practical communication systems \cite{PSK}. We
first propose a scheme to generate a PPS set $\Xi =\left\{ \lambda ^{\left(
1\right) },\lambda ^{\left( 2\right) },\ldots ,\lambda ^{\left( N\right)
}\right\} $ over GF($2$). $\lambda ^{\left( N\right) }$ is an all-zero
sequence and other sequences can be generated by using the method as follows:

(1) given a primitive polynomial of degree $s$ over GF($2$), a base sequence
of a length $2^{s}-1$ is generated by using LFSR;

(2) other sequences are obtained by cyclic shifting of the base sequence;

(3) by adding a zero-element to the end of each sequence, the occurrence
rates of all elements in all sequences are equal with each other;

(4) mapping the elements of the sequences to $\{0,\pi /2\}$: $0$ mapping $0$%
, $1$ mapping $\pi /2$.

\begin{figure}[tbph]
\centering\includegraphics[height=1.6942in, width=4.3059in]{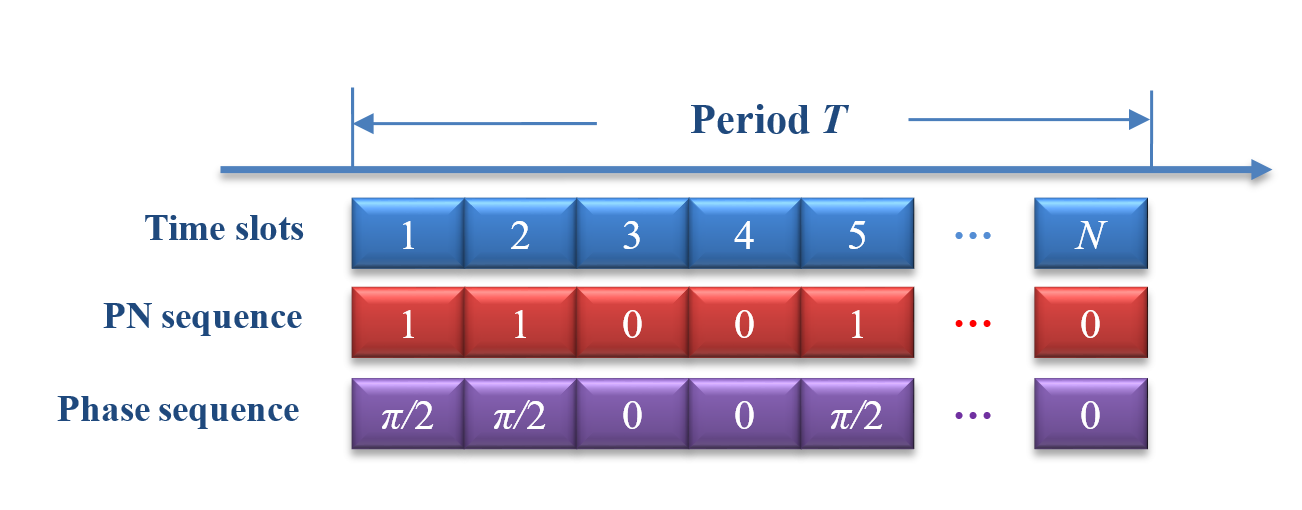}
\caption{The relationship between time slots, an m-sequence and PPS is
shown. }
\label{fig1}
\end{figure}

In Fig. \ref{fig1}, we demonstrate the relationship between time slots, an
m-sequence and phase sequence $\lambda ^{\left( i\right) }$ with $N$ phase
units $\lambda ^{\left( i\right) }=\left[ 
\begin{array}{cccc}
\lambda _{1}^{\left( i\right) } & \lambda _{2}^{\left( i\right) } & ... & 
\lambda _{N}^{\left( i\right) }%
\end{array}%
\right] $. For better understanding our scheme, the PPSs in the cases of
modulating $4$ and $8$ optical fields are illustrated below. An m-sequence
of length $2^{2}-1$\ is generated by a primitive polynomial of the lowest
degree over $GF(2)$, which is $\left[ 
\begin{array}{ccc}
1 & 0 & 1%
\end{array}%
\right] $. Then we obtain the set $\Xi $ that includes $4$ PPSs of length $4$%
: $\left\{ \lambda ^{\left( 1\right) },\lambda ^{\left( 2\right) },\lambda
^{\left( 3\right) },\lambda ^{\left( 4\right) }\right\} $, where $\lambda
^{\left( 4\right) }$ is the all-zero sequence. The PPSs can be used to
modulate up to $4$ optical fields as follows 
\begin{eqnarray}
\lambda ^{\left( 1\right) } &=&\left[ 
\begin{array}{cccc}
1 & 0 & 1 & 0%
\end{array}%
\right] \times \pi /2,  \label{1} \\
\lambda ^{\left( 2\right) } &=&\left[ 
\begin{array}{cccc}
1 & 1 & 0 & 0%
\end{array}%
\right] \times \pi /2,  \nonumber \\
\lambda ^{\left( 3\right) } &=&\left[ 
\begin{array}{cccc}
0 & 1 & 1 & 0%
\end{array}%
\right] \times \pi /2,  \nonumber \\
\lambda ^{\left( 4\right) } &=&\left[ 
\begin{array}{cccc}
0 & 0 & 0 & 0%
\end{array}%
\right] \times \pi /2.  \nonumber
\end{eqnarray}%
By using the same method, an m-sequence of length $2^{3}-1$\ is generated by
a primitive polynomial of the $2^{nd}$ lowest degree over $GF(2)$, which is $%
\left[ 
\begin{array}{ccccccc}
1 & 0 & 0 & 1 & 0 & 1 & 1%
\end{array}%
\right] $. Then we obtain the set $\Xi $ that includes $8$ PPSs of length $8$%
: $\left\{ \lambda ^{\left( 1\right) },\lambda ^{\left( 2\right) }\ldots
,\lambda ^{\left( 8\right) }\right\} $, where the PPSs are shown as follows%
\begin{eqnarray}
\lambda ^{\left( 1\right) } &=&\left[ 
\begin{array}{cccccccc}
1 & 0 & 0 & 1 & 0 & 1 & 1 & 0%
\end{array}%
\right] \times \pi /2,  \label{2} \\
\lambda ^{\left( 2\right) } &=&\left[ 
\begin{array}{cccccccc}
1 & 1 & 0 & 0 & 1 & 0 & 1 & 0%
\end{array}%
\right] \times \pi /2,  \nonumber \\
\lambda ^{\left( 3\right) } &=&\left[ 
\begin{array}{cccccccc}
1 & 1 & 1 & 0 & 0 & 1 & 0 & 0%
\end{array}%
\right] \times \pi /2,  \nonumber \\
\lambda ^{\left( 4\right) } &=&\left[ 
\begin{array}{cccccccc}
0 & 1 & 1 & 1 & 0 & 0 & 1 & 0%
\end{array}%
\right] \times \pi /2,  \nonumber \\
\lambda ^{\left( 5\right) } &=&\left[ 
\begin{array}{cccccccc}
1 & 0 & 1 & 1 & 1 & 0 & 0 & 0%
\end{array}%
\right] \times \pi /2,  \nonumber \\
\lambda ^{\left( 6\right) } &=&\left[ 
\begin{array}{cccccccc}
0 & 1 & 0 & 1 & 1 & 1 & 0 & 0%
\end{array}%
\right] \times \pi /2,  \nonumber \\
\lambda ^{\left( 7\right) } &=&\left[ 
\begin{array}{cccccccc}
0 & 0 & 1 & 0 & 1 & 1 & 1 & 0%
\end{array}%
\right] \times \pi /2,  \nonumber \\
\lambda ^{\left( 8\right) } &=&\left[ 
\begin{array}{cccccccc}
0 & 0 & 0 & 0 & 0 & 0 & 0 & 0%
\end{array}%
\right] \times \pi /2.  \nonumber
\end{eqnarray}

Further, we define a map $f:\lambda \rightarrow e^{i\lambda }$ on the set of 
$\Xi $, and obtain a new sequence set $\Omega =\left\{ \varphi ^{\left(
j\right) }\left\vert \varphi ^{\left( j\right) }=e^{i\lambda ^{\left(
j\right) }},\right. j=1,\ldots ,N\right\} $. The map $f$\ corresponds to the
phase modulations of the PPSs on optical fields. According to the properties
of m-sequence, we can obtain following properties of the set $\Omega $, (1)
the closure property: the product of any sequences remains in the same set,
in additional $\pi $ phase contributed by the power of a sequence; (2) the
balance property: in exception to $\varphi ^{\left( N\right) }$, any
sequences of the set $\Omega $ satisfy, 
\begin{equation}
\sum\limits_{k=1}^{N}e^{i\theta }\varphi _{k}^{\left( j\right)
}=\sum\limits_{k=1}^{N}e^{i\left( \lambda _{k}^{\left( j\right) }+\theta
\right) }=0,\forall \theta \in 
\mathbb{R}
;  \label{3}
\end{equation}%
(3) the orthogonal property: any two sequences satisfy the following
normalized correlation 
\begin{eqnarray}
E\left( {\varphi ^{\left( i\right) },\varphi ^{\left( j\right) }}\right) &=&%
\frac{1}{N}\sum\limits_{k=1}^{N}{\varphi _{k}^{\left( i\right) }\varphi
_{k}^{\left( j\right) \ast }}  \label{4} \\
&=&\left\{ 
\begin{array}{cc}
1, & i=j \\ 
0, & i\neq j%
\end{array}%
\right. .  \nonumber
\end{eqnarray}

In conclusion, according to the properties above, the optical fields
modulated with different PPSs become independent and distinguishable in any
case.

\subsection{Modulation and demodulation on optical fields with pseudorandom
phase sequences \label{Sec II.B}}

In this section, we mainly focus on the modulation and demodulation of
optical fields with single polarization mode. We first consider the
modulation process of an optical field with a PPS $\lambda ^{\left( i\right)
}$ from the set $\Xi $. For example, we choose $\lambda ^{\left( 1\right) }$
to modulate the optical field labeled as the signal optical (SO) field, its
electric field component is%
\begin{equation}
E_{S}\left( t\right) =A_{S}e^{-i\left( \omega t+\lambda _{k}^{\left(
1\right) }\right) },  \label{5}
\end{equation}%
where $A_{S},\omega $ are the amplitude and frequency of the optical field
respectively, and $\lambda _{k}^{\left( 1\right) }$ is the phase unit of $%
\lambda ^{\left( 1\right) }$ at the $k$-th time slot.

\begin{figure}[htbp]
\centering\includegraphics[height=1.8092in, width=2.9127in]{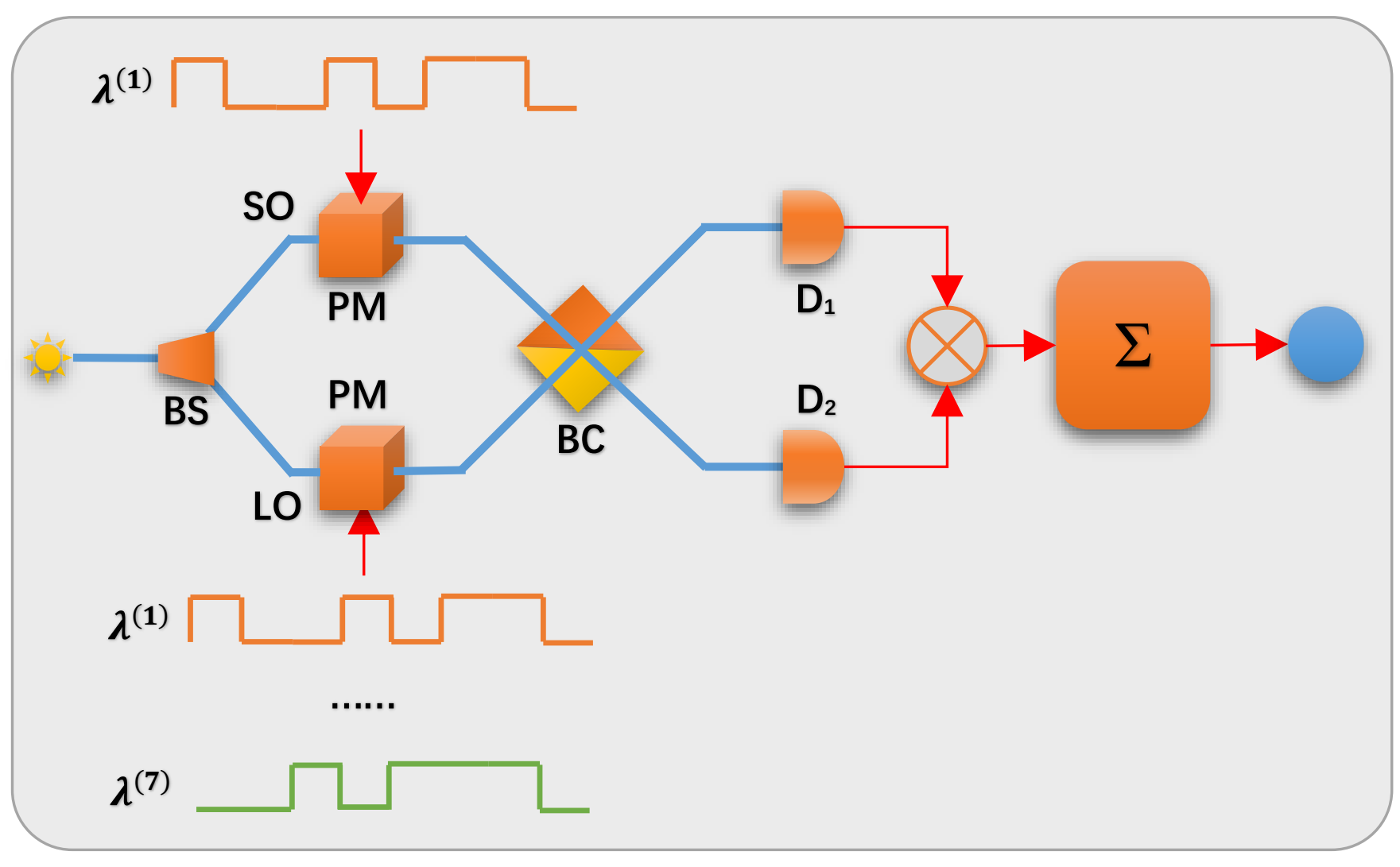}
\caption{The scheme of the coherent detection of pseudorandom phase sequence
is shown, where SO: signal optical field, LO: local optical field, BS: beam
splitter, PM: phase modulators, BC: beam coupler, $D_{1}$ and $D_{2}$:
photodetectors, $\otimes $: multiplier and $\Sigma $: integrator (integrate
over entire sequence period).}
\label{fig3}
\end{figure}

In order to perform the demodulation of PPS, we design a coherent detection
scheme as shown in Fig. \ref{fig3} that has been widely used in the coherent
communication \cite{Viterbi}. In the detection scheme, the local optical
(LO) beam and the SO beam interfere with each other through a beam coupler
(BC). In order to ensure the coherence of them, the two beams can be split
from the same optical source through a beam splitter (BS). The LO field can
be expressed as 
\begin{equation}
E_{L}\left( t\right) =A_{L}e^{-i\left( \omega t+\lambda _{k}^{\left(
n\right) }\right) },  \label{6}
\end{equation}%
where $\lambda ^{\left( n\right) }$ can be an arbitrary sequence of the set $%
\Xi $ and the amplitude $A_{L}=A_{S}$ assumed. After the coherent
superposition through the BC, the output fields can be expressed as 
\begin{equation}
\left( 
\begin{array}{c}
E_{1}\left( t\right) \\ 
E_{2}\left( t\right)%
\end{array}%
\right) =\frac{1}{\sqrt{2}}\left( 
\begin{array}{cc}
1 & i \\ 
-i & 1%
\end{array}%
\right) \left( 
\begin{array}{c}
E_{S}\left( t\right) \\ 
E_{L}\left( t\right)%
\end{array}%
\right) =\frac{A_{S}}{\sqrt{2}}\left( 
\begin{array}{c}
e^{-i\left( \omega t+\lambda _{k}^{\left( 1\right) }\right) }+ie^{-i\left(
\omega t+\lambda _{k}^{\left( n\right) }\right) } \\ 
-ie^{-i\left( \omega t+\lambda _{k}^{\left( 1\right) }\right) }+e^{-i\left(
\omega t+\lambda _{k}^{\left( n\right) }\right) }%
\end{array}%
\right) .  \label{7}
\end{equation}%
Then, the output electric signals of photodetectors $D_{1}$ and $D_{2}$ is
proportional to 
\begin{eqnarray}
D_{1} &=&\mu \left\vert E_{1}\left( t\right) \right\vert ^{2}=\mu \left\vert
A_{S}\right\vert ^{2}\left[ 1-\sin \left( \lambda _{k}^{\left( 1\right)
}-\lambda _{k}^{\left( n\right) }\right) \right] ,  \label{8} \\
D_{2} &=&\mu \left\vert E_{2}\left( t\right) \right\vert ^{2}=\mu \left\vert
A_{S}\right\vert ^{2}\left[ 1+\sin \left( \lambda _{k}^{\left( 1\right)
}-\lambda _{k}^{\left( n\right) }\right) \right] ,  \nonumber
\end{eqnarray}%
where $\mu $ is the parameter related to the sensitivity of photodetectors.
Finally, after correlation analysis of the two electric signals, we can
obtain as follow 
\begin{equation}
E=\left\langle D_{1}D_{2}\right\rangle =\frac{\mu ^{2}\left\vert
A_{S}\right\vert ^{4}\Delta T}{2}\sum\limits_{k=1}^{8}\left[ 1+\cos 2\left(
\lambda _{k}^{\left( 1\right) }-\lambda _{k}^{\left( n\right) }\right) %
\right] =\left\{ 
\begin{array}{c}
8\mu ^{2}\left\vert A_{S}\right\vert ^{4}\Delta T,\ n=1 \\ 
4\mu ^{2}\left\vert A_{S}\right\vert ^{4}\Delta T,\ n\neq 1%
\end{array}%
\right. ,  \label{9}
\end{equation}%
where $\Delta T$ is the PPS time slot. In addition to a constant, the result
satisfies the orthogonality of PPS.

\begin{figure}[htbp]
\centering\includegraphics[height=2.0263in, width=3.6867in]{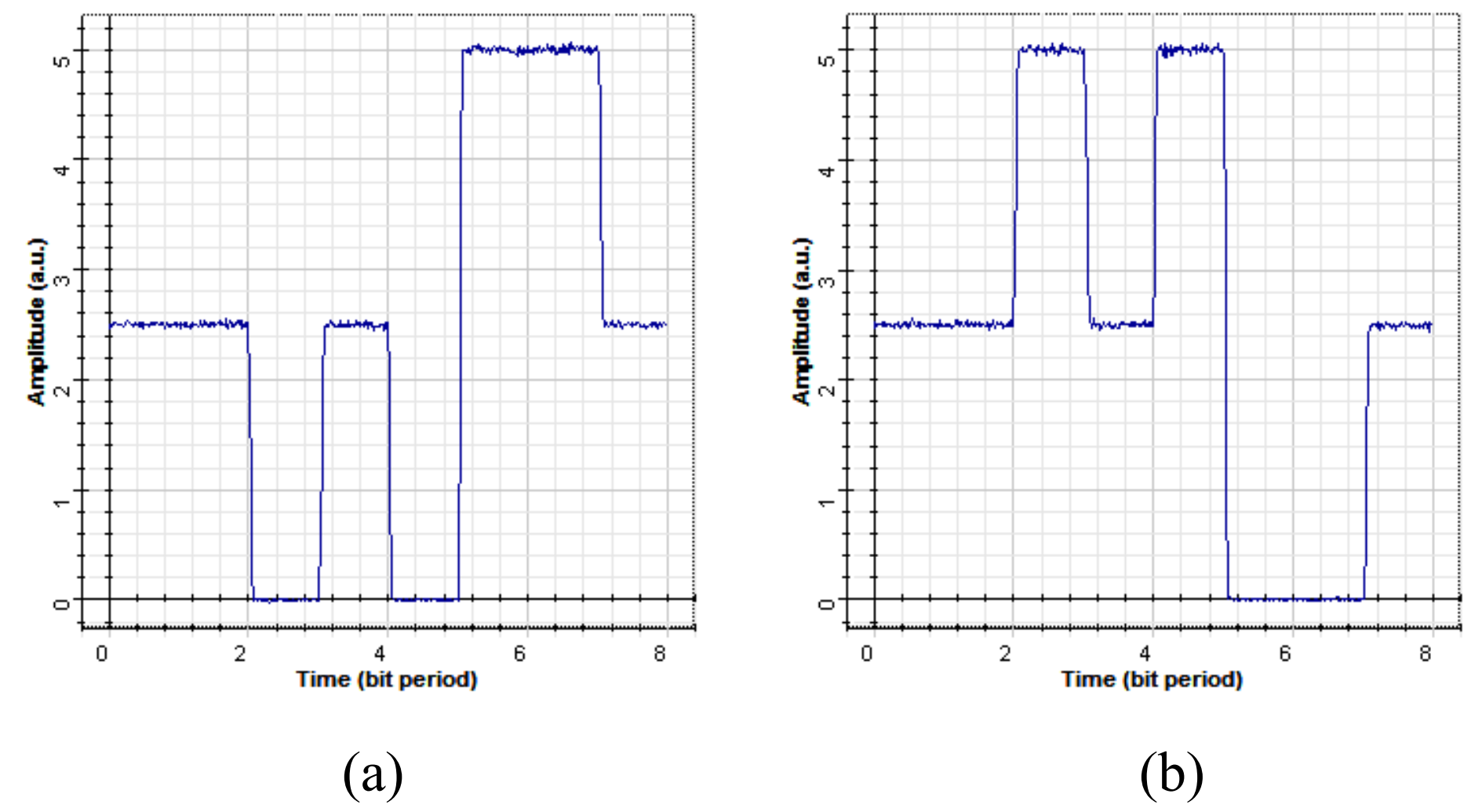}
\caption{The electric signals of $D_{1}$(a) and $D_{2}$(b) are shown when
the sequence of the LO field is $\protect\lambda ^{\left( 5\right) }$.}
\label{fig4}
\end{figure}

\begin{figure}[tbph]
\centering\includegraphics[height=1.9173in, width=3.979in]{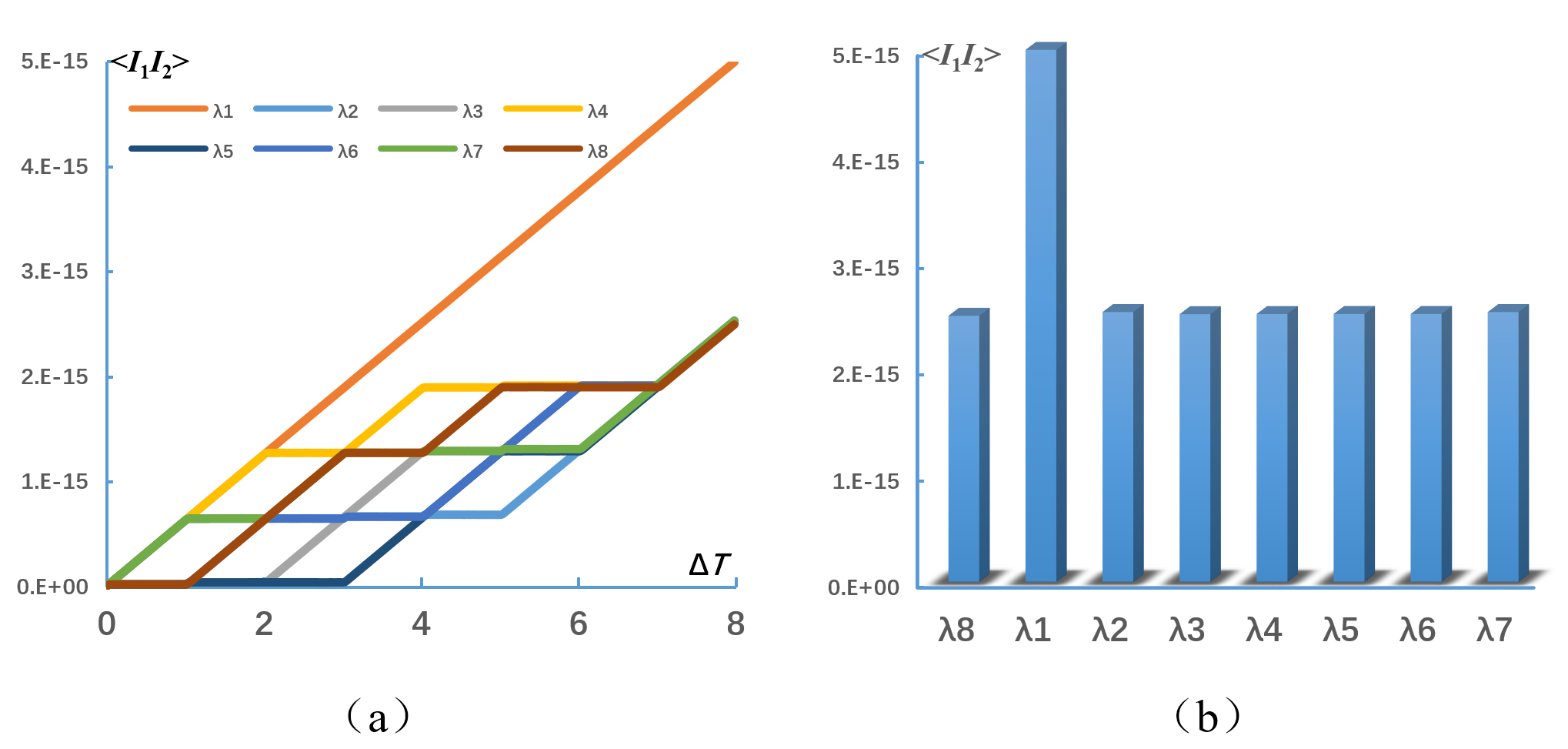}
\caption{The correlation analysis results are shown when the SO field with $%
\protect\lambda ^{\left( 1\right) }$ and the LO fields with\ $\protect%
\lambda ^{\left( 1\right) }\sim \protect\lambda ^{\left( 8\right) }$
(corresponding to $\protect\lambda 1\sim \protect\lambda 8$), where (a) is
the correlation intergrals vary with the sequence time slots $\Delta T$, and
(b) is the final correlation results.}
\label{fig5}
\end{figure}

To verify the above scheme, we utilize the software OPTISYSTEM to
numerically simulate it. Fig. \ref{fig4} shows the electric signals of two
photodetectors within a sequence period. Fig. \ref{fig5} shows the
correlation analysis results of the SO field and the LO field modulated with
different PPSs. We can find out the correlation result is the largest when
the SO and LO fields modulated with the same PPS. Hence, the orthogonality
of PPSs can be used to distinguish the optical fields with different PPSs.

\subsection{Similarities between an optical field and a single-particle
quantum state \label{Sec II.C}}

We note the similarities between Maxwell equation and Schr\"{o}dinger
equation \cite{Fu}. In fact, some properties utilized in quantum information
are wave properties, where the wave might not be a quantum wave \cite%
{Spreeuw}. Analogously to quantum states, optical fields also obey a
superposition principle, and can be transformed to any superposition state
by unitary transformations. Those analogous properties make possible the
analogies to quantum states using\textbf{\ }polarization or transverse modes
of optical fields \cite{Spreeuw,Fu,dragoman,Lee}\textbf{.}

We first consider two orthogonal modes (polarization or transverse) of
optical fields, as the optical analogies to quantum bits (qubits) $%
\left\vert 0\right\rangle $\ and\ $\left\vert 1\right\rangle $ 
\begin{equation}
\left\vert 0\right\rangle =\left( 
\begin{array}{c}
1 \\ 
0%
\end{array}%
\right) ,\left\vert 1\right\rangle =\left( 
\begin{array}{c}
0 \\ 
1%
\end{array}%
\right) .  \label{10}
\end{equation}%
Thus, any quantum state of a single particle\ can be imitated by the mode
superposition of optical field as follow 
\begin{equation}
\left\vert \psi \right\rangle =\alpha \left\vert 0\right\rangle +\beta
\left\vert 1\right\rangle ,  \label{11}
\end{equation}%
where $\left\vert \alpha \right\vert ^{2}+\left\vert \beta \right\vert
^{2}=1,\left( \alpha ,\beta \in 
\mathbb{C}
\right) $. Obviously, all the mode superposition can also span a Hilbert
space. We can transform a mode state to any other state using the unitary
transformation as follow \cite{Nielsen} 
\begin{equation}
\hat{U}\left( \alpha ,\beta ,\gamma ,\delta \right) =e^{i\alpha }\left( 
\begin{array}{cc}
e^{-i\beta /2} & 0 \\ 
0 & e^{i\beta /2}%
\end{array}%
\right) \left( 
\begin{array}{cc}
\cos \frac{\gamma }{2} & -\sin \frac{\gamma }{2} \\ 
\sin \frac{\gamma }{2} & \cos \frac{\gamma }{2}%
\end{array}%
\right) \left( 
\begin{array}{cc}
e^{-i\delta /2} & 0 \\ 
0 & e^{i\delta /2}%
\end{array}%
\right) ,  \label{12}
\end{equation}%
where $\alpha ,\beta ,\gamma ,\delta $ are real-valued. The modes $%
\left\vert 0\right\rangle $ and $\left\vert 1\right\rangle $ can be
transformed to any mode superpositions by using $\hat{U}\left( \alpha ,\beta
,\gamma ,\delta \right) $, respectively, as follows%
\begin{eqnarray}
\hat{U}\left( \alpha ,\beta ,\gamma ,\delta \right) \left\vert
0\right\rangle &=&e^{i\left( \alpha -\delta /2\right) }\left( e^{-i\beta
/2}\cos \frac{\gamma }{2}\left\vert 0\right\rangle +e^{i\beta /2}\sin \frac{%
\gamma }{2}\left\vert 1\right\rangle \right) ,  \label{13} \\
\hat{U}\left( \alpha ,\beta ,\gamma ,\delta \right) \left\vert
1\right\rangle &=&e^{i\left( \alpha +\delta /2\right) }\left( -e^{-i\beta
/2}\sin \frac{\gamma }{2}\left\vert 0\right\rangle +e^{i\beta /2}\cos \frac{%
\gamma }{2}\left\vert 1\right\rangle \right) .  \nonumber
\end{eqnarray}

Now, we consider some optical devices with one input and two outputs, such
as a beam splitter or a mode splitter, which split one input field $%
\left\vert \psi _{in}\right\rangle =\alpha \left\vert 0\right\rangle +\beta
\left\vert 1\right\rangle $ into two output fields $\left\vert \psi
_{out}^{\left( a\right) }\right\rangle $ and $\left\vert \psi _{out}^{\left(
b\right) }\right\rangle $. For the case of the beam splitter, the output
fields are $\left\vert \psi _{out}^{\left( a\right) }\right\rangle
=C_{a}\left( \alpha \left\vert 0\right\rangle +\beta e^{i\phi
_{a}}\left\vert 1\right\rangle \right) $ and $\left\vert \psi _{out}^{\left(
b\right) }\right\rangle =C_{b}\left( \alpha \left\vert 0\right\rangle +\beta
e^{i\phi _{b}}\left\vert 1\right\rangle \right) $ with an arbitrary power
ratio $\left\vert C_{a}\right\vert ^{2}:\left\vert C_{b}\right\vert ^{2}$
between the output beams, where $\phi _{a,b}$ are the additional phases due
to the splitter. For the case of the mode splitter, the output fields are $%
\left\vert \psi _{out}^{\left( a\right) }\right\rangle =\alpha e^{i\phi
_{a}}\left\vert 0\right\rangle $ and $\left\vert \psi _{out}^{\left(
b\right) }\right\rangle =\beta e^{i\phi _{b}}\left\vert 1\right\rangle $,
where $\phi _{a,b}$ are also the additional phases. Conversely, the devices
can act as a beam coupler or a mode coupler in which beams or modes from two
inputs are combined into the one output.

\section{Optical analogies to\textbf{\ }multiparticle quantum states \label%
{Sec III}}

In this section, we discuss optical analogies to multiparticle quantum
states using optical fields modulated with PPSs. We first demonstrate that $%
N $ optical fields modulated with $N$\ different PPSs can span a $N2^{N}$
dimensional Hilbert space that contains a tensor product structure \cite{Fu1}%
. Then, we introduce a phase ensemble model to imitate the quantum ensemble.
Further, by performing coherent demodulation scheme, we can obtain a mode
status matrix of the optical fields. Based on the mode status matrix, we
propose a sequential cycle permutation mechanism (SCPM) to imitate some
typical quantum states, such as the product state, Bell states, GHZ state
and W state.

\subsection{Ensemble model labeled by pseudorandom phase sequences \label%
{Sec III.A}}

In \cite{Fu1}, an effective simulation of quantum entanglement using optical
fields modulated with PPSs was discussed. In this paper, we will promote
this proposal further.

Referring from the concept of quantum ensemble, we propose a new concept of
a pseudorandom phase ensemble model. A phase ensemble is defined as a large
number of same optical fields modulated with different PPSs $\lambda ^{(i)}$%
, which are labeled by the phase units $\theta _{k}$ of $\lambda ^{(i)}$. A
phase ensemble is discrete if the phase unit $\theta _{k}$ is a uniformly
distributed discrete value within $[0,2\pi ]$. Then we can define that a
phase ensemble is complete if finite phase units are ergodic. According to
m-sequence theory, the occurrence of each value in a sequence is the same.
The phase ensemble is obviously ergodic in finite length. Clearly, we can
conclude that optical fields modulated with PPSs constitute a complete
discrete phase ensemble.

An ensemble average is defined as weighted average of any sequence $A^{(i)}$
within a sequence period as follow%
\begin{equation}
\bar{A}=\frac{1}{N}\sum\limits_{k=1}^{N}A^{(i)}=\frac{1}{N}%
\sum\limits_{k=1}^{N}A_{k}^{(i)},  \label{14}
\end{equation}%
where $A_{k}^{(i)}$ is a sequence unit labled by the phase unit $\theta _{k}$%
. A normalized correlation for two sequences $A^{(i)}$ and $A^{(j)}$ is
defined as%
\begin{equation}
E\left( A^{(i)},A^{(j)}\right) =\frac{1}{N}\sum%
\limits_{k=1}^{N}A_{k}^{(i)}A_{k}^{(j)\ast },  \label{15}
\end{equation}%
where $A_{k}^{(i)},A_{k}^{(j)}$ are the sequence units of $A^{(i)},A^{(j)}$
labled by the phase unit $\theta _{k}$, respectively.

\subsubsection{Hilbert space of basis states in the phase ensemble model 
\label{Sec III.A.1}}

Now we discuss a Hilbert space spanned by optical fields modulated with
PPSs. There are two orthogonal modes (polarization or transverse) of the
optical field, which are denoted by $\left\vert 0\right\rangle $ and $%
\left\vert 1\right\rangle $, respectively. Thus, a qubit state $\left\vert
\psi \right\rangle =\alpha \left\vert 0\right\rangle +\beta \left\vert
1\right\rangle $ can be expressed by the mode superposition, where $|\alpha
|^{2}+|\beta |^{2}=1,(\alpha ,\beta \in C)$. Obviously, each mode
superposition can span a two$\ $dimensional Hilbert space. Choosing any $N$
PPSs from the set $\Xi $ to modulate $N$ optical fields, we can obtain the
states expressed as follows%
\begin{equation}
\begin{array}{c}
\left\vert \psi _{1}\right\rangle =e^{i\lambda ^{(1)}}\left( \alpha
_{1}\left\vert 0\right\rangle +\beta _{1}\left\vert 1\right\rangle \right) ,
\\ 
\vdots \\ 
\left\vert \psi _{N}\right\rangle =e^{i\lambda ^{(N)}}\left( \alpha
_{N}\left\vert 0\right\rangle +\beta _{N}\left\vert 1\right\rangle \right) .%
\end{array}
\label{16}
\end{equation}%
According to the properties of PPSs and Hilbert space, we can define the
inner product of any two fields $\left\vert \psi _{a}\right\rangle $ and $%
\left\vert \psi _{b}\right\rangle $. We obtain the orthogonal property as
follow%
\begin{equation}
\left\langle \psi _{a}|\psi _{b}\right\rangle =\frac{1}{N}%
\sum\limits_{k=1}^{N}e^{i(\lambda _{k}^{(b)}-\lambda _{k}^{(a)})}(\alpha
_{a}^{\ast }\alpha _{b}+\beta _{a}^{\ast }\beta _{b})=\left\{ 
\begin{array}{c}
1,a=b, \\ 
0,a\neq b,%
\end{array}%
\right.  \label{17}
\end{equation}%
where $\lambda _{k}^{(a)},\lambda _{k}^{(b)}$ are the $k$-th units of $%
\lambda ^{(a)}$ and $\lambda ^{(b)}$, respectively. The orthogonal property
supports the tensor product structure of the multiple fields \cite{Fu1}.

A formal product state $\left\vert \Psi \right\rangle $ for the $N$ optical
fields is defined as being a direct product of $\left\vert \psi
_{n}\right\rangle $,%
\begin{equation}
\left\vert \Psi \right\rangle \equiv \left\vert \psi _{1}\right\rangle
\otimes \left\vert \psi _{2}\right\rangle \otimes \cdots \otimes \left\vert
\psi _{N}\right\rangle .  \label{18}
\end{equation}%
According to the definition, $N$ optical fields of Eq. (\ref{16}) can be
expressed as the following state%
\begin{equation}
\left\vert \Psi \right\rangle =e^{i\lambda ^{S}}\left( \alpha _{1}\left\vert
0\right\rangle +\beta _{1}\left\vert 1\right\rangle \right) \otimes \left(
\alpha _{2}\left\vert 0\right\rangle +\beta _{2}\left\vert 1\right\rangle
\right) \otimes \cdots \otimes \left( \alpha _{N}\left\vert 0\right\rangle
+\beta _{N}\left\vert 1\right\rangle \right) ,  \label{19}
\end{equation}%
where $\lambda ^{S}=\sum\nolimits_{n=1}^{N}\lambda ^{\left( n\right) }$. By
using an array of several mode transformation gates, the optical field $%
\left\vert \psi _{n}\right\rangle $ can be transformed from Eq. (\ref{16})
to the following general state 
\begin{equation}
\left\vert \psi _{n}\right\rangle =\sum\limits_{i=1}^{N}\alpha _{n}^{\left(
i\right) }e^{i\lambda ^{\left( i\right) }}\left\vert 0\right\rangle
+\sum\limits_{j=1}^{N}\beta _{n}^{\left( j\right) }e^{i\lambda ^{\left(
j\right) }}\left\vert 1\right\rangle .  \label{20}
\end{equation}%
Then, the formal product state $\left\vert \Psi \right\rangle $ can be
written as%
\begin{equation}
\left\vert \Psi \right\rangle =\left( \sum\limits_{i=1}^{N}\alpha
_{1}^{\left( i\right) }e^{i\lambda ^{\left( i\right) }}\left\vert
0\right\rangle +\sum\limits_{j=1}^{N}\beta _{1}^{\left( j\right)
}e^{i\lambda ^{\left( j\right) }}\left\vert 1\right\rangle \right) \otimes
\cdots \otimes \left( \sum\limits_{i=1}^{N}\alpha _{N}^{\left( i\right)
}e^{i\lambda ^{\left( i\right) }}\left\vert 0\right\rangle
+\sum\limits_{j=1}^{N}\beta _{N}^{\left( j\right) }e^{i\lambda ^{\left(
j\right) }}\left\vert 1\right\rangle \right) .  \label{21}
\end{equation}%
Further, we can obtain each item of the superposition of $\left\vert \Psi
\right\rangle $ as follows%
\begin{equation}
\begin{array}{c}
C_{00\cdots 0}\left\vert 00\cdots 0\right\rangle =\left[ \left(
\sum\limits_{i=1}^{N}\alpha _{1}^{\left( i\right) }e^{i\lambda ^{\left(
i\right) }}\right) \left( \sum\limits_{i=1}^{N}\alpha _{2}^{\left( i\right)
}e^{i\lambda ^{\left( i\right) }}\right) \cdots \left(
\sum\limits_{i=1}^{N}\alpha _{N}^{\left( i\right) }e^{i\lambda ^{\left(
i\right) }}\right) \right] \left\vert 00\cdots 0\right\rangle , \\ 
C_{00\cdots 1}\left\vert 00\cdots 1\right\rangle =\left[ \left(
\sum\limits_{i=1}^{N}\alpha _{1}^{\left( i\right) }e^{i\lambda ^{\left(
i\right) }}\right) \left( \sum\limits_{i=1}^{N}\alpha _{2}^{\left( i\right)
}e^{i\lambda ^{\left( i\right) }}\right) \cdots \left(
\sum\limits_{j=1}^{N}\beta _{N}^{\left( j\right) }e^{i\lambda ^{\left(
j\right) }}\right) \right] \left\vert 00\cdots 1\right\rangle , \\ 
\vdots \\ 
C_{11\cdots 1}\left\vert 11\cdots 1\right\rangle =\left[ \left(
\sum\limits_{j=1}^{N}\beta _{1}^{\left( j\right) }e^{i\lambda ^{\left(
j\right) }}\right) \left( \sum\limits_{j=1}^{N}\beta _{2}^{\left( j\right)
}e^{i\lambda ^{\left( i\right) }}\right) \cdots \left(
\sum\limits_{j=1}^{N}\beta _{N}^{\left( j\right) }e^{i\lambda ^{\left(
j\right) }}\right) \right] \left\vert 11\cdots 1\right\rangle ,%
\end{array}
\label{22}
\end{equation}%
where $\left\vert i_{1}\ldots i_{N}\right\rangle \equiv \left\vert
i_{1}\right\rangle \otimes \ldots \otimes \left\vert i_{N}\right\rangle
,\left( i_{n}=0\ or\ 1\right) $. According to the closure property, the PPSs
of $C_{i_{1}i_{2}\cdots i_{N}}$ remain in the set $\Xi $, which means $%
C_{i_{1}i_{2}\cdots i_{N}}=\sum\limits_{j=1}^{N}C_{i_{1}i_{2}\cdots
i_{N}}^{\left( j\right) }e^{i\lambda ^{\left( j\right) }}$. Therefore, we
obtain the following conclusion that the formal product state $\left\vert
\Psi \right\rangle $ can be expressed a linear superposition in the Hilbert
space with the basis $\left\{ \left. e^{i\lambda ^{\left( j\right)
}}\left\vert i_{1}i_{2}\cdots i_{N}\right\rangle \right\vert e^{i\lambda
^{\left( j\right) }}\in \Omega ,j=1\cdots N,i_{n}=0\;or\;1\right\} $ as
follows%
\begin{equation}
\left\vert \Psi \right\rangle =\sum\limits_{i_{1}=0}^{1}\cdots
\sum\limits_{i_{N}=0}^{1}\left[ \sum\limits_{j=1}^{N}C_{i_{1}i_{2}\cdots
i_{N}}^{\left( j\right) }e^{i\lambda ^{\left( j\right) }}\left\vert
i_{1}i_{2}\cdots i_{N}\right\rangle \right] ,  \label{23}
\end{equation}%
where $C_{i_{1}i_{2}\cdots i_{N}}^{\left( j\right) }$ denotes a total of $%
N2^{N}$ coefficients. Apparently, these fields span the $N2^{N}$ dimensional
Hilbert space.

\subsubsection{The state inseparability in the phase ensemble model \label%
{Sec III.A.2}}

The nonlocality correlation of quantum entanglement is demonstrated as the
inseparability of any two-party quantum states. The correlation depends on
the ensemble summaries of many measurement results. Similarly, we discuss
the state inseparability demonstrated in optical fields based on the phase
ensemble framework.

In order to research the properties, we first classify the subsets of the
formal product state according to the PPSs. A consensus PPS sub-state (CPSS)
is defined as being items with the same PPS in the formal product state $%
\left\vert \Psi \right\rangle $. A single PPS sub-state (SPSS) is defined as
being each of the items, except all consensus PPS sub-states, in the formal
product state $\left\vert \Psi \right\rangle $. For example, we assume that
the PPS $\lambda ^{\left( s_{1}\right) }$ corresponds to the first CPSS set $%
\left\{ \left\vert S_{1}^{\left( 1\right) }\right\rangle ,\left\vert
S_{2}^{\left( 1\right) }\right\rangle \cdots ,\left\vert S_{N_{1}}^{\left(
1\right) }\right\rangle \right\} $, ..., the PPS $\lambda ^{\left(
s_{p}\right) }$ corresponds to the $p$-th CPSS set $\left\{ \left\vert
S_{1}^{\left( p\right) }\right\rangle ,\left\vert S_{2}^{\left( p\right)
}\right\rangle \cdots ,\left\vert S_{N_{p}}^{\left( p\right) }\right\rangle
\right\} $, and other $N^{\prime }$ SPSSs in the formal product state $%
\left\vert \Psi \right\rangle $. Thus $\left\vert \Psi \right\rangle $ can
be expressed as%
\begin{equation}
\left\vert \Psi \right\rangle =e^{i\lambda ^{\left( s_{1}\right)
}}\sum\limits_{i=1}^{N_{1}}C_{i}^{\left( 1\right) }\left\vert S_{i}^{\left(
1\right) }\right\rangle +\cdots +e^{i\lambda ^{\left( s_{p}\right)
}}\sum\limits_{i=1}^{N_{p}}C_{i}^{\left( p\right) }\left\vert S_{i}^{\left(
p\right) }\right\rangle +\sum\limits_{j=1}^{N^{\prime }}C_{j}e^{i\lambda
^{(j)}}\left\vert x_{j}\right\rangle ,  \label{24}
\end{equation}%
where $\lambda ^{\left( s_{1}\right) },\cdots ,\lambda ^{\left( s_{p}\right)
}$, and $\lambda ^{\left( j\right) }$ are the distinct PPSs and $%
C_{i}^{\left( m\right) },C_{j}$ are the superposition coefficients of the
CPSSs and the SPSSs, respectively. Further, we introduce the definition of
the density matrix%
\begin{eqnarray}
\rho &\equiv &\left\vert \Psi \right\rangle \left\langle \Psi \right\vert
\label{25} \\
&=&\left( e^{i\lambda ^{(s_{1})}}\sum\limits_{i=1}^{N_{1}}C_{i}^{\left(
1\right) }\left\vert S_{i}^{\left( 1\right) }\right\rangle +\cdots
+e^{i\lambda ^{(s_{p})}}\sum\limits_{i=1}^{N_{p}}C_{i}^{\left( p\right)
}\left\vert S_{i}^{\left( p\right) }\right\rangle
+\sum\limits_{j=1}^{N^{\prime }}C_{j}e^{i\lambda ^{(j)}}\left\vert
x_{j}\right\rangle \right)  \nonumber \\
&&\times \left( e^{-i\lambda
^{(s_{1})}}\sum\limits_{i=1}^{N_{1}}C_{i}^{\left( 1\right) \ast
}\left\langle S_{i}^{\left( 1\right) }\right\vert +\cdots +e^{-i\lambda
^{(s_{p})}}\sum\limits_{i=1}^{N_{p}}C_{i}^{\left( p\right) \ast
}\left\langle S_{i}^{\left( p\right) }\right\vert
+\sum\limits_{j=1}^{N^{\prime }}C_{j}^{\ast }e^{-i\lambda
^{(j)}}\left\langle x_{j}\right\vert \right) ,  \nonumber
\end{eqnarray}%
which can be simplified to%
\begin{eqnarray}
\rho &=&\sum\limits_{n=1}^{2^{N}}\left\vert C_{n}\right\vert ^{2}\left\vert
x_{n}\right\rangle \left\langle x_{n}\right\vert
+\sum\limits_{m=1}^{p}\sum\limits_{i\neq i^{\prime }=1}^{N_{p}}\left(
C_{i}^{\left( m\right) }C_{i^{\prime }}^{\left( m\right) \ast }\left\vert
S_{i}^{\left( m\right) }\right\rangle \left\langle S_{i^{\prime }}^{\left(
m\right) }\right\vert +C_{i^{\prime }}^{\left( m\right) }C_{i}^{\left(
m\right) \ast }\left\vert S_{i^{^{\prime }}}^{\left( m\right) }\right\rangle
\left\langle S_{i}^{\left( m\right) }\right\vert \right)  \label{26} \\
&&+\sum\limits_{m\neq
n=1}^{p}\sum\limits_{i=1}^{N_{m}}\sum\limits_{j=1}^{N_{n}}\left(
C_{i}^{\left( m\right) }C_{j}^{\left( n\right) \ast }e^{i\left( \lambda
^{(s_{m})}-\lambda ^{(s_{n})}\right) }\left\vert S_{i}^{\left( m\right)
}\right\rangle \left\langle S_{j}^{\left( n\right) }\right\vert
+C_{j}^{\left( n\right) }C_{i}^{\left( m\right) \ast }e^{i\left( \lambda
^{(s_{n})}-\lambda ^{(s_{m})}\right) }\left\vert S_{j}^{\left( n\right)
}\right\rangle \left\langle S_{i}^{\left( m\right) }\right\vert \right) 
\nonumber \\
&&+\sum\limits_{m=1}^{p}\sum\limits_{i=1}^{N_{p}}\sum\limits_{j=1}^{N^{%
\prime }}\left( C_{j}C_{i}^{\left( m\right) \ast }e^{i\left( \lambda
^{(j)}-\lambda ^{(s_{m})}\right) }\left\vert x_{j}\right\rangle \left\langle
S_{i}^{\left( m\right) }\right\vert +C_{i}^{\left( m\right) }C_{j}^{\ast
}e^{i\left( \lambda ^{(s_{m})}-\lambda ^{(j)}\right) }\left\vert
S_{i}^{\left( m\right) }\right\rangle \left\langle x_{j}\right\vert \right) 
\nonumber \\
&&+\sum\limits_{j\neq j^{\prime }=1}^{N^{\prime }}\left( C_{j}C_{j^{\prime
}}^{\ast }e^{i\left( \lambda ^{(j)}-\lambda ^{(j^{\prime })}\right)
}\left\vert x_{j}\right\rangle \left\langle x_{j^{\prime }}\right\vert
+C_{j^{\prime }}C_{j}^{\ast }e^{i\left( \lambda ^{(j^{\prime })}-\lambda
^{(j)}\right) }\left\vert x_{j^{\prime }}\right\rangle \left\langle
x_{j}\right\vert \right) .  \nonumber
\end{eqnarray}%
where $C_{n}$ denote all coefficients of the CPSSs and SPSSs. Noteworthy,
the last three items must retain the PPSs due to the distinct PPSs and their
closure property. According to the definition of phase ensemble average Eq. (%
\ref{14}), an ensemble-averaged density matrix (EADM) can be defined $\tilde{%
\rho}\equiv \frac{1}{N}\sum\limits_{k=1}^{N}\rho $ and obtained as follow%
\begin{equation}
\tilde{\rho}=\sum\limits_{n=1}^{2^{N}}\left\vert C_{n}\right\vert
^{2}\left\vert x_{n}\right\rangle \left\langle x_{n}\right\vert
+\sum\limits_{m=1}^{p}\sum\limits_{i\neq i^{\prime }=1}^{N_{m}}\left(
C_{i}^{\left( m\right) }C_{i^{\prime }}^{\left( m\right) \ast }\left\vert
S_{i}^{\left( m\right) }\right\rangle \left\langle S_{i^{\prime }}^{\left(
m\right) }\right\vert +C_{i^{\prime }}^{\left( m\right) }C_{i}^{\left(
m\right) \ast }\left\vert S_{i^{\prime }}^{\left( m\right) }\right\rangle
\left\langle S_{i}^{\left( m\right) }\right\vert \right) ,  \label{27}
\end{equation}%
due to all PPSs satisfying $\sum\limits_{k=1}^{N}e^{i\lambda _{k}^{(m)}}=0$.
Note that all off-diagonal elements of the EADM are contributed from the
CPSSs after ensemble averaged. Also, it shows that the EADM $\tilde{\rho}$
might not be expressed in terms of direct products of the states $\left\vert
\psi _{n}\right\rangle $ due to only non-diagonal term $C_{i}^{\left(
m\right) }C_{i^{\prime }}^{\left( m\right) \ast }\left\vert S_{i}^{\left(
m\right) }\right\rangle \left\langle S_{i^{\prime }}^{\left( m\right)
}\right\vert +C_{i^{\prime }}^{\left( m\right) }C_{i}^{\left( m\right) \ast
}\left\vert S_{i^{\prime }}^{\left( m\right) }\right\rangle \left\langle
S_{i}^{\left( m\right) }\right\vert $ contributed from CPSSs remaining,
simialr to the case of quantum entanglement states.

In the phase ensemble, the expectation value of an arbitrary operator $\hat{P%
}$ can be defined as follow%
\begin{equation}
\bar{P}\equiv \frac{1}{N}\sum\limits_{k=1}^{N}tr\left( \rho \hat{P}\right) .
\label{28}
\end{equation}%
Further, according to the exchange of summation and matrix trace, the
expectation value can be simplified to%
\begin{eqnarray}
\bar{P} &=&tr\left[ \left( \frac{1}{N}\sum\limits_{k=1}^{N}\rho \right) \hat{%
P}\right] =tr\left( \tilde{\rho}\hat{P}\right)  \label{29} \\
&=&\sum\limits_{n=1}^{2^{N}}\left\vert C_{n}\right\vert ^{2}\left\langle
x_{n}\right\vert \hat{P}\left\vert x_{n}\right\rangle
+\sum\limits_{m=1}^{p}\sum\limits_{i\neq i^{\prime }=1}^{N_{m}}\left(
C_{i}^{\left( m\right) }C_{i^{\prime }}^{\left( m\right) \ast }\left\langle
S_{i^{\prime }}^{\left( m\right) }\right\vert \hat{P}\left\vert
S_{i}^{\left( m\right) }\right\rangle +C_{i^{\prime }}^{\left( m\right)
}C_{i}^{\left( m\right) \ast }\left\langle S_{i}^{\left( m\right)
}\right\vert \hat{P}\left\vert S_{i^{\prime }}^{\left( m\right)
}\right\rangle \right) .  \nonumber
\end{eqnarray}

It is worth noting that the inseparability is demostrated due to only
off-diagonal items contributed from CPSSs remaining in the correlation
measurement. Using this property, we can imitate a quantum state that
formally agrees with CPSSs in the formal product state under the phase
ensemble framework.

\subsubsection{The minimum complete phase ensemble \label{Sec III.A.3}}

In the phase ensemble model, we are interested in the simplest model that
requires minimal resources to be constructed. We define that a minimum
complete phase ensemble has the least CPSS set, in which the state has only
one CPSS set. By the definition, the state in Eq. (\ref{16}) is a type of
the minimal complete state. The CPSSs of the minimum complete state
correspond to one and only one PPS $\lambda
^{S}=\sum\limits_{n=1}^{N}\lambda ^{(n)}$, which is the sum of all used
PPSs. The simplest case is each field modulated with a different PPS as
follow 
\begin{equation}
\left( e^{i\lambda ^{(1)}}\left\vert i_{1}\right\rangle \right) \otimes
\left( e^{i\lambda ^{(2)}}\left\vert i_{2}\right\rangle \right) \cdots
\otimes \left( e^{i\lambda ^{(N)}}\left\vert i_{N}\right\rangle \right)
=e^{i\sum\nolimits_{n=1}^{N}\lambda ^{(n)}}\left\vert i_{1}i_{2}\cdots
i_{N}\right\rangle =e^{i\lambda ^{S}}\left\vert i_{1}i_{2}\cdots
i_{N}\right\rangle .  \label{30}
\end{equation}%
Now, we can express a minimum complete state $\left\vert \Psi \right\rangle $
as follow%
\begin{equation}
\left\vert \Psi \right\rangle =e^{i\lambda ^{S}}\sum\limits_{i=1}^{N^{\prime
}}C_{i}\left\vert x_{i}\right\rangle +\sum\limits_{j=1}^{N^{\prime \prime
}}C_{j}e^{i\lambda ^{(j)}}\left\vert x_{j}\right\rangle =e^{i\lambda
^{S}}\left( \sum\limits_{i=1}^{N^{\prime }}C_{i}\left\vert
x_{i}\right\rangle +\sum\limits_{j=1}^{N^{\prime \prime }}C_{j}e^{i\left(
\lambda ^{(j)}-\lambda ^{S}\right) }\left\vert x_{j}\right\rangle \right) ,
\label{31}
\end{equation}%
where $e^{i\lambda ^{(j)}}\left\vert x_{j}\right\rangle $ correspond to all
SPSSs. According to the analysis in the last subsection, the EADM can be
obtained%
\begin{equation}
\tilde{\rho}=\sum\limits_{n=1}^{2^{N}}\left\vert C_{n}\right\vert
^{2}\left\vert x_{n}\right\rangle \left\langle x_{n}\right\vert
+\sum\limits_{i\neq i^{\prime }=1}^{N^{\prime }}\left( C_{i}C_{i^{\prime
}}^{\ast }\left\vert x_{i}\right\rangle \left\langle x_{i^{\prime
}}\right\vert +C_{i^{\prime }}C_{i}^{\ast }\left\vert x_{i^{\prime
}}\right\rangle \left\langle x_{i}\right\vert \right) .  \label{32}
\end{equation}%
In conclusion, the minimum complete state satisfies the necessary conditions
for analogies to quantum states.

Considering all possible combinations, we can obtain $N!$ nonredundant
combinations of optical fields and PPSs. And all combinations can be
obtained in the same form $e^{i\lambda ^{S}}\left\vert i_{1}i_{2}\cdots
i_{N}\right\rangle $. Therefore, in the formal product state $\left\vert
\Psi \right\rangle $, a CPSS $e^{i\lambda ^{S}}\left\vert i_{1}i_{2}\cdots
i_{N}\right\rangle $ has $N!$ equivalent direct product decompositions.
There is a very important problem for our scheme that is not the lack of
resources but the redundancy of resources. In order to reduce the
redundancy, we have to introduce a simple and unique mechanism: a sequential
cycle permutation mechanism (SCPM) to realize the bijection imitation of
certain quantum states. The sequential cycle permutation is shown as follows%
\begin{eqnarray}
R_{1} &=&\left\{ \lambda ^{\left( 1\right) },\lambda ^{\left( 2\right)
},\cdots ,\lambda ^{\left( N\right) }\right\} ,  \label{33} \\
R_{2} &=&\left\{ \lambda ^{\left( 2\right) },\lambda ^{\left( 3\right)
},\cdots ,\lambda ^{\left( 1\right) }\right\} ,  \nonumber \\
&&\vdots  \nonumber \\
R_{N} &=&\left\{ \lambda ^{\left( N\right) },\lambda ^{\left( 1\right)
},\cdots ,\lambda ^{\left( N-1\right) }\right\} .  \nonumber
\end{eqnarray}%
It is clear that the sequential cycle permutation is a subset of the
sequential full permutations. According to the definition, the imitation
state obtained by using the sequential cycle permutation is obvious a
minimum complete state. According to Eq. (\ref{33}), each state corresponds
to the sequential cycle permutation as follows 
\begin{eqnarray}
R_{1} &:&\left( e^{i\lambda ^{(1)}}\left\vert i_{1}\right\rangle \right)
\otimes \left( e^{i\lambda ^{(2)}}\left\vert i_{2}\right\rangle \right)
\cdots \otimes \left( e^{i\lambda ^{(N)}}\left\vert i_{N}\right\rangle
\right) =e^{i\lambda ^{S}}\left\vert i_{1}i_{2}\cdots i_{N}\right\rangle ,
\label{34} \\
R_{2} &:&\left( e^{i\lambda ^{(2)}}\left\vert i_{1}\right\rangle \right)
\otimes \left( e^{i\lambda ^{(3)}}\left\vert i_{2}\right\rangle \right)
\cdots \otimes \left( e^{i\lambda ^{(1)}}\left\vert i_{N}\right\rangle
\right) =e^{i\lambda ^{S}}\left\vert i_{1}i_{2}\cdots i_{N}\right\rangle , 
\nonumber \\
&&\vdots  \nonumber \\
R_{N} &:&\left( e^{i\lambda ^{(N)}}\left\vert i_{1}\right\rangle \right)
\otimes \left( e^{i\lambda ^{(1)}}\left\vert i_{2}\right\rangle \right)
\cdots \otimes \left( e^{i\lambda ^{(N-1)}}\left\vert i_{N}\right\rangle
\right) =e^{i\lambda ^{S}}\left\vert i_{1}i_{2}\cdots i_{N}\right\rangle . 
\nonumber
\end{eqnarray}%
Hence, each sequential cycle permutation provides a subset of the minimum
complete states.

\subsection{Optical analogies to quantum entanglement \label{Sec III.B}}

Quantum entanglement is only defined for the Hilbert spaces that have a
rigorous tensor product structure in terms of subsystems \cite{Jozsa}. In
quantum mechanics, quantum entanglement cannot be expressed in terms of
direct products, but only is characterized by the correlation measurements.
The nonlocal correlations decided with Bell's inequality and GHZ's equality
criteria are the most fundamental property of quantum entanglement \cite%
{CHSH,GHZ}.

\subsubsection{Correlation analysis for optical analogies \label{Sec III.B.1}%
}

Here, it is necessary to introduce the correlation analysis analogy to
quantum measurement. In order to introduce correlation analysis, a
measurement operator $\hat{P}$ locally performed on $\left\vert \psi
\right\rangle $ is given%
\begin{equation}
\bar{P}(\theta )=\left\langle \psi \right\vert \hat{P}(\theta )\left\vert
\psi \right\rangle =\left( 
\begin{array}{cc}
\alpha ^{\ast } & \beta ^{\ast }%
\end{array}%
\right) \left( 
\begin{array}{cc}
0 & e^{i\theta } \\ 
e^{-i\theta } & 0%
\end{array}%
\right) \left( 
\begin{array}{c}
\alpha \\ 
\beta%
\end{array}%
\right) =\alpha ^{\ast }\beta e^{i\theta }+\alpha \beta ^{\ast }e^{-i\theta
}.  \label{39}
\end{equation}%
For convenience, coefficients $\alpha ,\beta $ are equal to $1/\sqrt{2}$,
yielding $\bar{P}(\theta )=\cos (\theta )$. Further we generalize $\hat{P}$
to the case of $N$ optical fields%
\begin{equation}
\hat{P}(\theta _{1},\ldots ,\theta _{N})=\hat{P}\left( \theta _{1}\right)
\otimes \hat{P}\left( \theta _{2}\right) \otimes \ldots \otimes \hat{P}%
\left( \theta _{N}\right) .  \label{40}
\end{equation}%
Then according to Eq. (\ref{29}), we obtain the correlation analysis for the
state of Eq. (\ref{31}) using $\hat{P}$ and the density matrix $\rho $ as
follow%
\begin{eqnarray}
E\left( \theta _{1},\ldots ,\theta _{N}\right) &=&\frac{1}{N}\sum_{k=1}^{N}Tr%
\left[ \rho \hat{P}(\theta _{1},\ldots ,\theta _{N})\right] =Tr\left[ \tilde{%
\rho}\hat{P}(\theta _{1},\ldots ,\theta _{N})\right]  \label{41} \\
&=&\sum_{n=1}^{2^{N}}|C_{n}|^{2}\left\langle x_{n}\right\vert \hat{P}%
\left\vert x_{n}\right\rangle +\sum_{i\neq i^{\prime }=1}^{N^{\prime
}}\left( C_{i}C_{i^{\prime }}^{\ast }\left\langle x_{i^{\prime }}\right\vert 
\hat{P}\left\vert x_{i}\right\rangle +C_{i^{\prime }}C_{i}^{\ast
}\left\langle x_{i}\right\vert \hat{P}\left\vert x_{i^{\prime
}}\right\rangle \right) .  \nonumber
\end{eqnarray}%
Eq. (\ref{41}) shows that only non-diagonal terms $\sum_{i\neq i^{\prime
}=1}^{N^{\prime }}\left( C_{i}C_{i^{\prime }}^{\ast }\left\langle
x_{i^{\prime }}\right\vert \hat{P}\left\vert x_{i}\right\rangle
+C_{i^{\prime }}C_{i}^{\ast }\left\langle x_{i}\right\vert \hat{P}\left\vert
x_{i^{\prime }}\right\rangle \right) $ contributed from CPSSs remain.

\subsubsection{\textit{Bell states of two particles}: Bell's inequality
criterion \label{Sec III.B.2}}

For convenience, we first consider two optical fields modulated with the
PPSs. Chosen any two PPSs of $\lambda ^{\left( a\right) }$ and $\lambda
^{\left( b\right) }$ from the set $\Xi $, two fields modulated with the PPSs
can be expressed as follows 
\begin{eqnarray}
\left\vert \psi _{a}\right\rangle &=&e^{i\lambda ^{\left( a\right) }}\left(
\alpha _{a}\left\vert 0\right\rangle +\beta _{a}\left\vert 1\right\rangle
\right) ,  \label{35} \\
\left\vert \psi _{b}\right\rangle &=&e^{i\lambda ^{\left( b\right) }}\left(
\alpha _{b}\left\vert 0\right\rangle +\beta _{b}\left\vert 1\right\rangle
\right) .  \nonumber
\end{eqnarray}%
where $\left\vert 0\right\rangle $ and $\left\vert 1\right\rangle $ are
assumed to be two orthogonal polarization modes, respectively. Here we
assume $\alpha _{a,b},\beta _{a,b}$ are equal to $1/\sqrt{2}$. The direct
product state of the two fields can be expressed as follows%
\begin{equation}
\left\vert \Psi \right\rangle =\left\vert \psi _{a}\right\rangle \otimes
\left\vert \psi _{b}\right\rangle =\frac{e^{i\left( \lambda ^{\left(
a\right) }+\lambda ^{\left( b\right) }\right) }}{2}\left( \left\vert
00\right\rangle +\left\vert 01\right\rangle +\left\vert 10\right\rangle
+\left\vert 11\right\rangle \right) ,  \label{38}
\end{equation}%
where $\lambda ^{\left( a\right) }+\lambda ^{\left( b\right) }$ remains in
the set $\Xi $ due to the closure property.

\begin{figure}[htbp]
\centering\includegraphics[height=1.7417in, width=3.1964in]{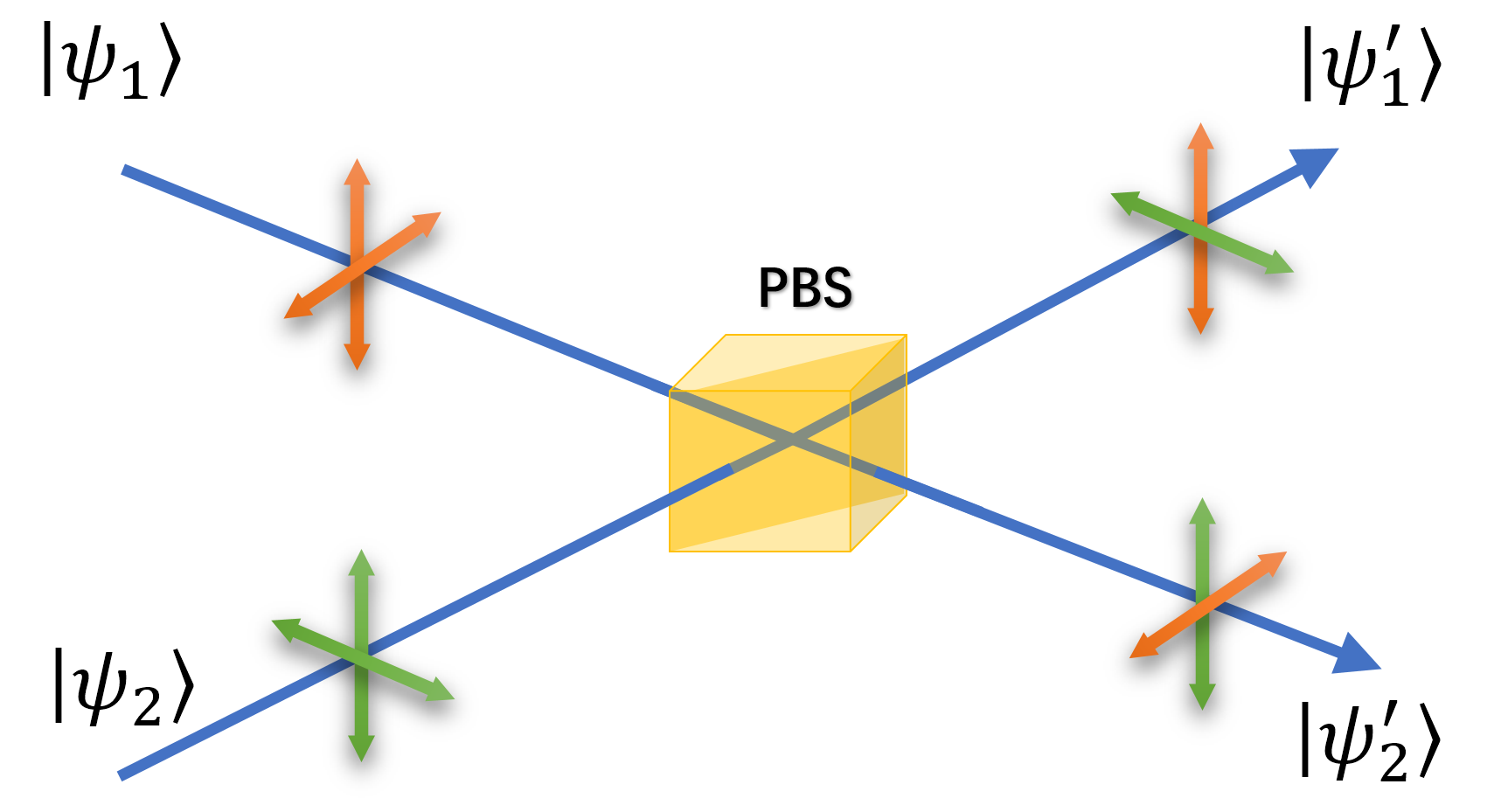}
\caption{The scheme to exchange modes of two optical fields of Eq. (\protect
\ref{35}) to imitate one of Bell states is shown, where PBS is$\ $%
polarization beam splitter.}
\label{fig7}
\end{figure}

By using a polarization beam splitter \cite{Fu,Fu1}, the modes $\left\vert
1\right\rangle $ of $\left\vert \psi _{a}\right\rangle $ and $\left\vert
\psi _{b}\right\rangle $ are exchanged as shown in Fig. \ref{fig7}. Then we
obtain the following fields%
\begin{eqnarray}
\left\vert \psi _{a}^{\prime }\right\rangle &=&\frac{1}{\sqrt{2}}\left(
e^{i\lambda ^{(a)}}\left\vert 0\right\rangle +e^{i\lambda ^{(b)}}\left\vert
1\right\rangle \right) =\frac{e^{i\lambda ^{(a)}}}{\sqrt{2}}\left(
\left\vert 0\right\rangle +e^{i\gamma ^{(a)}}\left\vert 1\right\rangle
\right) ,  \label{42} \\
\left\vert \psi _{b}^{\prime }\right\rangle &=&\frac{1}{\sqrt{2}}\left(
e^{i\lambda ^{(b)}}\left\vert 0\right\rangle +e^{i\lambda ^{(a)}}\left\vert
1\right\rangle \right) =\frac{e^{i\lambda ^{(b)}}}{\sqrt{2}}\left(
\left\vert 0\right\rangle +e^{i\gamma ^{(b)}}\left\vert 1\right\rangle
\right) ,  \nonumber
\end{eqnarray}%
where the relative phase sequences (RPSs) $\gamma ^{(a)}=-\gamma
^{(b)}=\lambda ^{(b)}-\lambda ^{(a)}$, and $\gamma ^{(a)}+\gamma ^{(b)}=0$.
The state $\left\vert \Psi \right\rangle $ is obtained%
\begin{equation}
\left\vert \Psi \right\rangle =\left\vert \psi _{a}^{\prime }\right\rangle
\otimes \left\vert \psi _{b}^{\prime }\right\rangle =\frac{e^{i\left(
\lambda ^{(a)}+\lambda ^{(b)}\right) }}{2}\left[ \left\vert 00\right\rangle
+\left\vert 11\right\rangle +e^{i\gamma ^{(a)}}\left\vert 10\right\rangle
+e^{i\gamma ^{(b)}}\left\vert 01\right\rangle \right] .  \label{43}
\end{equation}%
Due to $\sum\limits_{k=1}^{N}e^{i\gamma
_{k}^{(a)}}=\sum\limits_{k=1}^{N}e^{2i\gamma _{k}^{(a)}}=0$ and $%
\sum\limits_{k=1}^{N}e^{i\gamma _{k}^{(b)}}=\sum\limits_{k=1}^{N}e^{2i\gamma
_{k}^{(b)}}=0$, we obtain the EADM $\tilde{\rho}$ as follow%
\begin{equation}
\tilde{\rho}\equiv \frac{1}{N}\sum\limits_{k=1}^{N}\left\vert \Psi
\right\rangle \left\langle \Psi \right\vert =\left( 
\begin{array}{cccc}
1 & 0 & 0 & 1 \\ 
0 & 1 & 0 & 0 \\ 
0 & 0 & 1 & 0 \\ 
1 & 0 & 0 & 1%
\end{array}%
\right) .  \label{43a}
\end{equation}%
Apparently the EADM $\tilde{\rho}$ cannot be decomposed into the direct
products due to only non-diagonal term $\left\vert 00\right\rangle
\left\langle 11\right\vert +\left\vert 11\right\rangle \left\langle
00\right\vert $ remaining, which is similar to the Bell state $\left\vert
\Psi ^{+}\right\rangle =\frac{1}{\sqrt{2}}\left( \left\vert 00\right\rangle
+\left\vert 11\right\rangle \right) $.

Then we obtain the results $\bar{P}(\theta _{a},k)=\cos (\theta _{a}+\gamma
_{k}^{(a)})$ and $\bar{P}(\theta _{b},k)=\cos (\theta _{b}+\gamma
_{k}^{(b)}) $ for the measurement operators $\hat{P}\left( \theta
_{a}\right) $ and $\hat{P}\left( \theta _{b}\right) $ locally performed on
the fields $\left\vert \psi _{a}^{\prime }\right\rangle $ and$\ \left\vert
\psi _{b}^{\prime }\right\rangle $, where $\gamma _{k}^{(a)},\gamma
_{k}^{(b)}$ are the $k$-th units of the RPSs $\gamma ^{(a)}$ and $\gamma
^{(b)}$, respectively. Then the correlation function is%
\begin{equation}
E(\theta _{a},\theta _{b})=\frac{1}{NC}\sum_{k=1}^{N}\bar{P}(\theta _{a},k)%
\bar{P}(\theta _{b},k)=\cos (\theta _{a}+\theta _{b}),  \label{44}
\end{equation}%
where $C=1/2$ is the normalization coefficient. The fields in Eq. (\ref{42})
are considered to be the optical analogy to the Bell state $\left\vert \Psi
^{+}\right\rangle $. By substituting the above correlation functions into
Bell inequality (CHSH inequality) \cite{CHSH}%
\begin{equation}
\left\vert B\right\vert =\left\vert E(\theta _{a},\theta _{b})-E(\theta
_{a},\theta _{b}^{\prime })+E(\theta _{a}^{\prime },\theta _{b}^{\prime
})+E(\theta _{a}^{\prime },\theta _{b})\right\vert =2\sqrt{2}>2,  \label{45}
\end{equation}%
where $\theta _{a},\theta _{a}^{\prime },\theta _{b}$ and $\theta
_{b}^{\prime }$ are $\pi /4,-\pi /4,0$ and $\pi /2$, respectively, Bell's
inequality is maximally violated.

Bell state $\left\vert \Psi ^{+}\right\rangle $ differs from $\left\vert
\Psi ^{-}\right\rangle $ by $\pi $ phase. Similarly, the optical analogy to
the Bell state $\left\vert \Psi ^{-}\right\rangle $ is expressed as 
\begin{eqnarray}
\left\vert \psi _{a}^{\prime }\right\rangle &=&\frac{e^{i\lambda ^{(a)}}}{%
\sqrt{2}}\left( \left\vert 0\right\rangle +e^{i\gamma ^{(a)}}\left\vert
1\right\rangle \right) ,  \label{45a} \\
\left\vert \psi _{b}^{\prime }\right\rangle &=&\frac{e^{i\lambda ^{(b)}}}{%
\sqrt{2}}\left( \left\vert 0\right\rangle +e^{i\left( \gamma ^{(b)}+\pi
\right) }\left\vert 1\right\rangle \right) .  \nonumber
\end{eqnarray}%
By performing the transformation $\hat{\sigma}_{x}:\left\vert 0\right\rangle
\leftrightarrow \left\vert 1\right\rangle $ on $\left\vert \psi _{b}^{\prime
}\right\rangle $ of the state $\left\vert \Psi ^{\pm }\right\rangle $, we
obtain the optical analogy to the Bell state $\left\vert \Phi
^{+}\right\rangle $ expressed as 
\begin{eqnarray}
\left\vert \psi _{a}^{\prime }\right\rangle &=&\frac{e^{i\lambda ^{(a)}}}{%
\sqrt{2}}\left( \left\vert 0\right\rangle +e^{i\gamma ^{(a)}}\left\vert
1\right\rangle \right) ,  \label{45b} \\
\left\vert \psi _{b}^{\prime }\right\rangle &=&\frac{e^{i\lambda ^{(b)}}}{%
\sqrt{2}}\left( \left\vert 1\right\rangle +e^{i\gamma ^{(b)}}\left\vert
0\right\rangle \right) ,  \nonumber
\end{eqnarray}%
and of $\left\vert \Phi ^{-}\right\rangle $ expressed as 
\begin{eqnarray}
\left\vert \psi _{a}^{\prime }\right\rangle &=&\frac{e^{i\lambda ^{(a)}}}{%
\sqrt{2}}\left( \left\vert 0\right\rangle +e^{i\gamma ^{(a)}}\left\vert
1\right\rangle \right) ,  \label{45c} \\
\left\vert \psi _{b}^{\prime }\right\rangle &=&\frac{e^{i\lambda ^{(b)}}}{%
\sqrt{2}}\left( \left\vert 1\right\rangle +e^{i\left( \gamma ^{(b)}+\pi
\right) }\left\vert 0\right\rangle \right) .  \nonumber
\end{eqnarray}%
Then their correlation functions $E_{\Psi ^{-}}\left( \theta _{a},\theta
_{b}\right) =-\cos \left( \theta _{a}+\theta _{b}\right) ,E_{\Phi ^{\pm
}}\left( \theta _{a},\theta _{b}\right) =\pm \cos \left( \theta _{a}-\theta
_{b}\right) $ are obtained. To substitute the correlation functions into Eq.
(\ref{45}), we also obtain the maximal violation of Bell's inequality. The
violation of Bell's criterion demonstrates the nonlocal correlation of the
two optical fields in our scheme, which results from shared randomness of
the PPSs.

\subsubsection{\textit{GHZ states}: GHZ equality criterion \label{Sec
III.B.3}}

The nonlocality of the multipartite entangled GHZ states can in principle be
manifest in a new criterion and need not be statistical as the violation of
Bell inequality \cite{GHZ}. Preparing three fields $\left\vert \psi
_{a}\right\rangle ,\left\vert \psi _{b}\right\rangle $ and $\left\vert \psi
_{c}\right\rangle $ similar to Eq. (\ref{35}), and cyclically exchanging the
modes $\left\vert 1\right\rangle $ of the fields, we obtain as follows%
\begin{eqnarray}
\left\vert \psi _{a}^{\prime }\right\rangle &=&\frac{1}{\sqrt{2}}\left(
e^{i\lambda ^{(a)}}\left\vert 0\right\rangle +e^{i\lambda ^{(b)}}\left\vert
1\right\rangle \right) =\frac{e^{i\lambda ^{(a)}}}{\sqrt{2}}\left(
\left\vert 0\right\rangle +e^{i\gamma ^{(a)}}\left\vert 1\right\rangle
\right) ,  \label{46} \\
\left\vert \psi _{b}^{\prime }\right\rangle &=&\frac{1}{\sqrt{2}}\left(
e^{i\lambda ^{(b)}}\left\vert 0\right\rangle +e^{i\lambda ^{(c)}}\left\vert
1\right\rangle \right) =\frac{e^{i\lambda ^{(b)}}}{\sqrt{2}}\left(
\left\vert 0\right\rangle +e^{i\gamma ^{(b)}}\left\vert 1\right\rangle
\right) ,  \nonumber \\
\left\vert \psi _{c}^{\prime }\right\rangle &=&\frac{1}{\sqrt{2}}\left(
e^{i\lambda ^{(c)}}\left\vert 0\right\rangle +e^{i\lambda ^{(a)}}\left\vert
1\right\rangle \right) =\frac{e^{i\lambda ^{(c)}}}{\sqrt{2}}\left(
\left\vert 0\right\rangle +e^{i\gamma ^{(c)}}\left\vert 1\right\rangle
\right) ,  \nonumber
\end{eqnarray}%
where the RPSs $\gamma ^{(a)}=\lambda ^{(b)}-\lambda ^{(a)},\gamma
^{(b)}=\lambda ^{(c)}-\lambda ^{(b)},\gamma ^{(c)}=\lambda ^{(a)}-\lambda
^{(c)}$ and $\gamma ^{(a)}+\gamma ^{(b)}+\gamma ^{(c)}=0$. The fields are
considered to be the optical analogy to GHZ state $\left\vert \Psi
\right\rangle =\frac{1}{\sqrt{2}}\left( \left\vert 000\right\rangle
+\left\vert 111\right\rangle \right) $. We obtain the local measurement
results $\bar{P}(\theta _{a},k)=\cos (\theta _{a}+\gamma _{k}^{(a)}),\bar{P}%
(\theta _{b},k)=\cos (\theta _{b}+\gamma _{k}^{(b)}),\bar{P}(\theta
_{c},k)=\cos (\theta _{c}+\gamma _{k}^{(c)})$ for the fields $\left\vert
\psi _{a}^{\prime }\right\rangle ,\left\vert \psi _{b}^{\prime
}\right\rangle $ and $\left\vert \psi _{c}^{\prime }\right\rangle $,
respectively, and the correlation function can be obtained%
\begin{equation}
E(\theta _{a},\theta _{b},\theta _{c})=\frac{1}{NC}\sum\limits_{k=1}^{N}\bar{%
P}(\theta _{a},k)\bar{P}(\theta _{b},k)\bar{P}(\theta _{c},k)=\cos (\theta
_{a}+\theta _{b}+\theta _{c}),  \label{47}
\end{equation}%
where $C=1/4$ is the normalized coefficient. If $\theta _{a}+\theta
_{b}+\theta _{c}=0,E(\theta _{a},\theta _{b},\theta _{c})=1$. If $\theta
_{a}+\theta _{b}+\theta _{c}=\pi ,E(\theta _{a},\theta _{b},\theta _{c})=-1$%
. By using GHZ State, the family of simple proofs of Bell's theorem without
inequalities can be obtained \cite{GHZ}, which is different from the
criterion of CHSH inequality \cite{CHSH}. The sign of the correlation
function can be also treated as the criterion, such as the negative
correlation for nonlocal and the positive correlation for local when $\theta
_{a}=\pi /3,\theta _{b}=\pi /3,\theta _{c}=\pi /3$. We also obtain the
negative correlation using Eq. (\ref{47}). The results are similar to the
quantum case of GHZ states.

Further, the optical analogy to GHZ state could be generalized to the case
of $N$ particles. By preparing $N$ optical fields similar to Eq. (\ref{35})
and cyclically exchanging the modes $\left\vert 1\right\rangle $, the fields
can be obtained as follows 
\begin{eqnarray}
\left\vert \psi _{1}\right\rangle &=&e^{i\lambda ^{\left( 1\right)
}}\left\vert 0\right\rangle +e^{i\lambda ^{\left( 2\right) }}\left\vert
1\right\rangle =e^{i\lambda ^{\left( 1\right) }}\left( \left\vert
0\right\rangle +e^{i\gamma ^{\left( 1\right) }}\left\vert 1\right\rangle
\right) ,  \label{47a} \\
\left\vert \psi _{2}\right\rangle &=&e^{i\lambda ^{\left( 2\right)
}}\left\vert 0\right\rangle +e^{i\lambda ^{\left( 3\right) }}\left\vert
1\right\rangle =e^{i\lambda ^{\left( 2\right) }}\left( \left\vert
0\right\rangle +e^{i\gamma ^{\left( 2\right) }}\left\vert 1\right\rangle
\right) ,  \nonumber \\
&&\vdots  \nonumber \\
\left\vert \psi _{N}\right\rangle &=&e^{i\lambda ^{\left( N\right)
}}\left\vert 0\right\rangle +e^{i\lambda ^{\left( 1\right) }}\left\vert
1\right\rangle =e^{i\lambda ^{\left( N\right) }}\left( \left\vert
0\right\rangle +e^{i\gamma ^{\left( N\right) }}\left\vert 1\right\rangle
\right) .  \nonumber
\end{eqnarray}%
where the RPSs satisfy $\gamma ^{(1)}+\cdots +\gamma ^{(N)}=0$. We can
obtain the correlation function%
\begin{equation}
E(\theta _{1},\ldots ,\theta _{N})=\frac{1}{NC}\sum\limits_{k=1}^{N}\bar{P}%
(\theta _{1},k)\ldots \bar{P}(\theta _{N},k)=\cos \left( \theta _{1}+\cdots
+\theta _{N}\right) ,  \label{48}
\end{equation}%
where $\bar{P}(\theta _{i},k)=\cos (\theta _{i}+\gamma _{k}^{(i)})$ are the
local measurement results of the optical fields $\left\vert \psi
_{i}\right\rangle $, and $C=1/2^{N-1}$ is the normalized coefficient.

Using the same notion, we can obtain optical analogy results of other
quantum entanglement states. It should be pointed out that the phase
randomness provided by PPSs is different from the case of quantum mixed
states. Quantum mixed states result from decoherence and all coherent
superposition items (non-diagonal terms) disappear. In contract to the
decoherence, some coherent superposition items remain in the optical analogy
state due to the constraints of the RPSs, such as $\gamma ^{(a)}+\gamma
^{(b)}=0,\gamma ^{(a)}+\gamma ^{(b)}+\gamma ^{(c)}=0$ for the analogies to
Bell states and GHZ state, respectively. These remaining items make it
possible to imitate quantum entangled pure states.

\subsubsection{Numerical simulation for the optical analogy to quantum
entanglement \label{Sec III.B.4}}

\begin{figure}[htbp]
\centering\includegraphics[height=1.5878in, width=2.6152in]{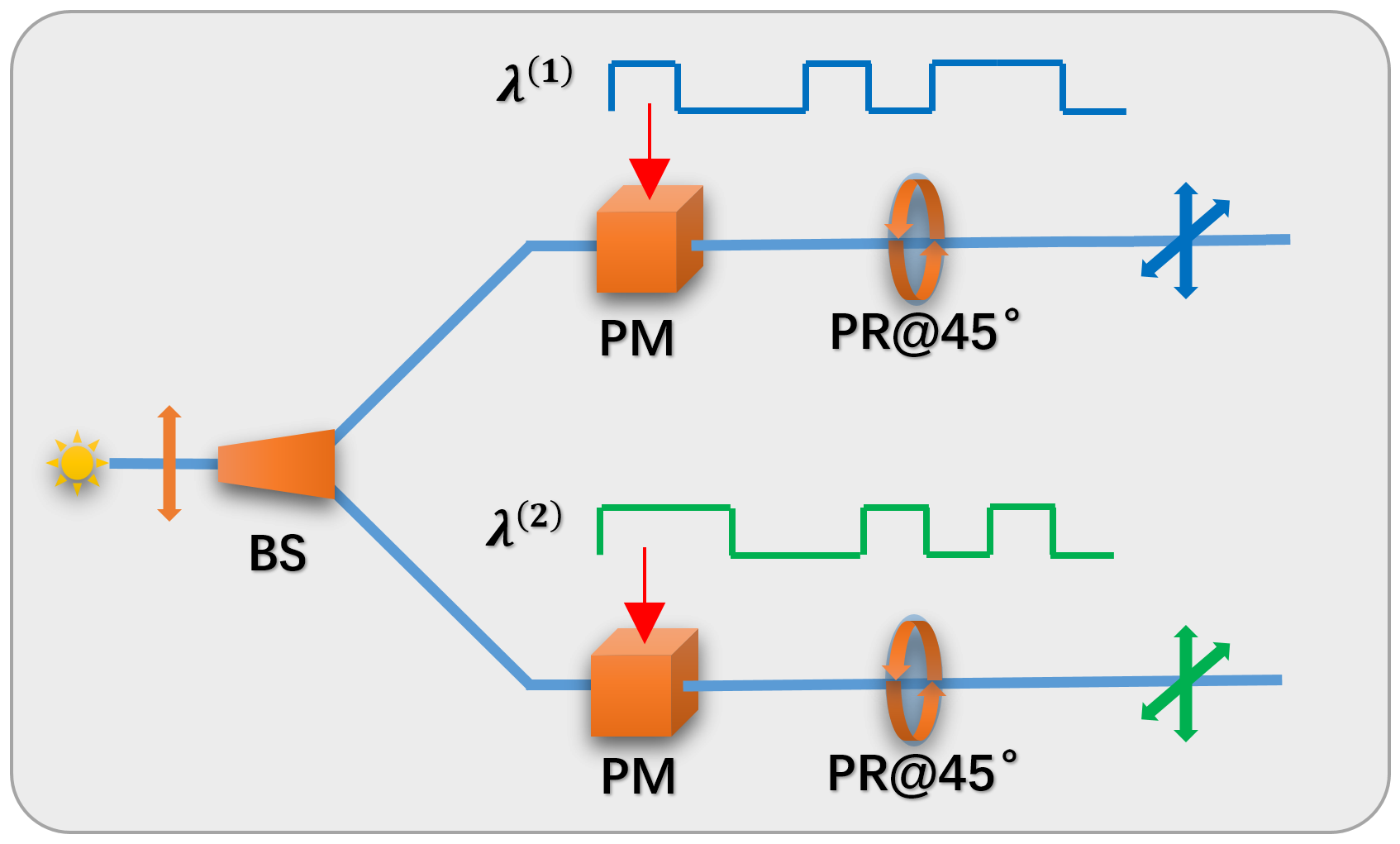}
\caption{The sheme to realize the simulation of quantum product state is
shown, where PR@$45^{\circ }$: $45^{\circ }$ polarization rotators.}
\label{fig6}
\end{figure}

Here we numerically simulate the optical analogy to quantum entanglement by
using the software OPTISYSTEM. First we propose the scheme to produce the
product state of Eq. (\ref{35}) shown in Fig. \ref{fig6}. In the scheme, we
choose two sequences $\lambda _{k}^{\left( 1\right) }$ and $\lambda
_{k}^{\left( 2\right) }$ from the set $\Xi $ to modulate the optical fields,
and obtain 
\begin{eqnarray}
E_{1}\left( t\right) &=&\left( A_{\uparrow }+A_{\rightarrow }\right)
e^{-i\left( \omega t+\lambda _{k}^{\left( 1\right) }\right) },  \label{49} \\
E_{2}\left( t\right) &=&\left( A_{\uparrow }+A_{\rightarrow }\right)
e^{-i\left( \omega t+\lambda _{k}^{\left( 2\right) }\right) },  \nonumber
\end{eqnarray}%
where $A_{\uparrow }$ and $A_{\rightarrow }$ denote the amplitudes of two
orthogonal polarization modes $\left\vert 0\right\rangle $ and $\left\vert
1\right\rangle $, respectively.

\begin{figure}[htbp]
\centering\includegraphics[height=2.0825in, width=4.5109in]{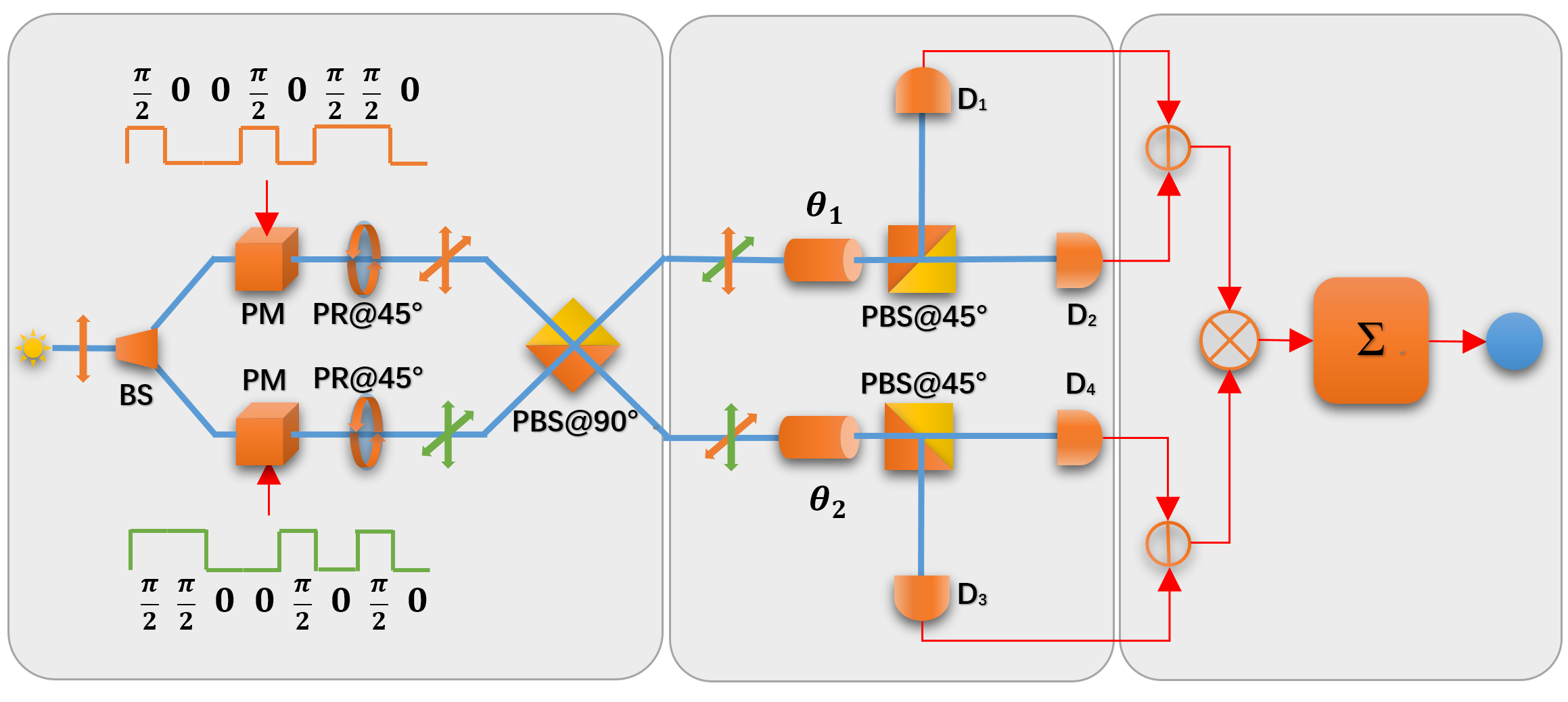}
\caption{The scheme to realize the correlation measurement of quantum
entanglement state is shown, where $\protect\theta _{1}$ and $\protect\theta %
_{2}$ are the modulators to produce phase differences between two
polarization modes; PBS$@45^{\circ }$ and PBS$@90^{\circ }$ polarization
beam splitters at $45^{\circ }$ and $90^{\circ }$ respectively; PR@$%
45^{\circ }$: $45^{\circ }$ polarization rotators.}
\label{fig7a}
\end{figure}

After mode exchanged as shown in Fig. \ref{fig7}, the optical fields can be
written as follows%
\begin{eqnarray}
E_{1}\left( t\right) &=&A_{\uparrow }e^{-i\left( \omega t+\lambda
_{k}^{\left( 1\right) }\right) }+A_{\rightarrow }e^{-i\left( \omega
t+\lambda _{k}^{\left( 2\right) }\right) },  \label{50} \\
E_{2}\left( t\right) &=&A_{\uparrow }e^{-i\left( \omega t+\lambda
_{k}^{\left( 2\right) }\right) }+A_{\rightarrow }e^{-i\left( \omega
t+\lambda _{k}^{\left( 1\right) }\right) }.  \nonumber
\end{eqnarray}%
The numerical simulation scheme of the correlation measurement is shown in
Fig. \ref{fig7a}. First, two modes of $E_{1}\left( t\right) $ and $%
E_{2}\left( t\right) $ are modulated with phase differences $\theta _{1}$
and $\theta _{2}$, respectively. The fields can be written as follows 
\begin{eqnarray}
E_{1}\left( t\right) &=&A_{\uparrow }e^{-i\left( \omega t+\lambda
_{k}^{\left( 1\right) }+\theta _{1}/2\right) }+A_{\rightarrow }e^{-i\left(
\omega t+\lambda _{k}^{\left( 2\right) }-\theta _{1}/2\right) },  \label{50a}
\\
E_{2}\left( t\right) &=&A_{\uparrow }e^{-i\left( \omega t+\lambda
_{k}^{\left( 2\right) }+\theta _{2}/2\right) }+A_{\rightarrow }e^{-i\left(
\omega t+\lambda _{k}^{\left( 1\right) }-\theta _{2}/2\right) }.  \nonumber
\end{eqnarray}%
Then the fields $E_{1}\left( t\right) $ and $E_{2}\left( t\right) $ are
split two beams by the polarization beamsplitters at angles $45^{\circ }$
and input the photodetectors $\left( D_{1},D_{2}\right) $ and $\left(
D_{3},D_{4}\right) $, respectively. The differential signals of
photodetectors are proportional to 
\begin{eqnarray}
I_{1} &=&\left\vert D_{1}\right\vert ^{2}-\left\vert D_{2}\right\vert
^{2}=\mu \left\vert A_{s}\right\vert ^{2}\cos \left( \lambda _{k}^{\left(
1\right) }-\lambda _{k}^{\left( 2\right) }+\theta _{1}\right) ,  \label{50b}
\\
I_{2} &=&\left\vert D_{3}\right\vert ^{2}-\left\vert D_{4}\right\vert
^{2}=\mu \left\vert A_{s}\right\vert ^{2}\cos \left( \lambda _{k}^{\left(
2\right) }-\lambda _{k}^{\left( 1\right) }+\theta _{2}\right) ,  \nonumber
\end{eqnarray}%
where $\left\vert D_{i}\right\vert ^{2}$ are the output electric signals of
the photodetectors, $A_{s}=A_{\uparrow }=A_{\rightarrow }$ assumed. Then we
calculate the correlation function $E(\theta _{1},\theta _{2})=\left\langle
I_{1}I_{2}\right\rangle /C=\cos \left( \theta _{1}+\theta _{2}\right) $,
where $C=4\mu ^{2}\left\vert A_{s}\right\vert ^{4}\Delta T$ is the
normalization coefficient. Finally, using the software OPTISYSTEM, we obtain
the numerical results of the\ normalized correlation function as shown in
Fig. \ref{fig8} (Detailed simulation data and OPTISYSTEM models will be
provided in the supplementary material). By substituting the above
correlation results into Bell inequality (CHSH inequality) \cite{CHSH}: $%
\left\vert B\right\vert =2.825\pm 0.001>2$ where $\theta _{1},\theta
_{1}^{\prime },\theta _{2}$ and $\theta _{2}^{\prime }$ are $\pi /4,-\pi
/4,0 $ and $\pi /2$, respectively.

\begin{figure}[htbp]
\centering\includegraphics[height=2.789in, width=3.7144in]{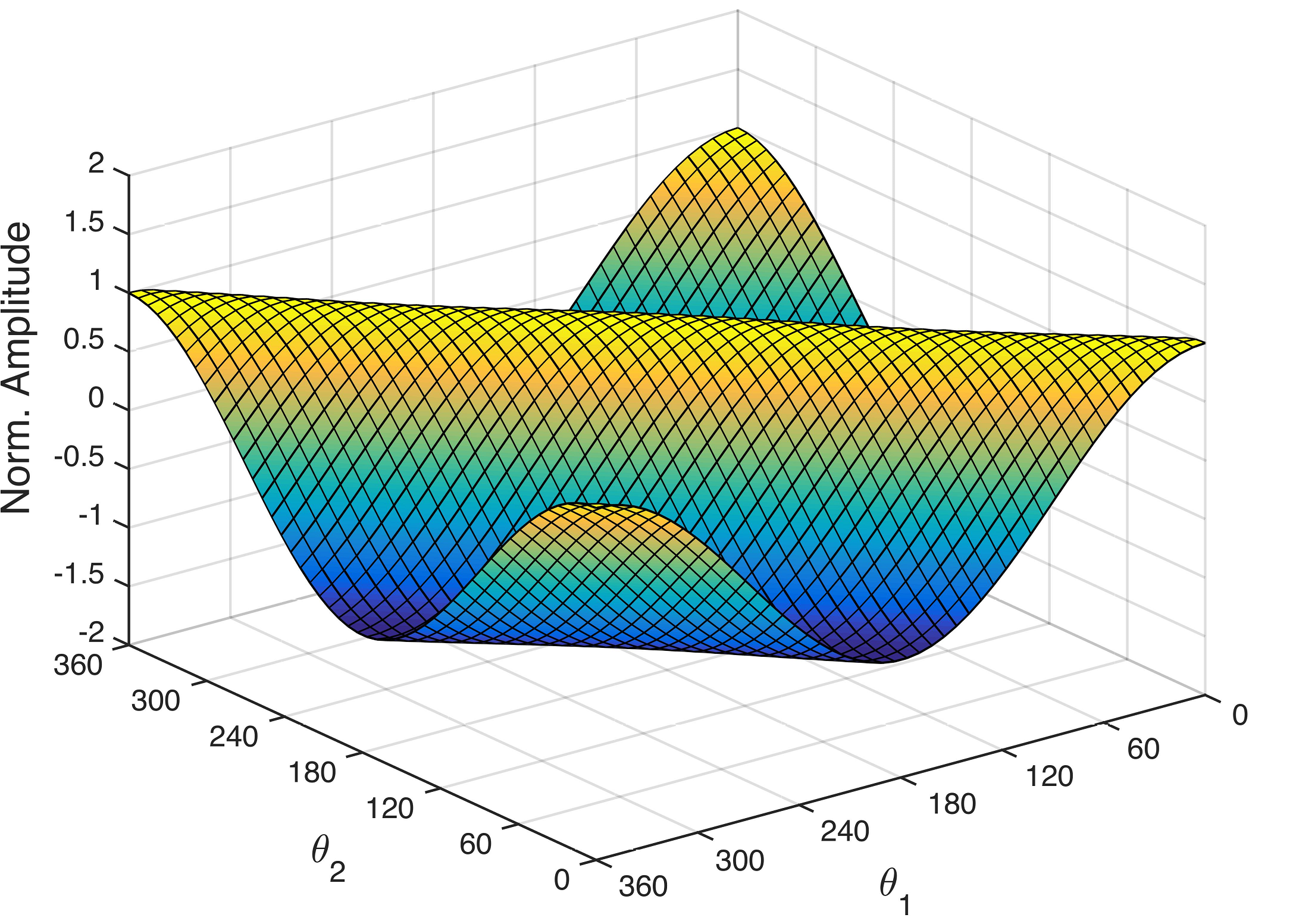}
\caption{The result of the correlation function for the analogy to one of
Bell states is shown.}
\label{fig8}
\end{figure}

\subsection{Optical analogies to quantum states of \textbf{multiple
particles }\label{Sec III.C}}

We have discussed the coherent demodulation process in Sec. \ref{Sec II.B}.
Here we discuss how to imitate quantum states with the help of the coherent
demodulation.

First, we consider the general form of $N$ optical fields modulated with
PPSs $\left\{ \lambda ^{\left( 1\right) },\ldots ,\lambda ^{\left( N\right)
}\right\} $ chosen from the set $\Xi $, and the states can be expressed as
Eq. (\ref{20}). It is noteworthy that although multiple PPSs are
superimposed on two orthogonal modes of the fields, all of the PPSs can be
demodulated and discriminated by respectively performing the coherent
demodulations on the two orthogonal modes, which has already been verified
by many actual communication systems \cite{Viterbi,Peterson,PSK}.

\begin{figure}[tbph]
\centering\includegraphics[height=2.2355in, width=3.6236in]{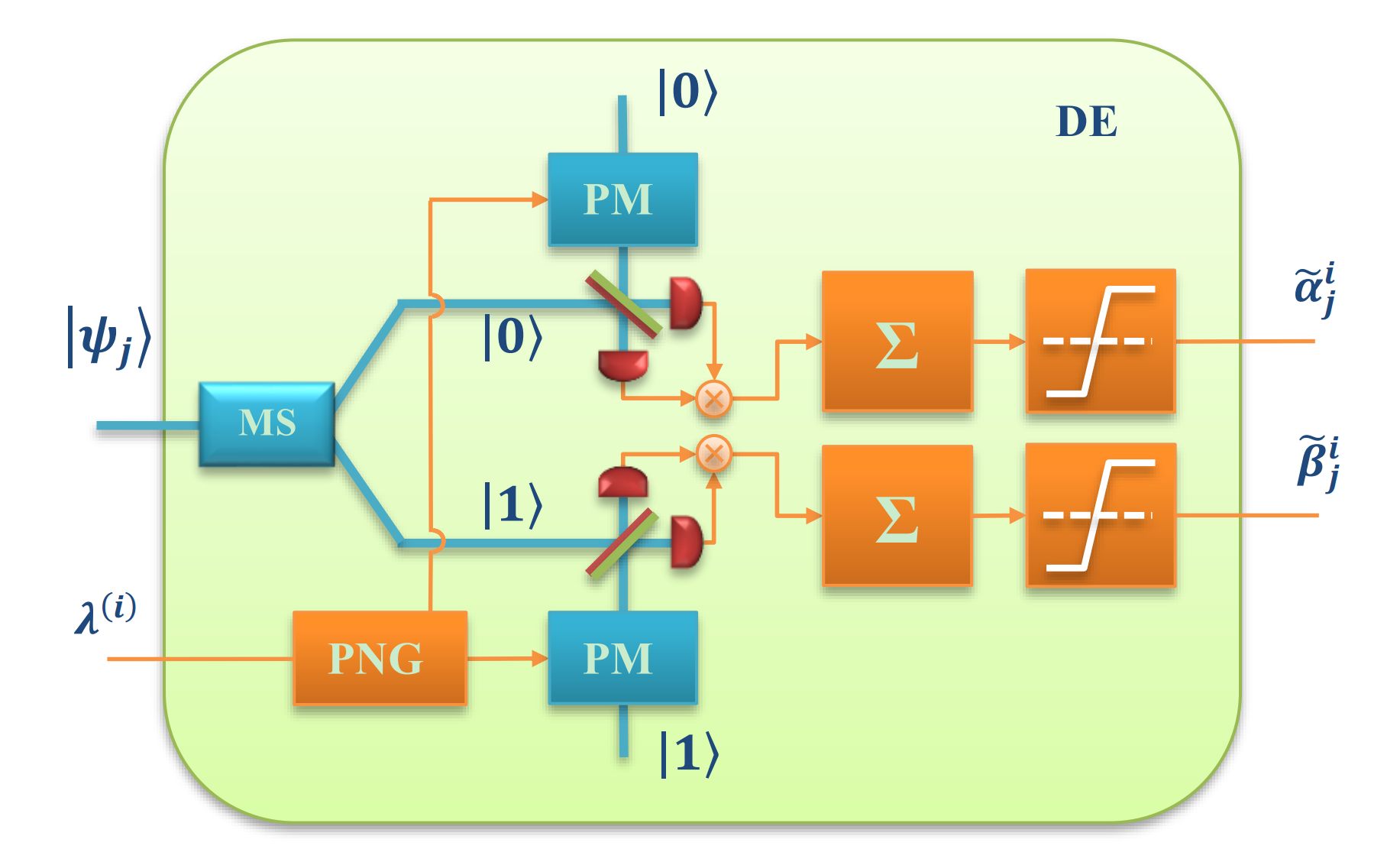}
\caption{The coherent demodulation scheme of pseudorandom phase sequence is
shown, where SO: signal optical fields, LO: local optical fields, MS: mode
splitters, PM: phase modulators, PNG: pseudorandom number generators, $%
\otimes $: multipliers and $\Sigma $: integrators (integrate over entire
sequence period).}
\label{fig9a}
\end{figure}

\begin{figure}[htbp]
\centering\includegraphics[height=1.8464in, width=4.7392in]{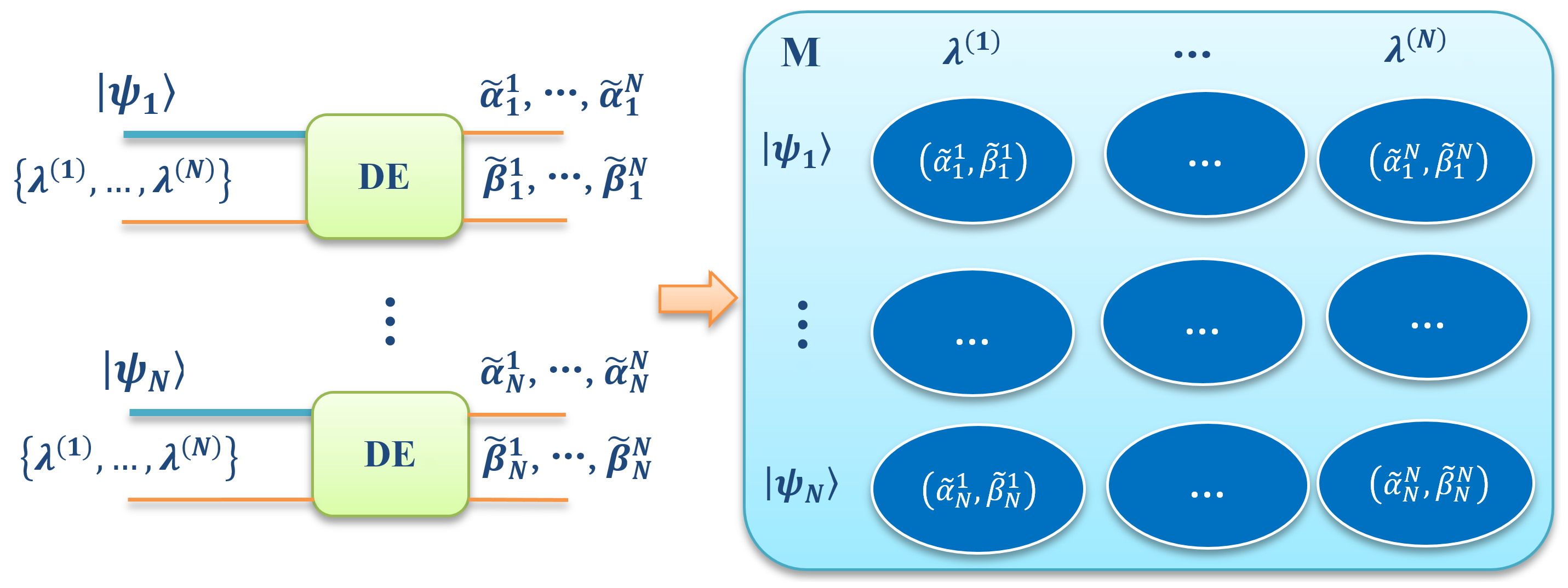}
\caption{The PPS coherent demodulation scheme for multiple input fields is
shown, where the DE block is shown in Fig. \protect\ref{fig9a}.}
\label{fig9}
\end{figure}

Now we propose a scheme, as shown in Fig. \ref{fig9a}, to perform the
coherent demodulation introduced in Sec. \ref{Sec II.B}. In the scheme, the
coherent demodulations are performed on each SO field and LO field modulated
with reference PPSs $\left\{ \lambda ^{\left( 1\right) },\ldots ,\lambda
^{\left( N\right) }\right\} $. Thus a mode status matrix $M\left( \tilde{%
\alpha}_{i}^{j},\tilde{\beta}_{i}^{j}\right) $, as shown in Fig. \ref{fig9},
can be obtained by performing $N$ coherent demodulations on each SO field,
where $\tilde{\alpha}_{i}^{j},\tilde{\beta}_{i}^{j}$ are the mode status of
the orthogonal polarization modes $\left\vert 0\right\rangle $ and $%
\left\vert 1\right\rangle $, respectively. For better fault tolerance, the
mode status output discrete values through the threshold discrimination and
binarization of the measurement results. Of the matrix $M\left( \tilde{\alpha%
}_{i}^{j},\tilde{\beta}_{i}^{j}\right) $, each element represents the mode
status of the $i$th optical field when the reference PPS is $\lambda
^{\left( j\right) }$, and takes one of four possible discrete values: $%
\left( 1,0\right) ,\left( 0,1\right) ,\left( 1,1\right) $ or $0$, denoting
that exists only mode $\left\vert 0\right\rangle $, only mode $\left\vert
1\right\rangle $, both $\left\vert 0\right\rangle $ and $\left\vert
1\right\rangle $, neither $\left\vert 0\right\rangle $ nor $\left\vert
1\right\rangle $, respectively. Thus we obtain a one-to-one correspondence
relationship between the $N$ optical fields and the matrix $M$. Thus we
consider the matrix $M$ as a bridge to connect the optical fields and the
quantum states.

\begin{figure}[tbph]
\centering\includegraphics[height=3.6426in, width=3.8891in]{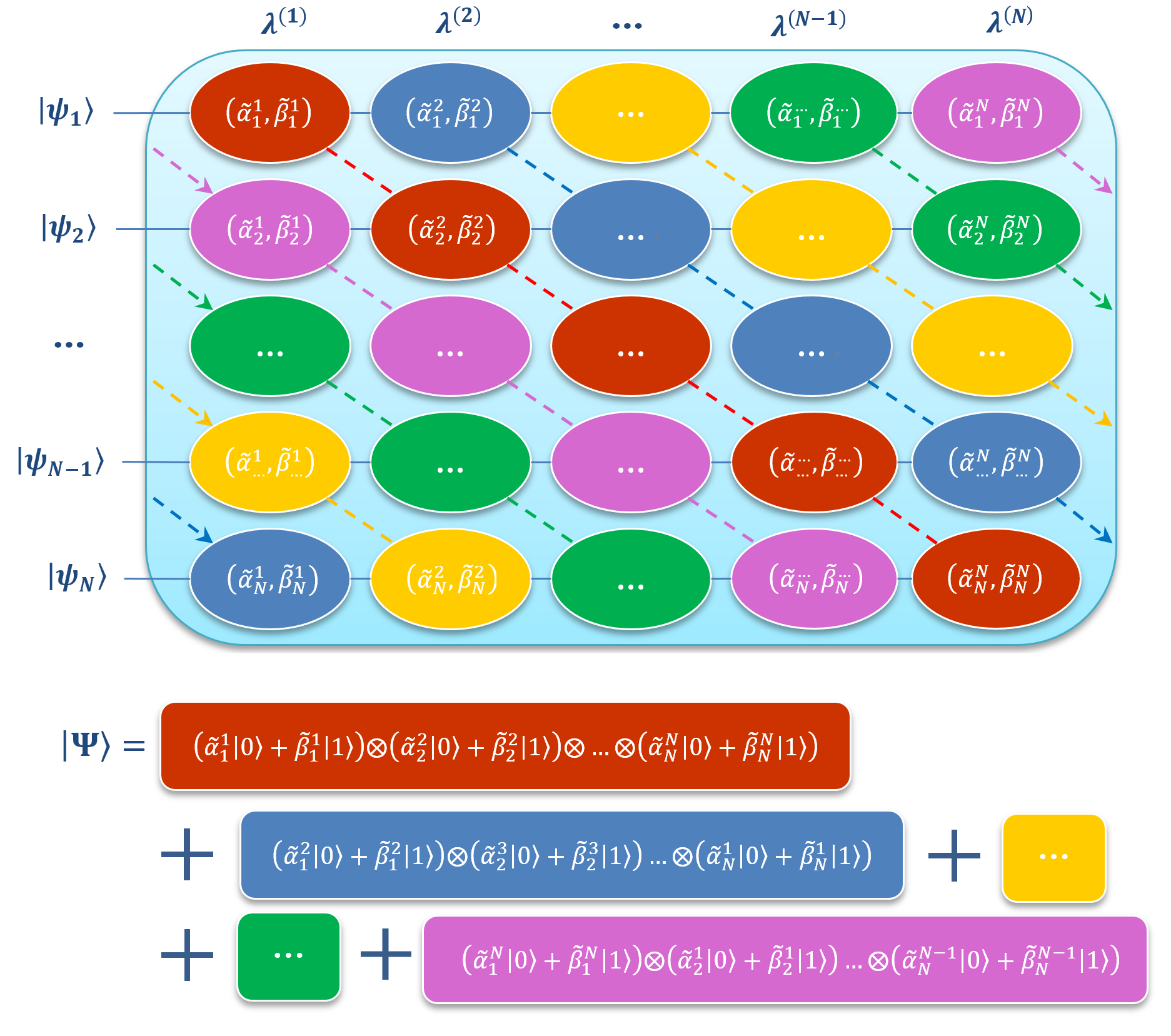}
\caption{The scheme to imitate quantum state is shown, the mode status
matrix $M\left( \tilde{\protect\alpha}_{i}^{j},\tilde{\protect\beta}%
_{i}^{j}\right) $ related to the $i$th optical field and the reference PPS $%
\protect\lambda ^{\left( j\right) }$, in which the mode status with the same
color for the same sequence permutation.}
\label{fig10}
\end{figure}

Now we discuss how to construct the imitation states based on the matrix $M$%
. In the subsection \ref{Sec III.A.3}, we propose the SCPM to reduce the
redundancy of sequential permutation. Here we apply the SCPM to imitate
quantum states based on the $M$ matrix. In order to clearly present the SCPM
in the $M$ matrix, as shown in Fig. \ref{fig10}, the matrix elements
belonging to the same permutation $R_{r}$ are labeled with the same color,
such as the red color corresponding to $R_{1}$, the blue color corresponding
to $R_{2}$, etc. Considering each permutation $R_{r}$ corresponding to
equivalent direct product decompositions, thus the imitated quantum state
must be a direct product of the elements belonging to the same $R_{r}$ and
then superposition of equivalent direct products corresponding to all
permutations. Therefore we obtain%
\begin{eqnarray}
\left\vert \Psi \right\rangle &=&\left( \tilde{\alpha}_{1}^{1}\left\vert
0\right\rangle +\tilde{\beta}_{1}^{1}\left\vert 1\right\rangle \right)
\otimes \ldots \otimes \left( \tilde{\alpha}_{N}^{N}\left\vert
0\right\rangle +\tilde{\beta}_{N}^{N}\left\vert 1\right\rangle \right)
+\left( \tilde{\alpha}_{1}^{2}\left\vert 0\right\rangle +\tilde{\beta}%
_{1}^{2}\left\vert 1\right\rangle \right) \otimes \ldots \otimes \left( 
\tilde{\alpha}_{N}^{1}\left\vert 0\right\rangle +\tilde{\beta}%
_{N}^{1}\left\vert 1\right\rangle \right)  \label{53} \\
&&+\ldots +\left( \tilde{\alpha}_{1}^{N}\left\vert 0\right\rangle +\tilde{%
\beta}_{1}^{N}\left\vert 1\right\rangle \right) \otimes \ldots \otimes
\left( \tilde{\alpha}_{N}^{N-1}\left\vert 0\right\rangle +\tilde{\beta}%
_{N}^{N-1}\left\vert 1\right\rangle \right) ,  \nonumber
\end{eqnarray}%
where $\tilde{\alpha}_{i}^{j},\tilde{\beta}_{i}^{j}$ is the mode discrete
status obtained from the matrix $M\left( \tilde{\alpha}_{i}^{j},\tilde{\beta}%
_{i}^{j}\right) $.

It is noteworthy that the SCPM to reduce the redundancy is one of the
feasible ways to imitate quantum states based on the matrix $M$. Other
mechanisms might also work, as long as a sequential ergodic ensemble can be
obtained. In order to prove the feasibility of the scheme, in the next
subsection, we will discuss the imitations of several typical quantum
states, including the product states, Bell states, GHZ states and W states.

\subsubsection{The imitation states of several typical quantum states \label%
{Sec III.C.1}}

In this subsection, we discuss optical analogies to several typical quantum
states and construct their imitation states applying the scheme proposed in
the last subsection, including the product state, Bell states, GHZ state and
W state.

\paragraph{The product state}

First, we discuss\ the optical analogy to the product state of $N$\ qubit.
The optical fields are shown as follows 
\begin{eqnarray}
\left\vert \psi _{1}\right\rangle &=&e^{i\lambda ^{\left( 1\right) }}\left(
\left\vert 0\right\rangle +\left\vert 1\right\rangle \right) ,  \label{54} \\
&&\vdots  \nonumber \\
\left\vert \psi _{N}\right\rangle &=&e^{i\lambda ^{\left( N\right) }}\left(
\left\vert 0\right\rangle +\left\vert 1\right\rangle \right) .  \nonumber
\end{eqnarray}%
By employing the scheme as shown in Fig. \ref{fig10}, we obtain the matrix 
\begin{equation}
M\left( \tilde{\alpha}_{i}^{j},\tilde{\beta}_{i}^{j}\right) =\left( 
\begin{array}{ccc}
\left( 1,1\right) &  &  \\ 
& \ddots &  \\ 
&  & \left( 1,1\right)%
\end{array}%
\right) ,  \label{55}
\end{equation}%
which demonstrates that each optical field is the superposition of two
orthogonal modes and no entanglement is involved. According to Eq. (\ref{53}%
), we obtain the imitation state as follow 
\begin{eqnarray}
\left\vert \Psi \right\rangle &=&\left( \left\vert 0\right\rangle
+\left\vert 1\right\rangle \right) \otimes \ldots \otimes \left( \left\vert
0\right\rangle +\left\vert 1\right\rangle \right)  \label{56} \\
&=&\left\vert 0\ldots 0\right\rangle +\left\vert 0\ldots 1\right\rangle
+\ldots +\left\vert 1\ldots 1\right\rangle ,  \nonumber
\end{eqnarray}%
which is same as the quantum product state expect a normalization factor.

\paragraph{Bell states}

Now we discuss the optical analogy to one of the four Bell states $%
\left\vert \Psi ^{+}\right\rangle =\frac{1}{\sqrt{2}}\left( \left\vert
00\right\rangle +\left\vert 11\right\rangle \right) $, which contains two
optical fields as Eq. (\ref{42}). By employing the scheme as shown in Fig. %
\ref{fig10}, we obtain the matrix%
\begin{equation}
M\left( \tilde{\alpha}_{i}^{j},\tilde{\beta}_{i}^{j}\right) =\left( 
\begin{array}{cc}
\left( 1,0\right) & \left( 0,1\right) \\ 
\left( 0,1\right) & \left( 1,0\right)%
\end{array}%
\right) .  \label{58}
\end{equation}%
According to the SCPM, we obtain that $R_{1}=\{\lambda ^{\left( 1\right)
},\lambda ^{\left( 2\right) }\}$ and $R_{2}=\{\lambda ^{\left( 2\right)
},\lambda ^{\left( 1\right) }\}$. Based on the matrix $M$, for the selection
of $R_{1}$, we obtain the modes of the fields are $\left\vert 0\right\rangle 
$ and $\left\vert 0\right\rangle $; for the selection of $R_{2}$, we obtain
the modes are $\left\vert 1\right\rangle $ and $\left\vert 1\right\rangle $.
If we randomly choose $R_{1}$ or $R_{2}$, we can randomly obtain the mode
status of $\left\vert 0\right\rangle \left\vert 0\right\rangle $ or $%
\left\vert 1\right\rangle \left\vert 1\right\rangle $, which is similar to
quantum measurements for the Bell state $\left\vert \Psi ^{+}\right\rangle $%
. According to Eq. (\ref{53}), we obtain the imitation state as follow%
\begin{eqnarray}
\left\vert \Psi \right\rangle &=&\left( \tilde{\alpha}_{1}^{1}\left\vert
0\right\rangle +\tilde{\beta}_{1}^{1}\left\vert 1\right\rangle \right)
\otimes \left( \tilde{\alpha}_{2}^{2}\left\vert 0\right\rangle +\tilde{\beta}%
_{2}^{2}\left\vert 1\right\rangle \right) +\left( \tilde{\alpha}%
_{1}^{2}\left\vert 0\right\rangle +\tilde{\beta}_{1}^{2}\left\vert
1\right\rangle \right) \otimes \left( \tilde{\alpha}_{2}^{1}\left\vert
0\right\rangle +\tilde{\beta}_{2}^{1}\left\vert 1\right\rangle \right)
\label{59} \\
&=&\left\vert 00\right\rangle +\left\vert 11\right\rangle ,  \nonumber
\end{eqnarray}%
which is same as the Bell state $\left\vert \Psi ^{+}\right\rangle $ expect
a normalization factor.

We discuss the optical analogy to another Bell state $\left\vert \Phi
^{+}\right\rangle $, which contains two optical fields as Eq. (\ref{45b}).
We then obtain the matrix%
\begin{equation}
M\left( \tilde{\alpha}_{i}^{j},\tilde{\beta}_{i}^{j}\right) =\left( 
\begin{array}{cc}
\left( 1,0\right) & \left( 0,1\right) \\ 
\left( 1,0\right) & \left( 0,1\right)%
\end{array}%
\right) .  \label{61}
\end{equation}%
If we randomly choose $R_{1}$ or $R_{2}$, we can also randomly obtain the
mode results $\left\vert 0\right\rangle \left\vert 1\right\rangle $ or $%
\left\vert 0\right\rangle \left\vert 1\right\rangle $, which is similar to
quantum measurement for the Bell state $\left\vert \Phi ^{+}\right\rangle $.
We can obtain the imitation state 
\begin{eqnarray}
\left\vert \Psi \right\rangle &=&\left( \tilde{\alpha}_{1}^{1}\left\vert
0\right\rangle +\tilde{\beta}_{1}^{1}\left\vert 1\right\rangle \right)
\otimes \left( \tilde{\alpha}_{2}^{2}\left\vert 0\right\rangle +\tilde{\beta}%
_{2}^{2}\left\vert 1\right\rangle \right) +\left( \tilde{\alpha}%
_{1}^{2}\left\vert 0\right\rangle +\tilde{\beta}_{1}^{2}\left\vert
1\right\rangle \right) \otimes \left( \tilde{\alpha}_{2}^{1}\left\vert
0\right\rangle +\tilde{\beta}_{2}^{1}\left\vert 1\right\rangle \right)
\label{62} \\
&=&\left\vert 10\right\rangle +\left\vert 01\right\rangle ,  \nonumber
\end{eqnarray}%
which is same as the Bell state $\left\vert \Phi ^{+}\right\rangle $ expect
a normalization factor.

\paragraph{GHZ state}

For tripartite systems there are only two different classes of genuine
tripartite entanglement, the GHZ class and the W class \cite{GHZ,Nielsen}.
First we discuss the optical analogy to GHZ state $\left\vert \Psi
_{GHZ}\right\rangle =\frac{1}{\sqrt{2}}\left( \left\vert 000\right\rangle
+\left\vert 111\right\rangle \right) $, which contains three optical fields
as Eq. (\ref{46}), where the superscripts of the PPSs become $1,2,3$. We
obtain the matrix%
\begin{equation}
M\left( \tilde{\alpha}_{i}^{j},\tilde{\beta}_{i}^{j}\right) =\left( 
\begin{array}{ccc}
\left( 1,0\right) & \left( 0,1\right) & 0 \\ 
0 & \left( 1,0\right) & \left( 0,1\right) \\ 
\left( 0,1\right) & 0 & \left( 1,0\right)%
\end{array}%
\right) .  \label{66}
\end{equation}%
According to the SCPM, we obtain that $R_{1}=\{\lambda ^{\left( 1\right)
},\lambda ^{\left( 2\right) },\lambda ^{\left( 3\right) }\}$, $%
R_{2}=\{\lambda ^{\left( 2\right) },\lambda ^{\left( 3\right) },\lambda
^{\left( 1\right) }\}$ and $R_{3}=\{\lambda ^{\left( 3\right) },\lambda
^{\left( 1\right) },\lambda ^{\left( 2\right) }\}$. Based on the matrix $M$,
for the selection of $R_{1}$, we obtain the mode status are $\left\vert
0\right\rangle \left\vert 0\right\rangle \left\vert 0\right\rangle $; for
the selection of $R_{2}$, we obtain the mode status are $\left\vert
1\right\rangle \left\vert 1\right\rangle \left\vert 1\right\rangle $; for
the selection of $R_{3}$, we obtain nothing. Thus we can obtain the
imitation state%
\begin{eqnarray}
\left\vert \Psi \right\rangle &=&\left( \tilde{\alpha}_{1}^{1}\left\vert
0\right\rangle +\tilde{\beta}_{1}^{1}\left\vert 1\right\rangle \right)
\otimes \left( \tilde{\alpha}_{2}^{2}\left\vert 0\right\rangle +\tilde{\beta}%
_{2}^{2}\left\vert 1\right\rangle \right) \otimes \left( \tilde{\alpha}%
_{3}^{3}\left\vert 0\right\rangle +\tilde{\beta}_{3}^{3}\left\vert
1\right\rangle \right)  \label{67} \\
&&+\left( \tilde{\alpha}_{1}^{2}\left\vert 0\right\rangle +\tilde{\beta}%
_{1}^{2}\left\vert 1\right\rangle \right) \otimes \left( \tilde{\alpha}%
_{2}^{3}\left\vert 0\right\rangle +\tilde{\beta}_{2}^{3}\left\vert
1\right\rangle \right) \otimes \left( \tilde{\alpha}_{3}^{1}\left\vert
0\right\rangle +\tilde{\beta}_{3}^{1}\left\vert 1\right\rangle \right) 
\nonumber \\
&=&\left\vert 000\right\rangle +\left\vert 111\right\rangle .  \nonumber
\end{eqnarray}

For $N$ quantum particles, we can obtain the analogy to GHZ state $%
\left\vert \Psi _{GHZ}\right\rangle =\frac{1}{\sqrt{2}}\left( \left\vert
0...0\right\rangle +\left\vert 1...1\right\rangle \right) $, which contains $%
N$ optical fields as Eq. (\ref{47a}). Performing the scheme as shown in Fig. %
\ref{fig10}, we obtain the matrix%
\begin{equation}
M\left( \tilde{\alpha}_{i}^{j},\tilde{\beta}_{i}^{j}\right) =\left( 
\begin{array}{ccccc}
\left( 1,0\right) & \left( 0,1\right) &  &  &  \\ 
& \left( 1,0\right) & \left( 0,1\right) &  &  \\ 
&  & \ddots & \ddots &  \\ 
&  &  & \left( 1,0\right) & \left( 0,1\right) \\ 
\left( 0,1\right) &  &  &  & \left( 1,0\right)%
\end{array}%
\right) .  \label{69}
\end{equation}%
Thus we can easily obtain the imitation state same as GHZ state $\left\vert
\Psi _{GHZ}\right\rangle $ expect a normalization factor.

In the following, we discuss any unitary transformation states of GHZ state.
Assuming the unitary transformation simplified from Eq. (\ref{13}), 
\begin{eqnarray}
\hat{U}\left\vert \psi \right\rangle &=&\hat{U}\left( \alpha \left\vert
0\right\rangle +\beta \left\vert 1\right\rangle \right)  \label{70} \\
&=&\alpha \left( C_{0}\left\vert 0\right\rangle +C_{1}\left\vert
1\right\rangle \right) +\beta \left( C_{0}^{\ast }\left\vert 1\right\rangle
-C_{1}^{\ast }\left\vert 0\right\rangle \right) ,  \nonumber
\end{eqnarray}%
we employ it on GHZ state of $N$ quantum particles and obtain as follow 
\begin{equation}
\hat{U}_{n}\left\vert \Psi _{GHZ}\right\rangle =\frac{1}{\sqrt{2}}\left[
C_{0}\left\vert 0...0_{n}...0\right\rangle +C_{1}\left\vert
0...1_{n}...0\right\rangle +C_{0}^{\ast }\left\vert
1...1_{n}...1\right\rangle -C_{1}^{\ast }\left\vert
1...0_{n}...1\right\rangle \right] .  \label{70a}
\end{equation}%
Similarly, we employ the transformation on the optical analogy fields as
follows 
\begin{eqnarray}
\left\vert \psi _{1}\right\rangle &=&e^{i\lambda ^{\left( 1\right)
}}\left\vert 0\right\rangle +e^{i\lambda ^{\left( 2\right) }}\left\vert
1\right\rangle ,  \label{70b} \\
&&\vdots  \nonumber \\
\hat{U}_{n}\left\vert \psi _{n}\right\rangle &=&e^{i\lambda ^{\left(
n\right) }}\left( C_{0}\left\vert 0\right\rangle +C_{1}\left\vert
1\right\rangle \right) +e^{i\lambda ^{\left( n+1\right) }}\left( C_{0}^{\ast
}\left\vert 1\right\rangle -C_{1}^{\ast }\left\vert 0\right\rangle \right) ,
\nonumber \\
&&\vdots  \nonumber \\
\left\vert \psi _{N}\right\rangle &=&e^{i\lambda ^{\left( N\right)
}}\left\vert 0\right\rangle +e^{i\lambda ^{\left( 1\right) }}\left\vert
1\right\rangle .  \nonumber
\end{eqnarray}%
We obtain the matrix%
\begin{equation}
M\left( \tilde{\alpha}_{i}^{j},\tilde{\beta}_{i}^{j}\right) =\left( 
\begin{array}{ccccc}
\left( 1,0\right) & \left( 0,1\right) &  &  &  \\ 
& \ddots & \ddots &  &  \\ 
&  & \left( C_{0},C_{1}\right) & \left( -C_{1}^{\ast },C_{0}^{\ast }\right)
&  \\ 
&  &  & \ddots & \ddots \\ 
\left( 0,1\right) &  &  &  & \left( 1,0\right)%
\end{array}%
\right) ,  \label{70c}
\end{equation}%
and the imitation state 
\begin{equation}
\left\vert \Psi \right\rangle =C_{0}\left\vert 0...0_{n}...0\right\rangle
+C_{1}\left\vert 0...1_{n}...0\right\rangle +C_{0}\left\vert
1...1_{n}...1\right\rangle -C_{1}^{\ast }\left\vert
1...0_{n}...1\right\rangle .  \label{70d}
\end{equation}%
Obviously, we can obtain the result completely similar to quantum states.
For a simple case, we discuss the NOT unitary transformation $\hat{U}%
_{NOT}\left\vert \psi _{n}\right\rangle \rightarrow \hat{U}_{NOT}\left(
\alpha _{n}\left\vert 0\right\rangle +\beta _{n}\left\vert 1\right\rangle
\right) =\alpha _{n}\left\vert 1\right\rangle +\beta _{n}\left\vert
0\right\rangle $ as follows 
\begin{eqnarray}
\left\vert \psi _{1}\right\rangle &=&e^{i\lambda ^{\left( 1\right)
}}\left\vert 0\right\rangle +e^{i\lambda ^{\left( 2\right) }}\left\vert
1\right\rangle ,  \nonumber \\
\left\vert \psi _{2}\right\rangle &=&e^{i\lambda ^{\left( 2\right)
}}\left\vert 0\right\rangle +e^{i\lambda ^{\left( 3\right) }}\left\vert
1\right\rangle ,  \nonumber \\
&&\vdots  \nonumber \\
\hat{U}_{NOT}\left\vert \psi _{n}\right\rangle &=&\hat{U}_{NOT}\left(
e^{i\lambda ^{\left( n\right) }}\left\vert 0\right\rangle +e^{i\lambda
^{\left( n+1\right) }}\left\vert 1\right\rangle \right) =e^{i\lambda
^{\left( n\right) }}\left\vert 1\right\rangle +e^{i\lambda ^{\left(
n+1\right) }}\left\vert 0\right\rangle ,  \nonumber \\
&&\vdots  \nonumber \\
\left\vert \psi _{N}\right\rangle &=&e^{i\lambda ^{\left( N\right)
}}\left\vert 0\right\rangle +e^{i\lambda ^{\left( 1\right) }}\left\vert
1\right\rangle ,  \nonumber
\end{eqnarray}%
and the transformed imitation state can be expressed as follow%
\begin{equation}
\left\vert \Psi \right\rangle =\left\vert 00...1_{n}...0\right\rangle
+\left\vert 11...0_{n}...1\right\rangle .  \label{72}
\end{equation}%
Similar to quantum computation, we can obtain a new state by using only one
step instead of $2$ steps as classical computation. This is a simple example
to demonstrate parallel computing capability.

\paragraph{W state}

Now we discuss the optical analogy to W state $\left\vert \Psi
_{W}\right\rangle =\frac{1}{\sqrt{3}}\left( \left\vert 100\right\rangle
+\left\vert 010\right\rangle +\left\vert 001\right\rangle \right) $, which
contains three optical fields as follows 
\begin{eqnarray}
\left\vert \psi _{1}\right\rangle &=&e^{i\lambda ^{\left( 1\right)
}}\left\vert 1\right\rangle +e^{i\lambda ^{\left( 2\right) }}\left\vert
0\right\rangle +e^{i\lambda ^{\left( 3\right) }}\left\vert 0\right\rangle ,
\label{74} \\
\left\vert \psi _{2}\right\rangle &=&e^{i\lambda ^{\left( 1\right)
}}\left\vert 1\right\rangle +e^{i\lambda ^{\left( 2\right) }}\left\vert
0\right\rangle +e^{i\lambda ^{\left( 3\right) }}\left\vert 0\right\rangle , 
\nonumber \\
\left\vert \psi _{3}\right\rangle &=&e^{i\lambda ^{\left( 1\right)
}}\left\vert 1\right\rangle +e^{i\lambda ^{\left( 2\right) }}\left\vert
0\right\rangle +e^{i\lambda ^{\left( 3\right) }}\left\vert 0\right\rangle . 
\nonumber
\end{eqnarray}%
Performing the same scheme in Fig. \ref{fig10}, we obtain the matrix%
\begin{equation}
M\left( \tilde{\alpha}_{i}^{j},\tilde{\beta}_{i}^{j}\right) =\left( 
\begin{array}{ccc}
\left( 0,1\right) & \left( 1,0\right) & \left( 1,0\right) \\ 
\left( 0,1\right) & \left( 1,0\right) & \left( 1,0\right) \\ 
\left( 0,1\right) & \left( 1,0\right) & \left( 1,0\right)%
\end{array}%
\right) .  \label{75}
\end{equation}%
According to the SCPM, we use $R_{1}$, $R_{2}$ and $R_{3}$ again. Based on
the matrix $M$, we obtain the mode status of $\left\vert 1\right\rangle
\left\vert 0\right\rangle \left\vert 0\right\rangle $, $\left\vert
0\right\rangle \left\vert 0\right\rangle \left\vert 1\right\rangle $, $%
\left\vert 0\right\rangle \left\vert 1\right\rangle \left\vert
0\right\rangle $ for the selection of $R_{1}$, $R_{2}$, $R_{3}$,
respectively. We find an interesting fact that if we need to lock the mode
status $\left\vert 1\right\rangle $ for the first field, $R_{1}$ must be
selected. This will lead to the mode status $\left\vert 0\right\rangle
\left\vert 0\right\rangle $ must be obtained from the other two fields.
Otherwise if the state status $\left\vert 0\right\rangle $ of the first
field is obtain, $R_{2}$ or $R_{3}$ can be selected. This will lead to the
other two fields are still in the state $\left\vert 0\right\rangle
\left\vert 1\right\rangle +\left\vert 1\right\rangle \left\vert
0\right\rangle $ similar to Bell state $\left\vert \Phi ^{+}\right\rangle $.
This fact is quite similar to the case of quantum measurement and the
collapse phenomenon for W state in quantum mechanics. We obtain the
imitation state as follow 
\begin{eqnarray}
\left\vert \Psi \right\rangle &=&\left( \tilde{\alpha}_{1}^{1}\left\vert
0\right\rangle +\tilde{\beta}_{1}^{1}\left\vert 1\right\rangle \right)
\otimes \left( \tilde{\alpha}_{2}^{2}\left\vert 0\right\rangle +\tilde{\beta}%
_{2}^{2}\left\vert 1\right\rangle \right) \otimes \left( \tilde{\alpha}%
_{3}^{3}\left\vert 0\right\rangle +\tilde{\beta}_{3}^{3}\left\vert
1\right\rangle \right)  \label{76} \\
&&+\left( \tilde{\alpha}_{1}^{2}\left\vert 0\right\rangle +\tilde{\beta}%
_{1}^{2}\left\vert 1\right\rangle \right) \otimes \left( \tilde{\alpha}%
_{2}^{3}\left\vert 0\right\rangle +\tilde{\beta}_{2}^{3}\left\vert
1\right\rangle \right) \otimes \left( \tilde{\alpha}_{3}^{1}\left\vert
0\right\rangle +\tilde{\beta}_{3}^{1}\left\vert 1\right\rangle \right) 
\nonumber \\
&&+\left( \tilde{\alpha}_{1}^{3}\left\vert 0\right\rangle +\tilde{\beta}%
_{1}^{3}\left\vert 1\right\rangle \right) \otimes \left( \tilde{\alpha}%
_{2}^{1}\left\vert 0\right\rangle +\tilde{\beta}_{2}^{1}\left\vert
1\right\rangle \right) \otimes \left( \tilde{\alpha}_{3}^{2}\left\vert
0\right\rangle +\tilde{\beta}_{3}^{2}\left\vert 1\right\rangle \right) 
\nonumber \\
&=&\left\vert 100\right\rangle +\left\vert 010\right\rangle +\left\vert
001\right\rangle .  \nonumber
\end{eqnarray}

For $N$ quantum particles, we can obtain the analogy to W state $\left\vert
\Psi _{W}\right\rangle =\frac{1}{\sqrt{N}}\left( \left\vert
10...0\right\rangle +\left\vert 01...0\right\rangle +...+\left\vert
00...1\right\rangle \right) $, which contains $N$ optical fields as follows%
\begin{eqnarray}
\left\vert \psi _{1}\right\rangle &=&e^{i\lambda ^{\left( 1\right)
}}\left\vert 1\right\rangle +e^{i\lambda ^{\left( 2\right) }}\left\vert
0\right\rangle +...+e^{i\lambda ^{\left( N\right) }}\left\vert
0\right\rangle ,  \label{77} \\
\left\vert \psi _{2}\right\rangle &=&e^{i\lambda ^{\left( 1\right)
}}\left\vert 1\right\rangle +e^{i\lambda ^{\left( 2\right) }}\left\vert
0\right\rangle +...+e^{i\lambda ^{\left( N\right) }}\left\vert
0\right\rangle ,  \nonumber \\
&&\vdots  \nonumber \\
\left\vert \psi _{N}\right\rangle &=&e^{i\lambda ^{\left( 1\right)
}}\left\vert 1\right\rangle +e^{i\lambda ^{\left( 2\right) }}\left\vert
0\right\rangle +...+e^{i\lambda ^{\left( N\right) }}\left\vert
0\right\rangle .  \nonumber
\end{eqnarray}%
Now we discuss a transformed state of W state. Applying the NOT unitary
transformation $\hat{U}_{NOT}:\left\vert 0\right\rangle \leftrightarrow
\left\vert 1\right\rangle $ to $\left\vert \psi _{n}\right\rangle $, we can
obtain the transformed optical fields as follows 
\begin{eqnarray}
\left\vert \psi _{1}\right\rangle &=&e^{i\lambda ^{\left( 1\right)
}}\left\vert 1\right\rangle +e^{i\lambda ^{\left( 2\right) }}\left\vert
0\right\rangle +...+e^{i\lambda ^{\left( N\right) }}\left\vert
0\right\rangle , \\
&&\vdots  \nonumber \\
\hat{U}_{NOT}\left\vert \psi _{n}\right\rangle &=&\hat{U}_{NOT}\left(
e^{i\lambda ^{\left( 1\right) }}\left\vert 1\right\rangle +e^{i\lambda
^{\left( 2\right) }}\left\vert 0\right\rangle +...+e^{i\lambda ^{\left(
N\right) }}\left\vert 0\right\rangle \right) =e^{i\lambda ^{\left( 1\right)
}}\left\vert 0\right\rangle +e^{i\lambda ^{\left( 2\right) }}\left\vert
1\right\rangle +...+e^{i\lambda ^{\left( N\right) }}\left\vert
1\right\rangle ,  \nonumber \\
&&\vdots  \nonumber \\
\left\vert \psi _{N}\right\rangle &=&e^{i\lambda ^{\left( 1\right)
}}\left\vert 1\right\rangle +e^{i\lambda ^{\left( 2\right) }}\left\vert
0\right\rangle +...+e^{i\lambda ^{\left( N\right) }}\left\vert
0\right\rangle .  \nonumber
\end{eqnarray}%
and the transformed imitation state can be expressed as%
\begin{equation}
\left\vert \Psi \right\rangle =\left\vert 10...1_{n}...00\right\rangle
+...+\left\vert 00...0_{n}...00\right\rangle +...+\left\vert
00...1_{n}...01\right\rangle .  \label{79}
\end{equation}%
Similar to quantum computation, we can obtain a new state by using only one
step instead of $N$ steps as classical computation. This is another example
to demonstrate parallel computing capability.

\subsubsection{Numerical simulations of the optical analogies to the
three-particle quantum states \label{Sec III.C.2}}

\begin{figure}[htbp]
\centering\includegraphics[height=1.7365in, width=3.5284in]{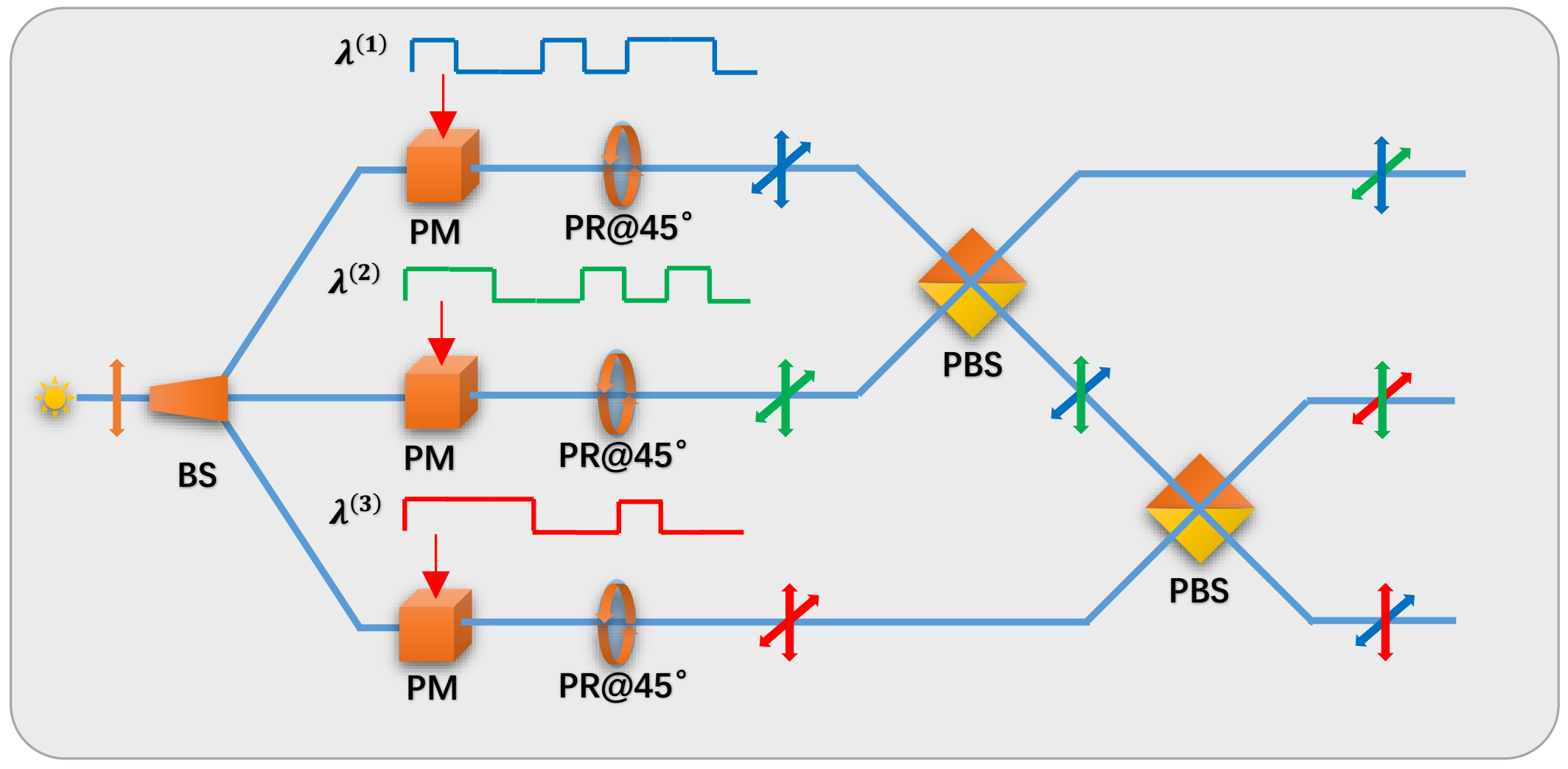}
\caption{The scheme to realize the optical analogy to quantum GHZ state is
shown, where PBS: polarization beam splitters, PR@$45^{\circ }$: $45^{\circ
} $ polarization rotators.}
\label{fig11}
\end{figure}

\begin{figure}[htbp]
\centering\includegraphics[height=1.7781in, width=3.5328in]{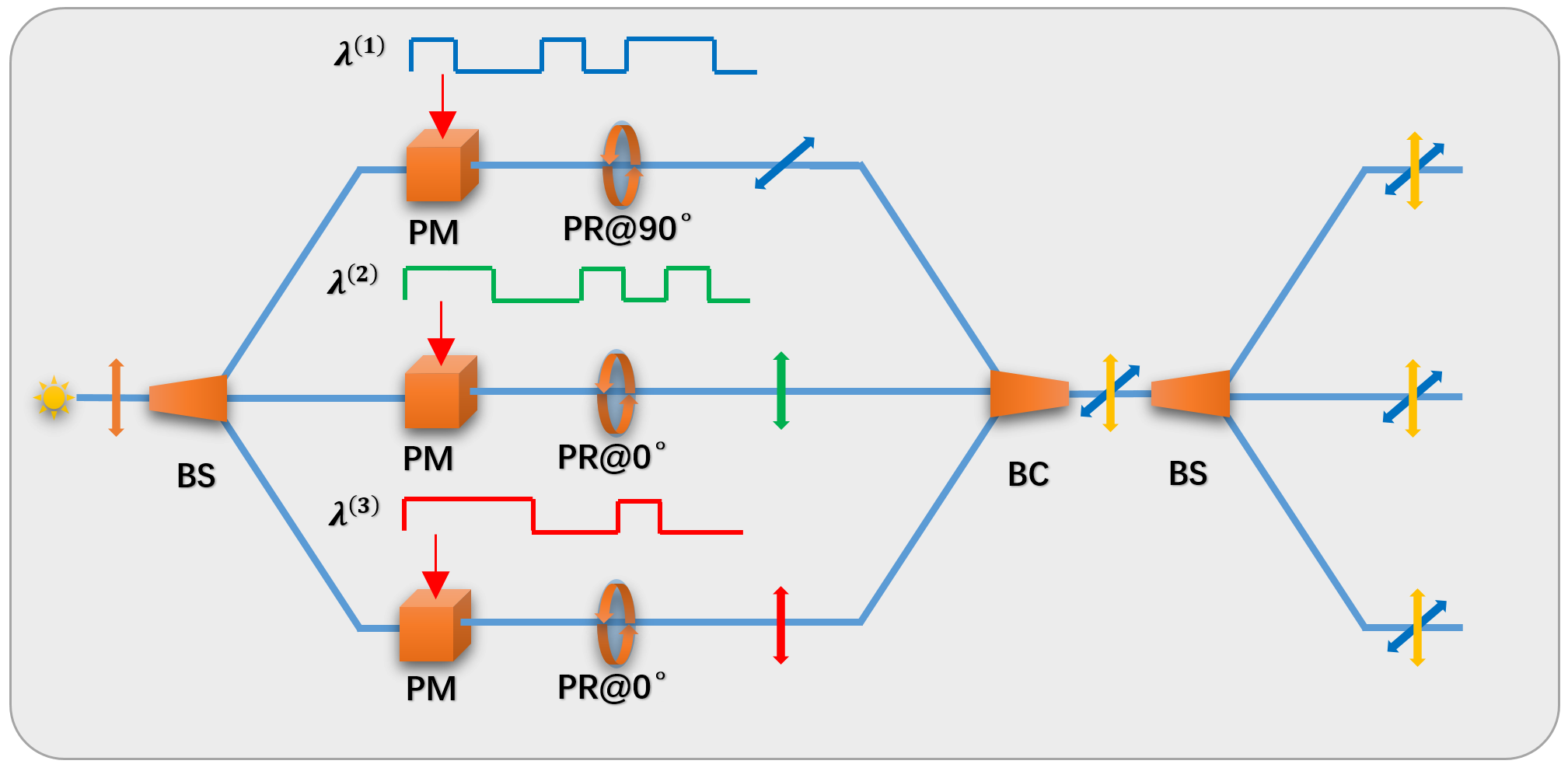}
\caption{The sheme to realize the optical analogy to quantum W state is
shown, where BC: beam couplers, BS: beam splitters, PR@$0^{\circ }$: $%
0^{\circ }$ polarization rotators, PR@$90^{\circ }$: $90^{\circ }$
polarization rotators.}
\label{fig12}
\end{figure}

In last subsection, we demonstrate the optical analogies of quantum states,
which are the product state, Bell states, GHZ state and W state. In this
subsection, we discuss numerical simulations of two optical analogies using
the software OPTISYSTEM. To construct the product state, we first choose
three PPSs $\lambda ^{\left( 1\right) },\lambda ^{\left( 2\right) }$ and $%
\lambda ^{\left( 3\right) }$ to modulate the optical fields, and obtain%
\begin{eqnarray}
E_{1}\left( t\right) &=&\left( A_{\uparrow }+A_{\rightarrow }\right)
e^{-i\left( \omega t+\lambda _{k}^{\left( 1\right) }\right) },  \label{80} \\
E_{2}\left( t\right) &=&\left( A_{\uparrow }+A_{\rightarrow }\right)
e^{-i\left( \omega t+\lambda _{k}^{\left( 2\right) }\right) },  \nonumber \\
E_{3}\left( t\right) &=&\left( A_{\uparrow }+A_{\rightarrow }\right)
e^{-i\left( \omega t+\lambda _{k}^{\left( 3\right) }\right) }.  \nonumber
\end{eqnarray}%
According to Eq. (\ref{46}), the optical analogies to GHZ state can be
written as follows%
\begin{eqnarray}
E_{1}\left( t\right) &=&A_{\uparrow }e^{-i\left( \omega t+\lambda
_{k}^{\left( 1\right) }\right) }+A_{\rightarrow }e^{-i\left( \omega
t+\lambda _{k}^{\left( 2\right) }\right) },  \label{81} \\
E_{2}\left( t\right) &=&A_{\uparrow }e^{-i\left( \omega t+\lambda
_{k}^{\left( 2\right) }\right) }+A_{\rightarrow }e^{-i\left( \omega
t+\lambda _{k}^{\left( 3\right) }\right) },  \nonumber \\
E_{3}\left( t\right) &=&A_{\uparrow }e^{-i\left( \omega t+\lambda
_{k}^{\left( 3\right) }\right) }+A_{\rightarrow }e^{-i\left( \omega
t+\lambda _{k}^{\left( 1\right) }\right) },  \nonumber
\end{eqnarray}%
which can be realized by mode exchange of the produce state by using
polarization beam splitters, as shown in Fig. \ref{fig11}. Then we can
express the optical analogies to W state as follows 
\begin{eqnarray}
E_{1}\left( t\right) &=&A_{\rightarrow }e^{-i\left( \omega t+\lambda
_{k}^{\left( 1\right) }\right) }+A_{\uparrow }e^{-i\left( \omega t+\lambda
_{k}^{\left( 2\right) }\right) }+A_{\uparrow }e^{-i\left( \omega t+\lambda
_{k}^{\left( 3\right) }\right) },  \label{82} \\
E_{2}\left( t\right) &=&A_{\rightarrow }e^{-i\left( \omega t+\lambda
_{k}^{\left( 1\right) }\right) }+A_{\uparrow }e^{-i\left( \omega t+\lambda
_{k}^{\left( 2\right) }\right) }+A_{\uparrow }e^{-i\left( \omega t+\lambda
_{k}^{\left( 3\right) }\right) },  \nonumber \\
E_{3}\left( t\right) &=&A_{\rightarrow }e^{-i\left( \omega t+\lambda
_{k}^{\left( 1\right) }\right) }+A_{\uparrow }e^{-i\left( \omega t+\lambda
_{k}^{\left( 2\right) }\right) }+A_{\uparrow }e^{-i\left( \omega t+\lambda
_{k}^{\left( 3\right) }\right) },  \nonumber
\end{eqnarray}%
which can be realized by using the beam coupler and splitter, as shown in
Fig. \ref{fig12}.

Further, we make use of the coherent demodulation method mentioned in
Section \ref{Sec II} to obtain the matrix $M$. Because each field of the
imitations of GHZ state and W state has two orthogonal polarization modes,
the coherent demodulation scheme need two LO fields with same orthogonal
modes as the SO fields, as shown in Fig. \ref{fig9a}. By using the software
OPTISYSTEM, we construct the numerical simulation model as shown in Fig. \ref%
{fig13} for the coherent demodulation as mensioned in Fig. \ref{fig9a}. In
Fig. \ref{fig14}, the electric signals of PDs are shown when the SO field is 
$E_{1}\left( t\right) $ of Eq. (\ref{81}) and the LO fields are modulated
with PPSs $\lambda ^{\left( 1\right) }$, $\lambda ^{\left( 2\right) }$, $%
\lambda ^{\left( 3\right) }$ respectively. Finally, by performing
correlation analysis as mentioned in Section \ref{Sec II.B}, we can obtain
the results for the three fields of Eq. (\ref{81}), as shown in Fig. \ref%
{fig15}. After disposing of constant and normalization, we can express the
measurement result as the matrix $M$ mentioned in Eq. (\ref{66}). For the
imitations of W state, the same results are shown in Fig. \ref{fig16} and
Fig. \ref{fig17}. After disposing of constant and normalization, we can also
express the result as the matrix $M$ mentioned in Eq. (\ref{75}).

\begin{figure}[htbp]
\centering\includegraphics[height=1.9329in, width=3.6123in]{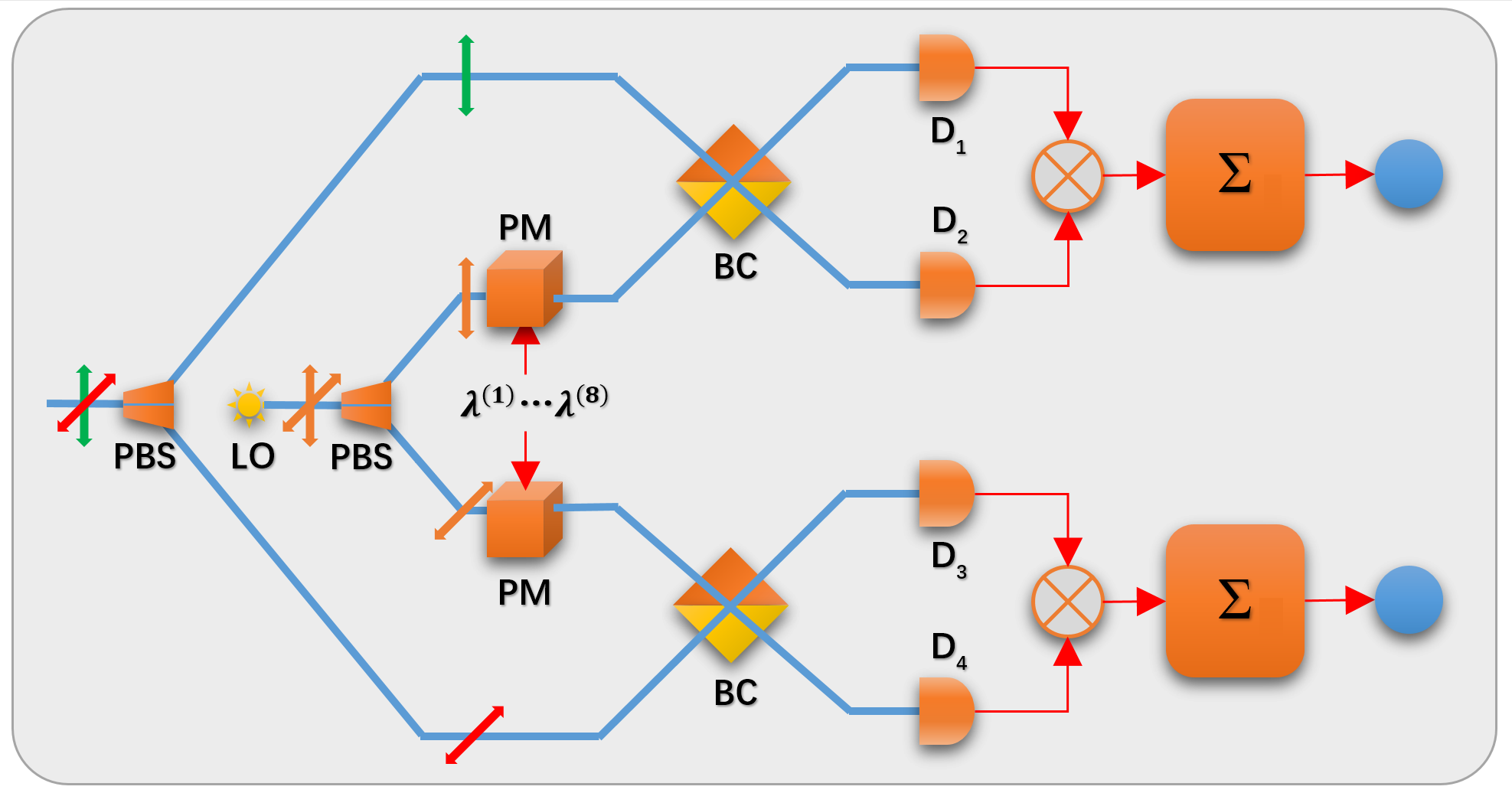}
\caption{The numerical simulation model of the coherent demodulation as
mentioned in Fig. \protect\ref{fig9a} is shown, where PBS: polarization beam
splitters and BC: beam couplers.}
\label{fig13}
\end{figure}

\begin{figure}[htbp]
\centering\includegraphics[height=3.8173in, width=3.8735in]{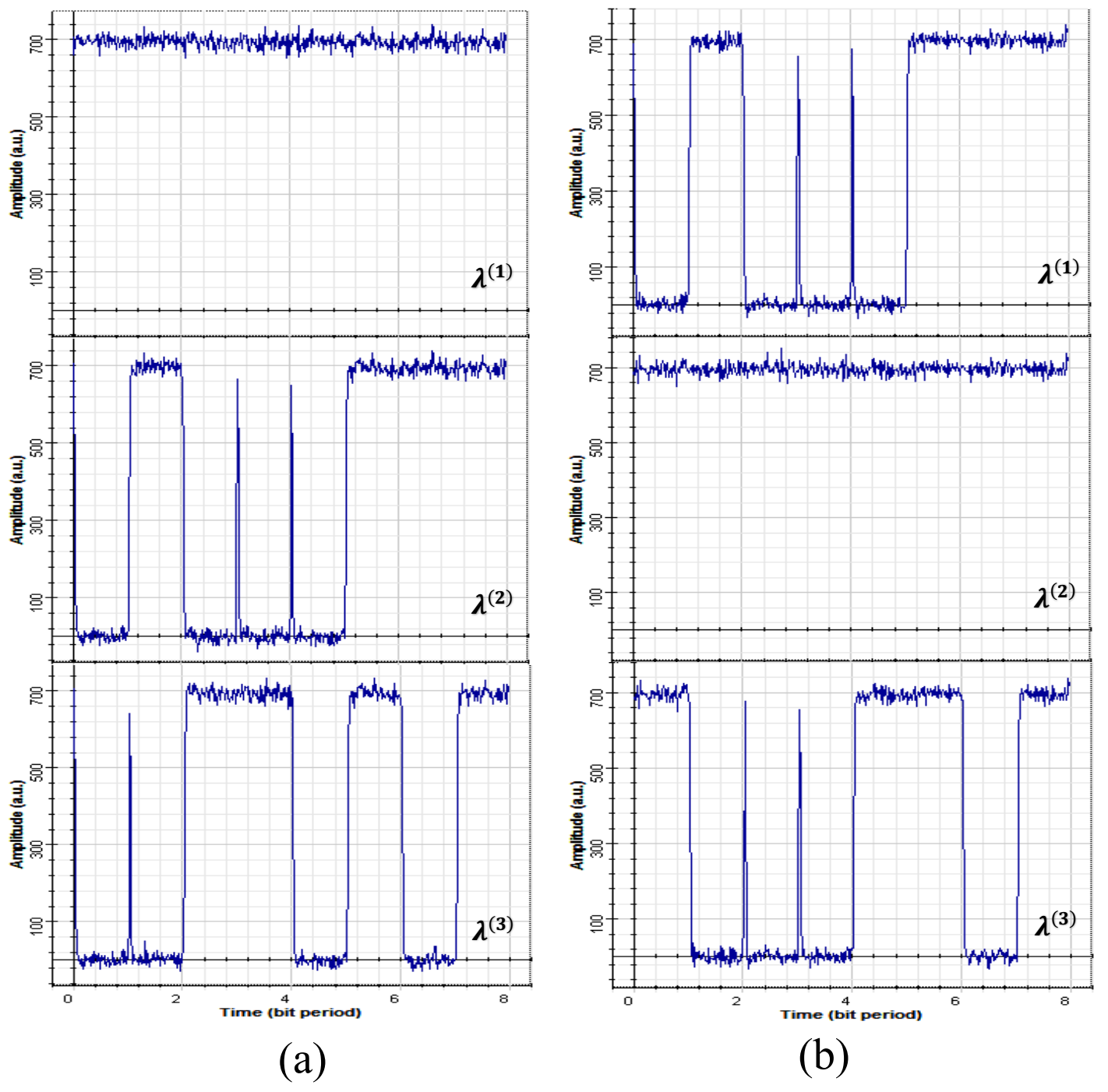}
\caption{The electric signals for the field $E_{1}\left( t\right) $ of the
GHZ imitation state with LO fields modulated with $\protect\lambda ^{\left(
1\right) }$, $\protect\lambda ^{\left( 2\right) }$ and $\protect\lambda %
^{\left( 3\right) }$ are shown, where (a) and (b) represent the two
orthogonal modes $\left\vert 0\right\rangle $ and $\left\vert 1\right\rangle 
$ respectively.}
\label{fig14}
\end{figure}

\begin{figure}[htbp]
\centering\includegraphics[height=2.0392in, width=3.9245in]{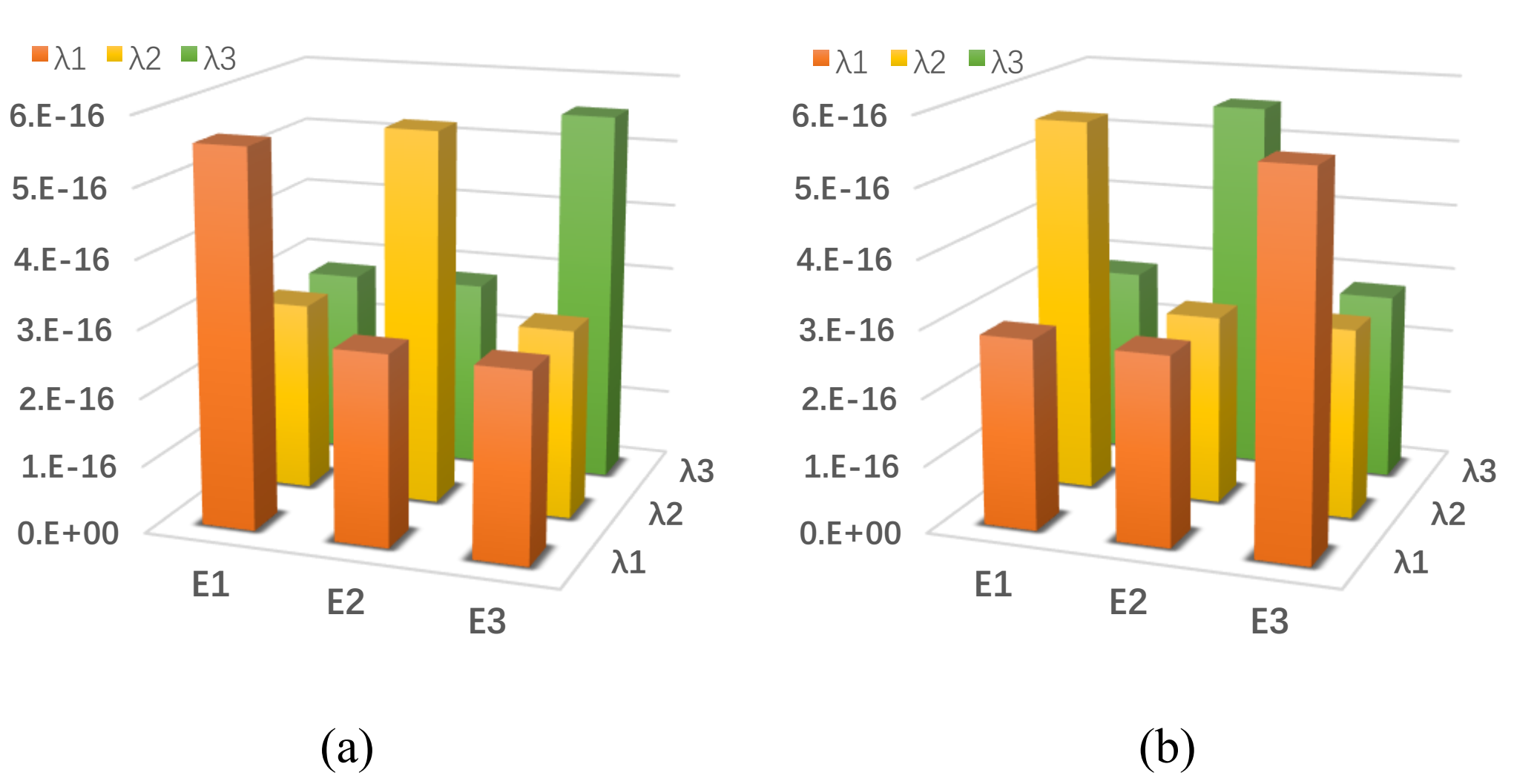}
\caption{The correlation analysis results of GHZ state are shown, where (a)
and (b) represent two orthogonal modes $A_{\uparrow }$ and $A_{\rightarrow }$
respectively, $E1$, $E2$, $E3$ three fields and $\protect\lambda 1$, $%
\protect\lambda 2$, $\protect\lambda 3$ the sequences $\protect\lambda %
^{\left( 1\right) }$, $\protect\lambda ^{\left( 2\right) }$, $\protect%
\lambda ^{\left( 3\right) }$ modulating on LO.}
\label{fig15}
\end{figure}

\begin{figure}[htbp]
\centering\includegraphics[height=3.6132in, width=3.7516in]{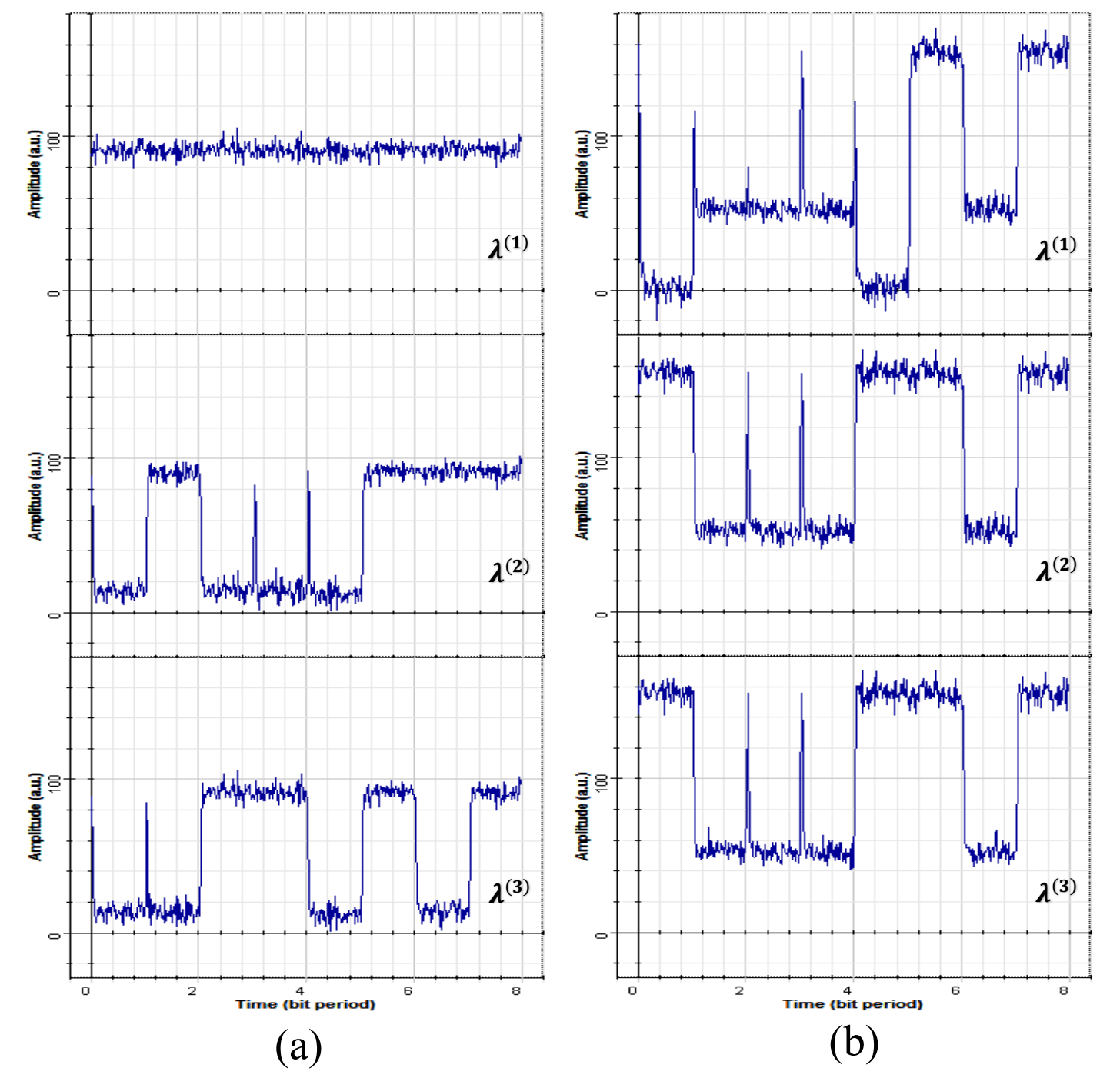}
\caption{The electric signals for the field $E_{1}\left( t\right) $ of the W
imitation state with LO fields modulated with $\protect\lambda ^{\left(
1\right) }$, $\protect\lambda ^{\left( 2\right) }$ and $\protect\lambda %
^{\left( 3\right) }$ are shown, where (a) and (b) represent two orthogonal
modes $A_{\rightarrow }$ and $A_{\uparrow }$, respectively.}
\label{fig16}
\end{figure}

\begin{figure}[htbp]
\centering\includegraphics[height=1.8827in, width=3.704in]{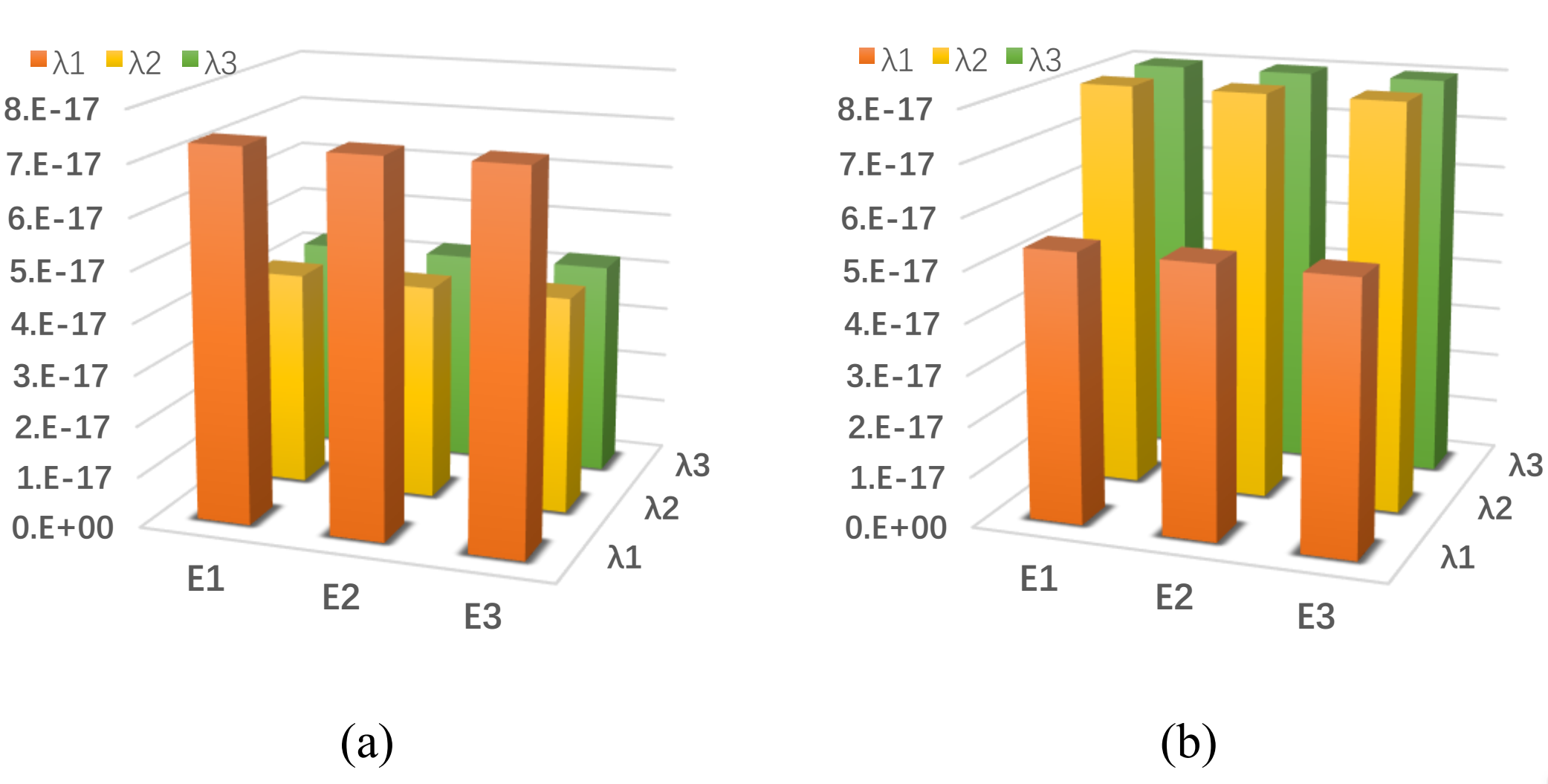}
\caption{The correlation analysis results of W state are shown, where (a)
and (b) represent two orthogonal modes $A_{\rightarrow }$ and $A_{\uparrow }$
respectively, $E1$, $E2$, $E3$ three fields and $\protect\lambda 1$, $%
\protect\lambda 2$, $\protect\lambda 3$ the sequences $\protect\lambda %
^{\left( 1\right) }$, $\protect\lambda ^{\left( 2\right) }$, $\protect%
\lambda ^{\left( 3\right) }$ modulating on LO.}
\label{fig17}
\end{figure}

\section{Imitation of quantum computation\label{Sec IV}}

In this section, we will propose a gate array model to imitate quantum
computation. In quantum computation, any quantum state can be obtained from
an initial state by using a gate array constructed with universal CNOT gate
and other single qubit gate \cite{Nielsen}. Similarly, we can construct gate
array models to produce the imitations of all kinds of quantum states, such
as GHZ state and W state, even very sophisticated states like the results of
Shor's algorithm. We consider the gate array models can be employed as the
imitations of quantum computation. Based on this understanding, we construct
some gate array models to imitate Shor's algorithm, Grover's algorithm and
quantum Fourier algorithm.

\subsection{Gate array model to imitate quantum computation \label{Sec IV.A}}

\begin{figure}[tbph]
\centering\includegraphics[height=1.6751in, width=4.4201in]{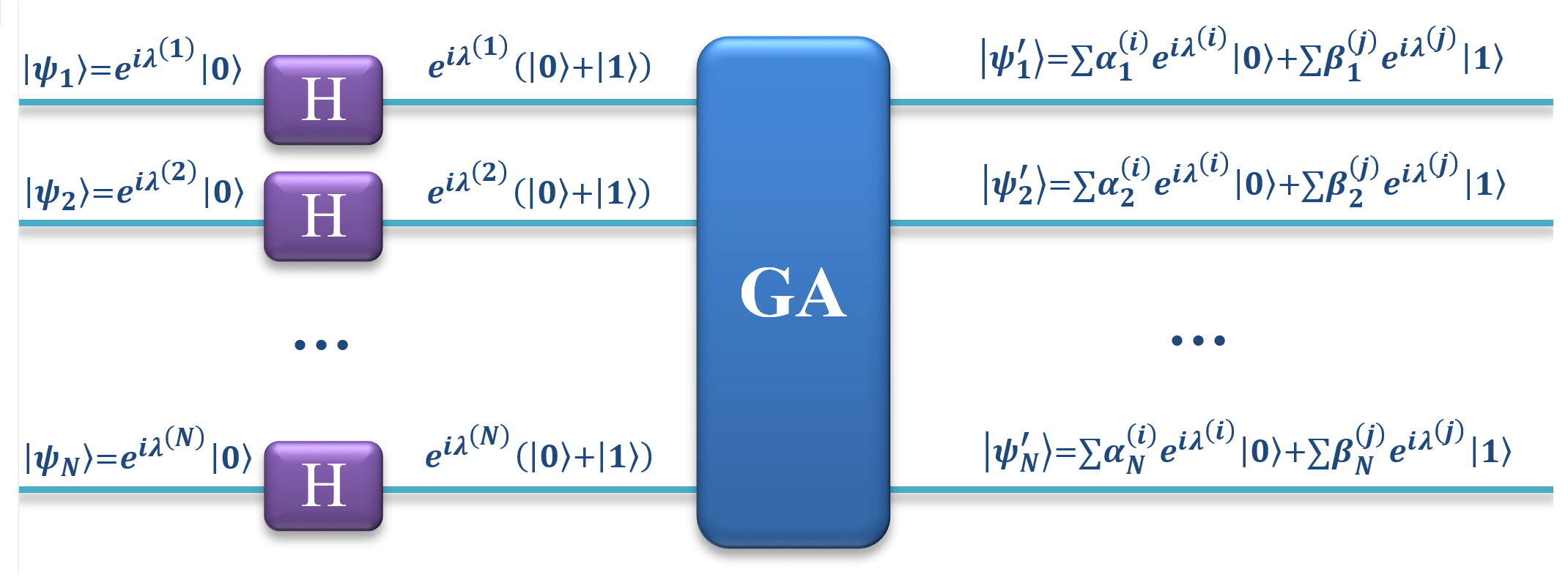}
\caption{A GA model to imitate quantum computation is shown, where GA
denotes the gate array.}
\label{fig18}
\end{figure}

In Ref. \cite{Fu1}, a constructure pathway of the imitation states is shown.
Here we use the same model as shown Fig. \ref{fig18}, however the gate array
model does not always achieve unitary transformations similar to quantum
computation. Now we discuss some basic units of the gate array model besides
the unitary transformation mensioned in Eq. (\ref{12}) and the mode
exchanger shown in Fig. \ref{fig7}.

(1) Combiner and splitter

\begin{figure}[htbp]
\centering\includegraphics[height=1.0352in, width=4.1753in]{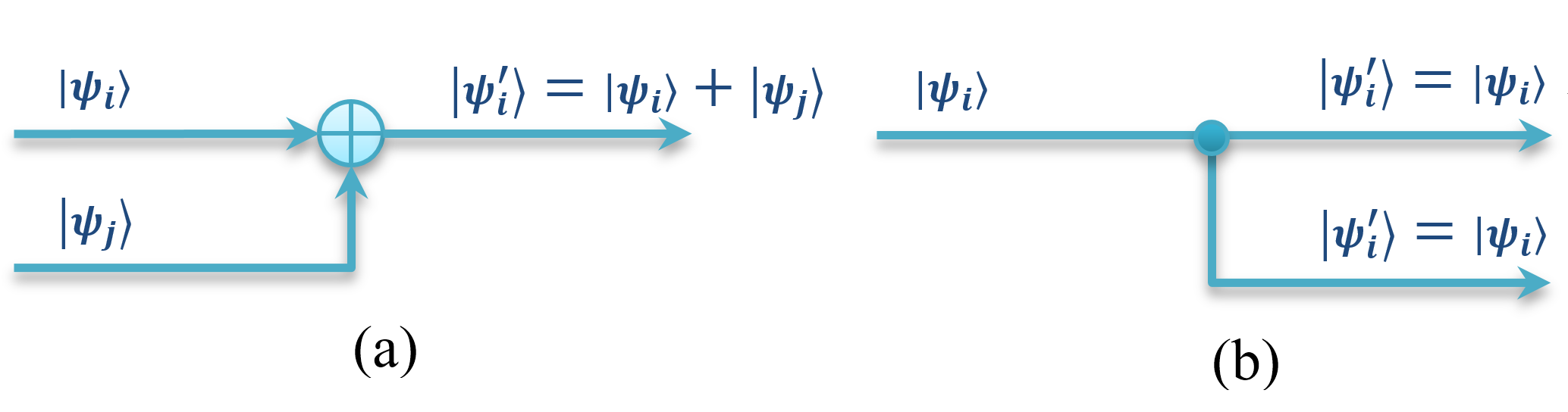}
\caption{Two basic devices (a) coupler and (b) splitter are shown.}
\label{fig19}
\end{figure}

Different from quantum state, we can conveniently combine and split an
optical field by using an optical coupler/splitter device, which principles
are discussed in Sec. \ref{Sec II.C}. The two basic devices are shown in
Fig. \ref{fig19} (a) and (b), respectively.

(b) Mode control gates

\begin{figure}[htbp]
\centering\includegraphics[height=2.0453in, width=3.6642in]{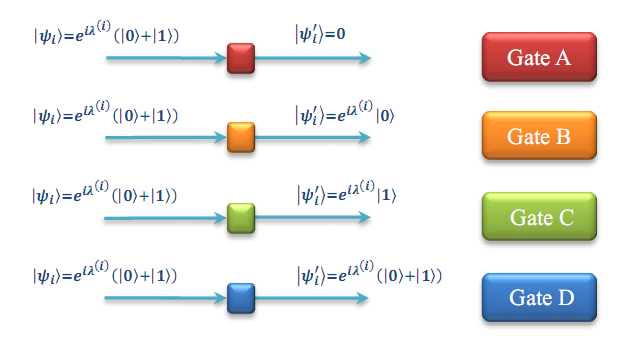}
\caption{Mode control gates as selective mode transit devices are shown.}
\label{fig20}
\end{figure}

Further, we define $4$ kinds of mode control gates as selective mode transit
devices with one input and one output as shown in Fig. \ref{fig20}. They are
defined as follows 
\begin{eqnarray}
\mathbf{Gate\ }A &:&\left\vert \psi _{i}\right\rangle =e^{i\lambda ^{\left(
i\right) }}\left( \left\vert 0\right\rangle +\left\vert 1\right\rangle
\right) \rightarrow \left\vert \psi _{i}^{\prime }\right\rangle =0,
\label{85} \\
\mathbf{Gate\ }B &:&\left\vert \psi _{i}\right\rangle =e^{i\lambda ^{\left(
i\right) }}\left( \left\vert 0\right\rangle +\left\vert 1\right\rangle
\right) \rightarrow \left\vert \psi _{i}^{\prime }\right\rangle =e^{i\lambda
^{\left( i\right) }}\left\vert 0\right\rangle ,  \nonumber \\
\mathbf{Gate\ }C &:&\left\vert \psi _{i}\right\rangle =e^{i\lambda ^{\left(
i\right) }}\left( \left\vert 0\right\rangle +\left\vert 1\right\rangle
\right) \rightarrow \left\vert \psi _{i}^{\prime }\right\rangle =e^{i\lambda
^{\left( i\right) }}\left\vert 1\right\rangle ,  \nonumber \\
\mathbf{Gate\ }D &:&\left\vert \psi _{i}\right\rangle =e^{i\lambda ^{\left(
i\right) }}\left( \left\vert 0\right\rangle +\left\vert 1\right\rangle
\right) \rightarrow \left\vert \psi _{i}^{\prime }\right\rangle =e^{i\lambda
^{\left( i\right) }}\left( \left\vert 0\right\rangle +\left\vert
1\right\rangle \right) .  \nonumber
\end{eqnarray}

Now we discuss one of basic structures of gate array models. According to
Sec. \ref{Sec III}, we can in principle imitate all quantum states by using
the SCPM. Similar to field programmable gate array (FPGA), we propose a
simple structure of gate array to satisfy the SCPM as shown in Fig. \ref%
{fig21}. Gate array $G_{kj}$ constituted by the basic units can transform $%
\left\vert \psi _{k}\right\rangle $ to achieve certain $\left\vert \psi
_{k}^{\prime }\right\rangle $. It is easy to know that a sequence
permutation with circulation of $p+1$ needs at least $p$ combiner devices
and $2p$ control gates. We believe that many imitation states can be
constructed by applying this structure that will be strictly proved in
future paper.

\begin{figure}[htbp]
\centering\includegraphics[height=1.3984in, width=4.7997in]{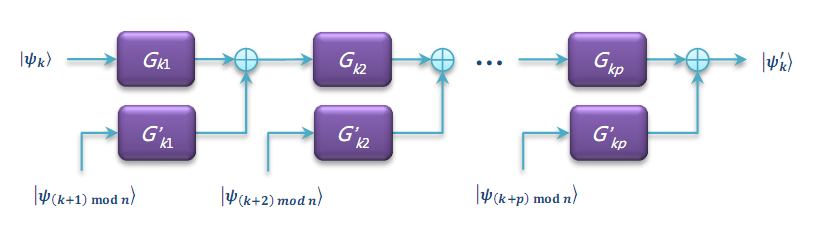}
\caption{One of basic structrues of gate array model is shown.}
\label{fig21}
\end{figure}

\begin{figure}[htbp]
\centering\includegraphics[height=1.4563in, width=4.9485in]{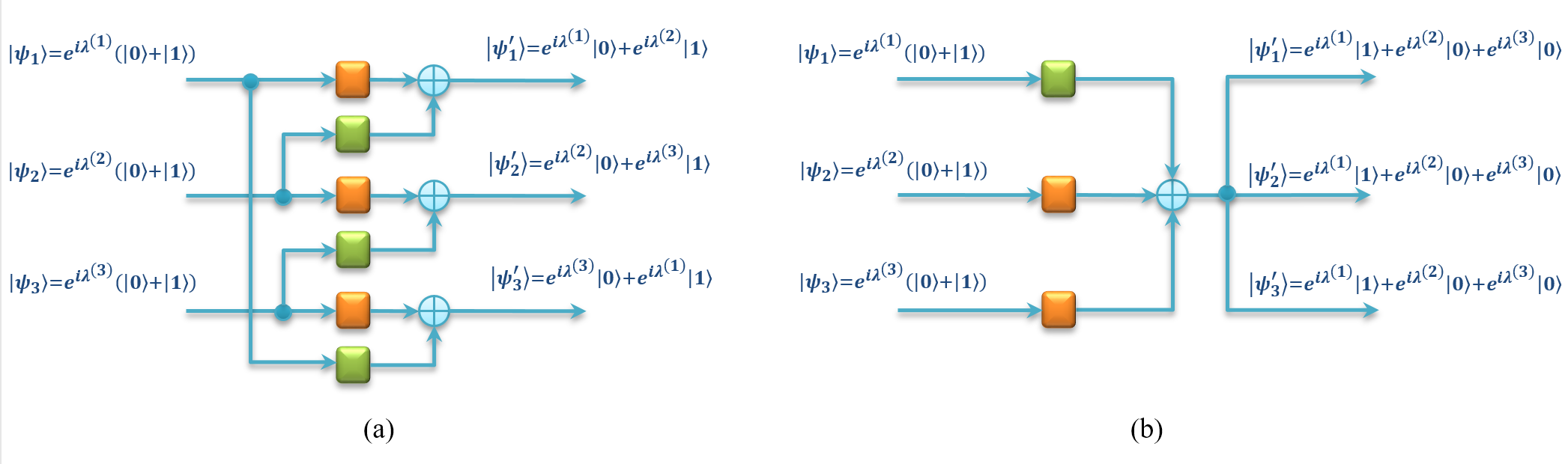}
\caption{The gate array models to transform the product states to (a) GHZ
state and (b) W state are shown.}
\label{fig22}
\end{figure}

Finally, we illustrate two gate array models to transform the product state
to GHZ state and W state as shown in Fig. \ref{fig22}.

\subsection{Imitation of quantum algorithm \label{Sec IV.B}}

\subsubsection{Analogy to Shor's Algorithm \label{Sec IV.B.1}}

Shor's algorithm is the most famous quantum algorithm for integer
factorization that runs only in polynomial time on a quantum computer \cite%
{Shor}. Specifically it takes time and quantum gates of order $O((\log
L)^{2}(\log \log L)(\log \log \log L))$ using fast multiplication,
demonstrating that the integer factorization problem can be efficiently
solved on a quantum computer and is thus in the complexity class bounded
error quantum polynomial time problem (BQP).

In this section, we propose a gate array model employing $8$ optical fields
produces optical analogies that imitates the result state of Shor's
algorithm to factor $L=15$ into $3\times 5$. First, we chose a random number 
$a$ coprime with $15$, for example $a=7$. We define a function as follow 
\begin{equation}
f\left( x\right) =a^{x} mod L=7^{x} mod 15.  \label{86}
\end{equation}%
The key step of the Shor's algorithm is to obtain the period $r$ to satisfy 
\begin{equation}
f\left( x+r\right) =7^{x+r} mod 15=7^{x} mod 15=f\left( x\right) .
\label{87}
\end{equation}%
In order to construct $f\left( x\right) $, we prepare $8$ optical fields
modulated with $8$ PPSs 
\begin{equation}
\left\vert \psi _{n}\right\rangle =e^{i\lambda ^{\left( n\right) }}\left(
\left\vert 0\right\rangle +\left\vert 1\right\rangle \right) ,n=1\ldots 8.
\label{88}
\end{equation}%
We can express the product state as follows 
\begin{equation}
\left\vert \Psi \right\rangle =\left\vert \psi _{1}\right\rangle \otimes
\ldots \otimes \left\vert \psi _{8}\right\rangle
=\sum\limits_{j=0}^{2^{8}-1}\left\vert j\right\rangle .  \label{89}
\end{equation}

\begin{figure}[htbp]
\centering\includegraphics[height=2.9162in, width=4.9147in]{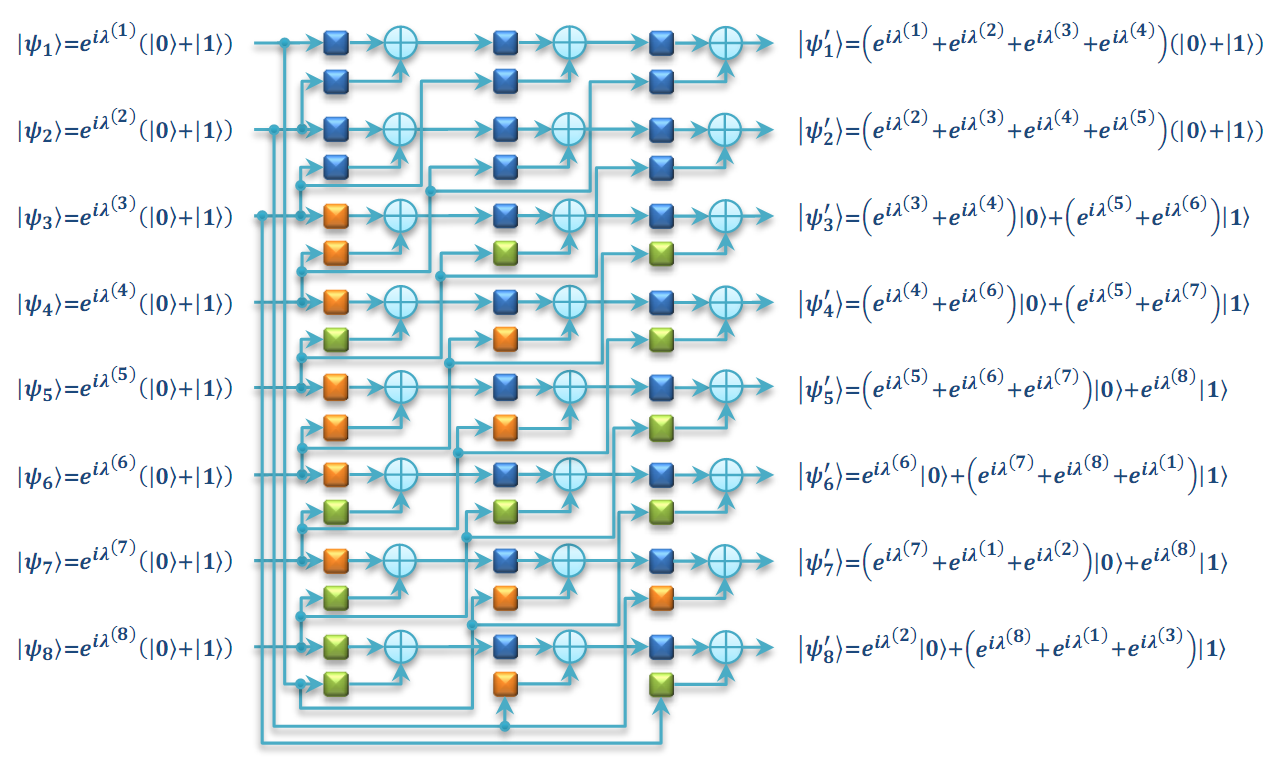}
\caption{A gate array model to imitate Shor's algorithm is shown.}
\label{fig23}
\end{figure}

Further, we construct the gate array model as shown in Fig. \ref{fig23}.
After passing through the gate array, the optical fields become the
following forms 
\begin{eqnarray}
\left\vert \psi _{1}^{\prime }\right\rangle &=&\left( e^{i\lambda ^{\left(
1\right) }}+e^{i\lambda ^{\left( 2\right) }}+e^{i\lambda ^{\left( 3\right)
}}+e^{i\lambda ^{\left( 4\right) }}\right) \left( \left\vert 0\right\rangle
+\left\vert 1\right\rangle \right) ,  \label{90} \\
\left\vert \psi _{2}^{\prime }\right\rangle &=&\left( e^{i\lambda ^{\left(
2\right) }}+e^{i\lambda ^{\left( 3\right) }}+e^{i\lambda ^{\left( 4\right)
}}+e^{i\lambda ^{\left( 5\right) }}\right) \left( \left\vert 0\right\rangle
+\left\vert 1\right\rangle \right) ,  \nonumber \\
\left\vert \psi _{3}^{\prime }\right\rangle &=&\left( e^{i\lambda ^{\left(
3\right) }}+e^{i\lambda ^{\left( 4\right) }}\right) \left\vert
0\right\rangle +\left( e^{i\lambda ^{\left( 5\right) }}+e^{i\lambda ^{\left(
6\right) }}\right) \left\vert 1\right\rangle ,  \nonumber \\
\left\vert \psi _{4}^{\prime }\right\rangle &=&\left( e^{i\lambda ^{\left(
4\right) }}+e^{i\lambda ^{\left( 6\right) }}\right) \left\vert
0\right\rangle +\left( e^{i\lambda ^{\left( 5\right) }}+e^{i\lambda ^{\left(
7\right) }}\right) \left\vert 1\right\rangle ,  \nonumber \\
\left\vert \psi _{5}^{\prime }\right\rangle &=&\left( e^{i\lambda ^{\left(
5\right) }}+e^{i\lambda ^{\left( 6\right) }}+e^{i\lambda ^{\left( 7\right)
}}\right) \left\vert 0\right\rangle +e^{i\lambda ^{\left( 8\right)
}}\left\vert 1\right\rangle ,  \nonumber \\
\left\vert \psi _{6}^{\prime }\right\rangle &=&e^{i\lambda ^{\left( 6\right)
}}\left\vert 0\right\rangle +\left( e^{i\lambda ^{\left( 7\right)
}}+e^{i\lambda ^{\left( 8\right) }}+e^{i\lambda ^{\left( 1\right) }}\right)
\left\vert 1\right\rangle ,  \nonumber \\
\left\vert \psi _{7}^{\prime }\right\rangle &=&\left( e^{i\lambda ^{\left(
7\right) }}+e^{i\lambda ^{\left( 1\right) }}+e^{i\lambda ^{\left( 2\right)
}}\right) \left\vert 0\right\rangle +e^{i\lambda ^{\left( 8\right)
}}\left\vert 1\right\rangle ,  \nonumber \\
\left\vert \psi _{8}^{\prime }\right\rangle &=&e^{i\lambda ^{\left( 2\right)
}}\left\vert 0\right\rangle +\left( e^{i\lambda ^{\left( 8\right)
}}+e^{i\lambda ^{\left( 1\right) }}+e^{i\lambda ^{\left( 3\right) }}\right)
\left\vert 1\right\rangle .  \nonumber
\end{eqnarray}

Using the coherent demodulation, we can obtain the matrix 
\begin{equation}
M\left( \tilde{\alpha}_{i}^{j},\tilde{\beta}_{i}^{j}\right) =\left( 
\begin{array}{cccccccc}
\left( 1,1\right)  & \left( 1,1\right)  & \left( 1,1\right)  & \left(
1,1\right)  & 0 & 0 & 0 & 0 \\ 
0 & \left( 1,1\right)  & \left( 1,1\right)  & \left( 1,1\right)  & \left(
1,1\right)  & 0 & 0 & 0 \\ 
0 & 0 & \left( 1,0\right)  & \left( 1,0\right)  & \left( 0,1\right)  & 
\left( 0,1\right)  & 0 & 0 \\ 
0 & 0 & 0 & \left( 1,0\right)  & \left( 0,1\right)  & \left( 1,0\right)  & 
\left( 0,1\right)  & 0 \\ 
0 & 0 & 0 & 0 & \left( 1,0\right)  & \left( 1,0\right)  & \left( 1,0\right) 
& \left( 0,1\right)  \\ 
\left( 0,1\right)  & 0 & 0 & 0 & 0 & \left( 1,0\right)  & \left( 0,1\right) 
& \left( 0,1\right)  \\ 
\left( 1,0\right)  & \left( 1,0\right)  & 0 & 0 & 0 & 0 & \left( 1,0\right) 
& \left( 0,1\right)  \\ 
\left( 0,1\right)  & \left( 1,0\right)  & \left( 0,1\right)  & 0 & 0 & 0 & 0
& \left( 0,1\right) 
\end{array}%
\right) .  \label{91}
\end{equation}%
Using the scheme mentioned in Sec. \ref{Sec III.C}, we can obtain the
imitated state 
\begin{eqnarray}
\left\vert \Psi ^{\prime }\right\rangle  &=&\left\vert
7^{x}mod15\right\rangle \left\vert x\right\rangle =\left( \left\vert
0\right\rangle +\left\vert 4\right\rangle +\left\vert 8\right\rangle
+\left\vert 12\right\rangle \right) \left\vert 1\right\rangle   \label{92} \\
&&+\left( \left\vert 1\right\rangle +\left\vert 5\right\rangle +\left\vert
9\right\rangle +\left\vert 13\right\rangle \right) \left\vert 7\right\rangle 
\nonumber \\
&&+\left( \left\vert 2\right\rangle +\left\vert 6\right\rangle +\left\vert
10\right\rangle +\left\vert 14\right\rangle \right) \left\vert
4\right\rangle   \nonumber \\
&&+\left( \left\vert 3\right\rangle +\left\vert 7\right\rangle +\left\vert
11\right\rangle +\left\vert 15\right\rangle \right) \left\vert
13\right\rangle .  \nonumber
\end{eqnarray}%
where the state $\left\vert 7^{x}mod15\right\rangle $ is represented with
the optical fields $\left\vert \psi _{1}^{\prime }\right\rangle \sim
\left\vert \psi _{4}^{\prime }\right\rangle $ and $\left\vert x\right\rangle 
$ is represented with the optical fields $\left\vert \psi _{5}^{\prime
}\right\rangle \sim \left\vert \psi _{8}^{\prime }\right\rangle $. There are
four kinds of superposition classified from last four fields containing the
values of $f\left( x\right) $ ($\left\vert 1\right\rangle ,\left\vert
7\right\rangle ,\left\vert 4\right\rangle $ and $\left\vert 13\right\rangle $%
) in the imitation state, which means the period of $f\left( x\right)
=7^{x}mod15$ is $r=4$. It is worth noting that, different from quantum
computing, we might obtain the expected period of without operating quantum
Fourier transformation, at least for relatively small integer factorization.
The remaining task is much easier. Because $L=15,a=7,r=4$, we obtain 
\begin{equation}
\gcd \left( a^{\frac{r}{2}}\pm 1,L\right) =\gcd \left( 7^{\frac{r}{2}}\pm
1,15\right) =\gcd \left( 49\pm 1,15\right) ,  \label{93}
\end{equation}%
where $\gcd (48,15)=3$, and $\gcd (50,15)=5$. Finally, we can deduced that $%
L\left( 15\right) =3\times 5$.

After further research the relation of the factorized integer and the gate
array, we believe this might become a true scheme to imitate quantum Shor's
algorithm. Now the computation cost of the model is analyzed. The operation
steps to get the imitation state include $3\times 8=24$ beam splitting
operations, $6\times 8=48$ polarization mode operations, $3\times 8=24$ beam
coupling operations. It can be seen that the number of operations is
increased with the number of optical fields growth in linear growth.
Considering each PPS has $8$ phase units, the total number of operations is
proportional to $8^{2}$. In conclusion, it takes time and gates of order $%
O(\left( \log L\right) ^{2})$.

\subsubsection{Numerical simulation of the imitation \label{Sec IV.B.2}}

\begin{figure}[htbp]
\centering\includegraphics[height=2.1854in, width=4.0214in]{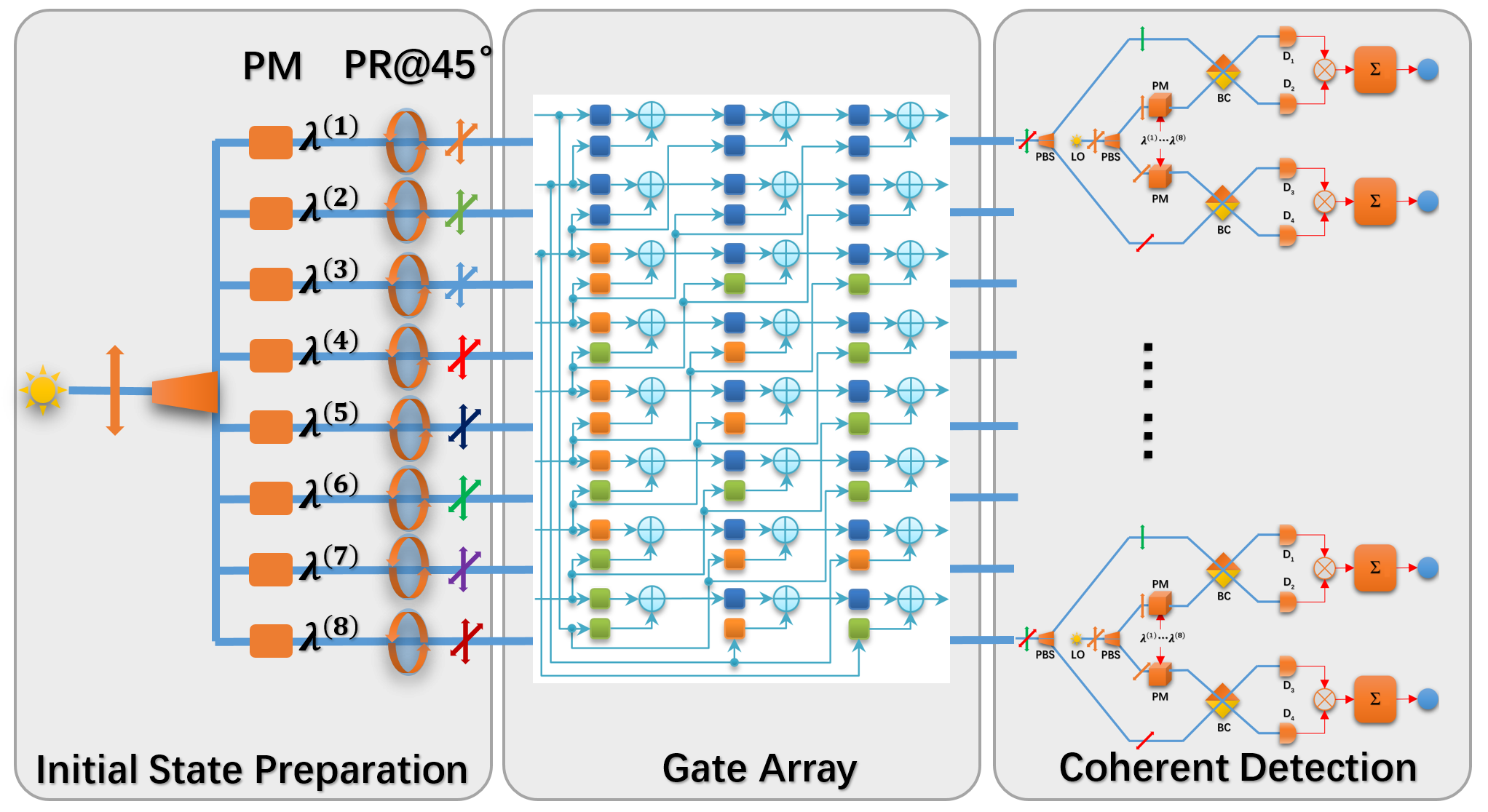}
\caption{The numerical simulation scheme to factorized $15=3\times 5$ using
the software OPTISYSTEM is shown.}
\label{fig24}
\end{figure}

\begin{figure}[htbp]
\centering\includegraphics[height=1.7755in, width=4.1156in]{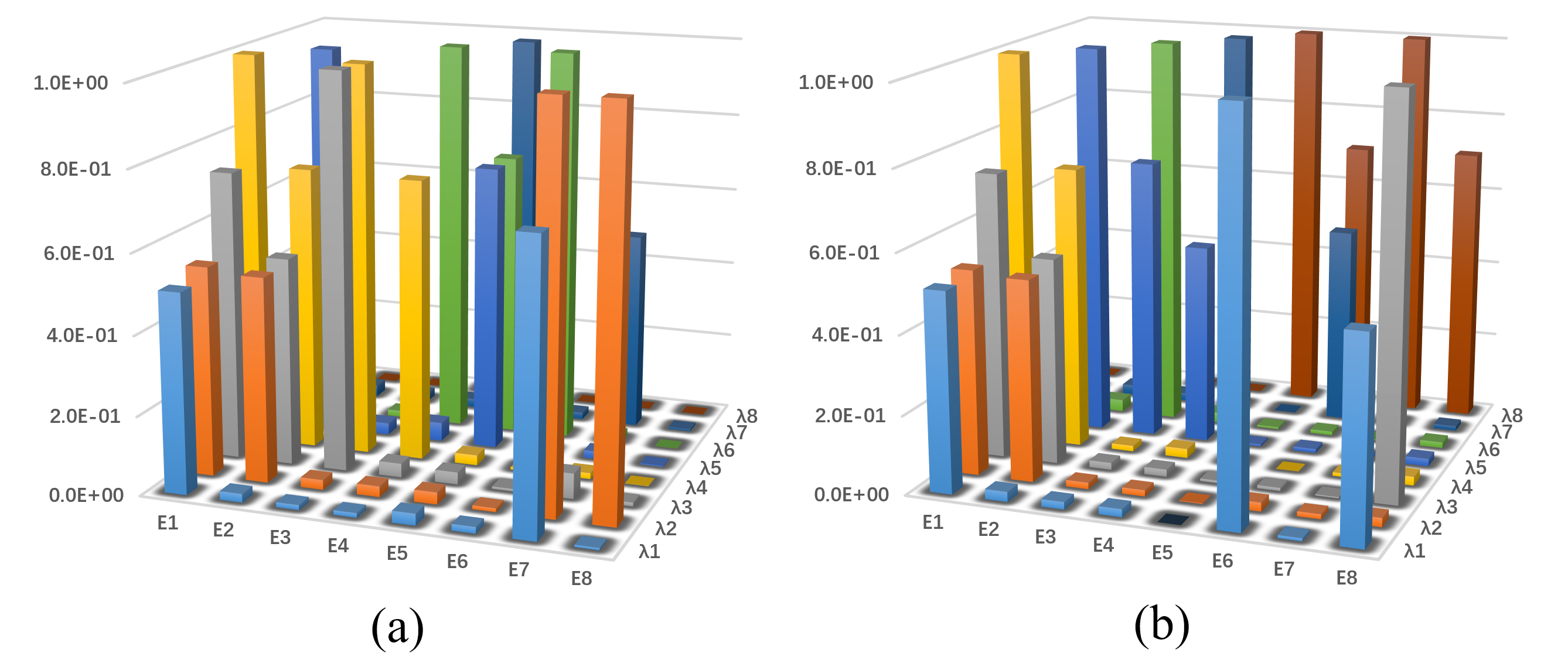}
\caption{The correlation measurement results of the superposition state of
the factorizing algorithm: (a) for mode $A_{\uparrow }$, and (b) for mode $%
A_{\rightarrow }$ are shown, where $E1\sim E8$ represent $\protect\psi %
_{1}\sim \protect\psi _{8}$ optical fields, and $\protect\lambda 1\sim 
\protect\lambda 8$ represent PPSs $\protect\lambda ^{\left( 1\right) }\sim 
\protect\lambda ^{\left( 8\right) }$.}
\label{fig25}
\end{figure}

In this subsection, we discuss the numerical simulation of the gate model as
shown Fig. \ref{fig23} using the software OPTISYSTEM. The schematic diagram
of numerical simulation is shown in Fig. \ref{fig24}. Firstly, the initial
state can be prepared by the method as shown in Fig. \ref{fig6}, that is,
modulate $8$ optical fields $E_{1}\sim E_{8}$ with $8$ PPSs respectively,
and then rotate the polarization of each field by $45^{\circ }$ and evolve
into the states in Eq. (\ref{88}). After numerically simulating the complex
gate array, we obtain the output fields and the electric signals of PDs
after the interference between each fields and LO fields. Finally, the
correlation results are obtained, substracted by the constant part and
normalized, as shown in Fig. \ref{fig25}. After the threshold discrimination
and binarization of results, we can express the measurement results as the $%
M $ matrix of Eq. (\ref{91}) and the imitated state of Eq. (\ref{92}).
(Detailed the numerical calculus, derivation and OPTISYSTEM models will be
provided in the supplementary materials). Using the SCPM mentioned in Sec. %
\ref{Sec II.B}, we can obtain the imitated state as shown Eq. (\ref{90}).

\subsubsection{Grover's Algorithm \label{Sec IV.B.3}}

Grover's algorithm is the quantum algorithm for searching an unsorted
database with $2^{N}$ entries in $O(\sqrt{2^{N}})$ time and using $O(\log
2^{N})$ storage space \cite{Grover}, that is faster than all classical
computations. In fact its time complexity $O(\sqrt{2^{N}})$ is
asymptotically the fastest possible for searching the unsorted database in
the linear quantum model, however, it only provides a quadratic speedup
rather than exponential speedup over their classical counterparts.

There are two key factors for searching an unsorted database: (1) to encode $%
2^{N}$ data into the superposition states of $N$ qubits and form a database;
(2) to verify the existence of a specified number in the superposition
states by measurements. In quantum computation, the first condition is very
easy to be satisfied, but the second is difficult to achieve. Because the
collapse of the superposition states to a specified state is completely
uncontrollable in the quantum measurement. In Grover's algorithm, the
specified transformation is repeated to transform the superposition states
for increasing the probability of the specified state, and after $O(\sqrt{%
2^{N}})$ repetitions the measuring probability will be close to $1/2$.

We now discuss the optical imitation of Grover's algorithm. In our scheme,
we can use $N$ optical fields modulated with PPSs. In principle, we can
encode any $2^{N}$ data as the superposition state of $N$ optical fields to
form a database. Different from the quantum computation, we can control the
superposition state to output any specified state using a mode control gate
array related to the specified state. Let $\left\vert S\right\rangle $
denote the superposition state of $N$ optical fields, 
\begin{equation}
\left\vert S\right\rangle =\sum\limits_{i=1}^{2^{N}}\left\vert
x_{i}\right\rangle ,  \label{94}
\end{equation}%
where $\left\vert x_{i}\right\rangle $ can be encoded by any $2^{N}$ data.
For example, we assume $\left\vert S\right\rangle $ as a superposition state
of $13$ random numbers 
\begin{eqnarray}
\left\vert S\right\rangle &=&\left\vert 59\right\rangle +\left\vert
61\right\rangle +\left\vert 63\right\rangle +\left\vert 76\right\rangle
+\left\vert 117\right\rangle +\left\vert 125\right\rangle +\left\vert
140\right\rangle +\left\vert 142\right\rangle +\left\vert 148\right\rangle
\label{95} \\
&&+\left\vert 187\right\rangle +\left\vert 212\right\rangle +\left\vert
238\right\rangle +\left\vert 247\right\rangle  \nonumber \\
&=&\left\vert 00111011\right\rangle +\left\vert 00111101\right\rangle
+\left\vert 00111111\right\rangle +\left\vert 01001100\right\rangle
+\left\vert 01110101\right\rangle  \nonumber \\
&&+\left\vert 01111101\right\rangle +\left\vert 10001100\right\rangle
+\left\vert 10001110\right\rangle +\left\vert 10010100\right\rangle 
\nonumber \\
&&+\left\vert 10111011\right\rangle +\left\vert 11010100\right\rangle
+\left\vert 11101110\right\rangle  \nonumber \\
&&+\left\vert 11110111\right\rangle .  \nonumber
\end{eqnarray}%
We choose $8$ optical fields modulated with $8$ PPSs and after passing
through a suitable gate array, that become the following form 
\begin{eqnarray}
\left\vert \psi _{1}^{\prime }\right\rangle &=&\left( e^{i\lambda ^{\left(
1\right) }}+e^{i\lambda ^{\left( 2\right) }}+e^{i\lambda ^{\left( 5\right)
}}+e^{i\lambda ^{\left( 8\right) }}\right) \left\vert 0\right\rangle +\left(
e^{i\lambda ^{\left( 3\right) }}+e^{i\lambda ^{\left( 4\right)
}}+e^{i\lambda ^{\left( 5\right) }}+e^{i\lambda ^{\left( 6\right)
}}+e^{i\lambda ^{\left( 7\right) }}\right) \left\vert 1\right\rangle ,
\label{96} \\
\left\vert \psi _{2}^{\prime }\right\rangle &=&\left( e^{i\lambda ^{\left(
2\right) }}+e^{i\lambda ^{\left( 4\right) }}+e^{i\lambda ^{\left( 5\right)
}}+e^{i\lambda ^{\left( 6\right) }}\right) \left\vert 0\right\rangle +\left(
e^{i\lambda ^{\left( 1\right) }}+e^{i\lambda ^{\left( 3\right)
}}+e^{i\lambda ^{\left( 5\right) }}+e^{i\lambda ^{\left( 7\right)
}}+e^{i\lambda ^{\left( 8\right) }}\right) \left\vert 1\right\rangle , 
\nonumber \\
\left\vert \psi _{3}^{\prime }\right\rangle &=&\left( e^{i\lambda ^{\left(
2\right) }}+e^{i\lambda ^{\left( 5\right) }}+e^{i\lambda ^{\left( 6\right)
}}\right) \left\vert 0\right\rangle +\left( e^{i\lambda ^{\left( 1\right)
}}+e^{i\lambda ^{\left( 3\right) }}+e^{i\lambda ^{\left( 4\right)
}}+e^{i\lambda ^{\left( 7\right) }}+e^{i\lambda ^{\left( 8\right) }}\right)
\left\vert 1\right\rangle ,  \nonumber \\
\left\vert \psi _{4}^{\prime }\right\rangle &=&\left( e^{i\lambda ^{\left(
1\right) }}+e^{i\lambda ^{\left( 3\right) }}+e^{i\lambda ^{\left( 6\right)
}}+e^{i\lambda ^{\left( 8\right) }}\right) \left\vert 0\right\rangle +\left(
e^{i\lambda ^{\left( 2\right) }}+e^{i\lambda ^{\left( 4\right)
}}+e^{i\lambda ^{\left( 5\right) }}+e^{i\lambda ^{\left( 7\right) }}\right)
\left\vert 1\right\rangle ,  \nonumber \\
\left\vert \psi _{5}^{\prime }\right\rangle &=&\left( e^{i\lambda ^{\left(
3\right) }}+e^{i\lambda ^{\left( 6\right) }}+e^{i\lambda ^{\left( 8\right)
}}\right) \left\vert 0\right\rangle +\left( e^{i\lambda ^{\left( 1\right)
}}+e^{i\lambda ^{\left( 2\right) }}+e^{i\lambda ^{\left( 4\right)
}}+e^{i\lambda ^{\left( 5\right) }}+e^{i\lambda ^{\left( 6\right)
}}+e^{i\lambda ^{\left( 7\right) }}\right) \left\vert 1\right\rangle , 
\nonumber \\
\left\vert \psi _{6}^{\prime }\right\rangle &=&e^{i\lambda ^{\left( 2\right)
}}\left\vert 0\right\rangle +\left( e^{i\lambda ^{\left( 1\right)
}}+e^{i\lambda ^{\left( 3\right) }}+e^{i\lambda ^{\left( 4\right)
}}+e^{i\lambda ^{\left( 5\right) }}+e^{i\lambda ^{\left( 6\right)
}}+e^{i\lambda ^{\left( 7\right) }}+e^{i\lambda ^{\left( 8\right) }}\right)
\left\vert 1\right\rangle ,  \nonumber \\
\left\vert \psi _{7}^{\prime }\right\rangle &=&\left( e^{i\lambda ^{\left(
1\right) }}+e^{i\lambda ^{\left( 2\right) }}+e^{i\lambda ^{\left( 6\right)
}}+e^{i\lambda ^{\left( 7\right) }}+e^{i\lambda ^{\left( 8\right) }}\right)
\left\vert 0\right\rangle +\left( e^{i\lambda ^{\left( 1\right)
}}+e^{i\lambda ^{\left( 3\right) }}+e^{i\lambda ^{\left( 4\right)
}}+e^{i\lambda ^{\left( 5\right) }}+e^{i\lambda ^{\left( 7\right) }}\right)
\left\vert 1\right\rangle ,  \nonumber \\
\left\vert \psi _{8}^{\prime }\right\rangle &=&\left( e^{i\lambda ^{\left(
2\right) }}+e^{i\lambda ^{\left( 3\right) }}+e^{i\lambda ^{\left( 5\right)
}}+e^{i\lambda ^{\left( 7\right) }}\right) \left\vert 0\right\rangle +\left(
e^{i\lambda ^{\left( 1\right) }}+e^{i\lambda ^{\left( 4\right)
}}+e^{i\lambda ^{\left( 6\right) }}+e^{i\lambda ^{\left( 8\right) }}\right)
\left\vert 1\right\rangle .  \nonumber
\end{eqnarray}

\begin{figure}[htbp]
\centering\includegraphics[height=2.6602in, width=4.1857in]{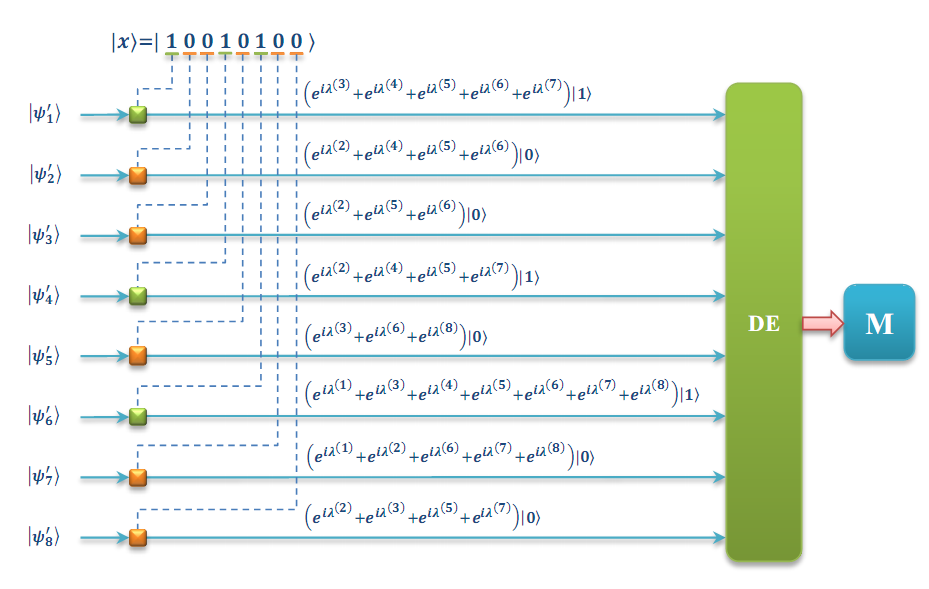}
\caption{A gate array model to select $\left\vert x\right\rangle $ from $%
\left\vert S\right\rangle $ is shown.}
\label{fig26}
\end{figure}

Due to the SCPM mentioned in Sec. \ref{Sec III}, any imitated state must
correspond to a certain sequencial cycle permutation. Therefore, the problem
to determine whether $\left\vert x\right\rangle $ exists in $\left\vert
S\right\rangle $ become that to search the corresponding sequencial cycle
permutation. For example, we search whether the number $\left\vert
x\right\rangle =\left\vert 148\right\rangle =\left\vert
10010100\right\rangle $ is in $\left\vert S\right\rangle $. First, the
optical fields of state $\left\vert S\right\rangle $ pass through the gate
array controlled by $\left\vert x\right\rangle $ as shown in Fig. \ref{fig26}%
. Then we obtain the matrix $M$ by using the coherent demodulation 
\begin{equation}
M\left( \tilde{\alpha}_{i}^{j},\tilde{\beta}_{i}^{j}\right) =\left( 
\begin{array}{cccccccc}
0 & 0 & \left( 0,1\right) & \left( 0,1\right) & \left( 0,1\right) & \left(
0,1\right) & \left( 0,1\right) & 0 \\ 
0 & \left( 1,0\right) & 0 & \left( 1,0\right) & \left( 1,0\right) & \left(
1,0\right) & 0 & 0 \\ 
0 & \left( 1,0\right) & 0 & 0 & \left( 1,0\right) & \left( 1,0\right) & 0 & 0
\\ 
0 & \left( 0,1\right) & 0 & \left( 0,1\right) & \left( 0,1\right) & 0 & 
\left( 0,1\right) & 0 \\ 
0 & 0 & \left( 1,0\right) & 0 & 0 & \left( 1,0\right) & 0 & \left( 1,0\right)
\\ 
\left( 0,1\right) & 0 & \left( 0,1\right) & \left( 0,1\right) & \left(
0,1\right) & \left( 0,1\right) & \left( 0,1\right) & \left( 0,1\right) \\ 
\left( 1,0\right) & \left( 1,0\right) & 0 & 0 & 0 & \left( 1,0\right) & 
\left( 1,0\right) & \left( 1,0\right) \\ 
0 & \left( 1,0\right) & \left( 1,0\right) & 0 & \left( 1,0\right) & 0 & 
\left( 1,0\right) & 0%
\end{array}%
\right)  \label{97}
\end{equation}%
Finally, it is easy to search only the corresponding sequence permutation $%
R_{4}=\left\{ \lambda ^{\left( 4\right) },\lambda ^{\left( 5\right)
},\lambda ^{\left( 6\right) },\lambda ^{\left( 7\right) },\lambda ^{\left(
8\right) },\lambda ^{\left( 1\right) },\lambda ^{\left( 2\right) },\lambda
^{\left( 3\right) }\right\} $ within $O(N^{2})$ operation steps. If we
choose $\left\vert x\right\rangle =\left\vert 240\right\rangle =\left\vert
11110000\right\rangle $, we can obtain the matrix 
\begin{equation}
M\left( \tilde{\alpha}_{i}^{j},\tilde{\beta}_{i}^{j}\right) =\left( 
\begin{array}{cccccccc}
0 & 0 & \left( 0,1\right) & \left( 0,1\right) & \left( 0,1\right) & \left(
0,1\right) & \left( 0,1\right) & 0 \\ 
\left( 0,1\right) & 0 & \left( 0,1\right) & 0 & \left( 0,1\right) & 0 & 
\left( 0,1\right) & \left( 0,1\right) \\ 
\left( 0,1\right) & 0 & \left( 0,1\right) & \left( 0,1\right) & 0 & 0 & 
\left( 0,1\right) & \left( 0,1\right) \\ 
0 & \left( 0,1\right) & 0 & \left( 0,1\right) & \left( 0,1\right) & 0 & 
\left( 0,1\right) & 0 \\ 
0 & 0 & \left( 1,0\right) & 0 & 0 & \left( 1,0\right) & 0 & \left( 1,0\right)
\\ 
0 & \left( 1,0\right) & 0 & 0 & 0 & 0 & 0 & 0 \\ 
\left( 1,0\right) & \left( 1,0\right) & 0 & 0 & 0 & \left( 1,0\right) & 
\left( 1,0\right) & \left( 1,0\right) \\ 
0 & \left( 1,0\right) & \left( 1,0\right) & 0 & \left( 1,0\right) & 0 & 
\left( 1,0\right) & 0%
\end{array}%
\right) .  \label{98}
\end{equation}%
In the matrix $M$, we can not search any corresponding sequencial cycle
permutation. Therefore we can conclude $\left\vert x\right\rangle
=\left\vert 240\right\rangle $ does not exist in $\left\vert S\right\rangle $%
.

Different from Grover's algorithm, this algorithm for searching an unsorted
database with $2^{N}$ entries in $O(N^{2})$ operation steps and using $%
O(N^{2})$ storage space.

\subsection{Optical analogy to quantum Fourier algorithm \label{Sec IV.C}}

Quantum Fourier algorithm is one of the most important tools of quantum
computation, and one of the algorithms which can bring about exponential
speedup \cite{Nielsen}. Shor's algorithm, hidden subgroup problem and
solving systems of linear equations make use of quantum Fourier algorithm 
\cite{Williams}. Quantum Fourier algorithm utilizes the superposition of
quantum state, whereby the required time and space for computation can be
notably reduced from $2^{N}$ to $N$. Hence, the implementation of quantum
Fourier algorithm is crucial to exponential speedup in quantum computation 
\cite{Deutsch}. In this section, we first propose an optical Fourier
algorithm to imitate quantum Fourier algorithm, then investigate the
required computational resources, and at last demonstrate the algorithm
applying to three optical fields as examples to verify its feasibility.

\subsubsection{Quantum Fourier transform \label{Sec IV.C.1}}

Generally, quantum Fourier transform takes as input a vector of complex
numbers, $f(0),f(1)\ldots ,f(2^{N}-1)$, and output a new vector of complex
numbers $\tilde{f}(0),\tilde{f}(1)\ldots ,\tilde{f}(2^{N}-1)$ as follow%
\begin{equation}
\tilde{f}\left( k\right) =\frac{1}{\sqrt{2^{N}}}\sum%
\limits_{j=0}^{2^{N}-1}e^{\frac{2\pi i}{2^{N}}jk}f\left( j\right) .
\label{99}
\end{equation}
This calculation involves the additions and multiplications of $2^{N}$
complex numbers, leading to an increase of computational complexity with the
increase of the number of vector components. Classically, the most effective
algorithm, fast Fourier transform is in time $O(2^{N}\log 2^{N})$. On the
contrary, the quantum Fourier transform can be defined as a unitary
transformation on $N$ qubits \cite{Nielsen}, which is%
\begin{equation}
\hat{F}\left\vert j\right\rangle =\frac{1}{\sqrt{2^{N}}}\sum%
\limits_{k=0}^{2^{N}-1}e^{\frac{2\pi i}{2^{N}}jk}\left\vert k\right\rangle .
\label{100}
\end{equation}%
Furthermore, the quantum Fourier transform of arbitrary state $\left\vert
\Psi \right\rangle =C_{0}\left\vert 0\right\rangle +\cdots
+C_{2^{N}-1}\left\vert 2^{N}-1\right\rangle $ can be expressed as%
\begin{eqnarray}
\left\vert \Psi \right\rangle _{F} &\equiv &\hat{F}\left\vert \Psi
\right\rangle =C_{0}\hat{F}\left\vert 0\right\rangle +C_{1}\hat{F}\left\vert
1\right\rangle +\cdots +C_{2^{N}-1}\hat{F}\left\vert 2^{N}-1\right\rangle
\label{101} \\
&=&\frac{1}{\sqrt{2^{N}}}\sum\limits_{k=0}^{2^{N}-1}\left[ C_{0}\omega
^{0\ast k}+C_{1}\omega ^{1\ast k}+\cdots +C_{2^{N}-1}\omega ^{\left(
2^{N}-1\right) \ast k}\right] \left\vert k\right\rangle ,  \nonumber
\end{eqnarray}%
where $\omega =2\pi i/2^{N}$. Then, we expand $\left\vert \Psi \right\rangle
_{F}$ into%
\begin{equation}
\left\vert \Psi \right\rangle _{F}=\sum\limits_{j_{1}=0}^{1}\cdots
\sum\limits_{j_{N}=0}^{1}D_{j_{N-1}\cdots j_{0}}\left\vert j_{N-1}\cdots
j_{0}\right\rangle ,  \label{102}
\end{equation}%
where the coefficients satisfy the following equation%
\begin{equation}
\left( 
\begin{array}{c}
D_{0} \\ 
D_{1} \\ 
\vdots \\ 
D_{2^{N}-1}%
\end{array}%
\right) =\frac{1}{\sqrt{2^{N}}}\left( 
\begin{array}{cccc}
1 & 1 & \cdots & 1 \\ 
1 & \omega & \cdots & \omega ^{2^{N}-1} \\ 
\vdots & \vdots & \ddots & \vdots \\ 
1 & \omega ^{2^{N}-1} & \cdots & \omega ^{\left( 2^{N}-1\right) ^{2}}%
\end{array}%
\right) \left( 
\begin{array}{c}
C_{0} \\ 
C_{1} \\ 
\vdots \\ 
C_{2^{N}-1}%
\end{array}%
\right) .  \label{103}
\end{equation}%
In quantum Fourier transform, after the Hadamard gate and controlled-phase
gate, we can obtain the final state $\left\vert j\right\rangle =\left\vert
j_{N-1}j_{N-2}\cdots j_{0}\right\rangle $ of quantum Fourier transform%
\begin{equation}
\hat{F}\left\vert j\right\rangle =\frac{1}{\sqrt{2^{N}}}\left( \left\vert
1\right\rangle +e^{2\pi i0.j_{0}}\left\vert 1\right\rangle \right) \left(
\left\vert 1\right\rangle +e^{2\pi i0.j_{1}j_{0}}\left\vert 1\right\rangle
\right) \cdots \left( \left\vert 1\right\rangle +e^{2\pi
i0.j_{N-1}j_{N-2}\cdots j_{0}}\left\vert 1\right\rangle \right) .
\label{104}
\end{equation}%
There are $N$ Hadamard gates and $N(N-1)/2$ controlled-phase gates on $N$
qubit registers, which means the quantum Fourier transform takes $O(N^{2})$
basic gate operations. Nevertheless, the quantum Fourier transform cannot
output precise result of final states directly, but the probability of every
state by repeated measurements, which can output the final result of Fourier
transform at a certain accuracy \cite{Nielsen}.

\subsubsection{Algorithm for the optical analogies \label{Sec IV.C.3}}

As mentioned in Sec. \ref{Sec III}, a general form of $\left\vert \psi
_{k}\right\rangle $ for $N$ fields can be constructed from Eq. (\ref{20}) by
using a gate array model,%
\begin{eqnarray}
\left\vert \psi _{n}\right\rangle &=&\sum\limits_{i=1}^{N}\alpha
_{n}^{\left( i\right) }e^{i\lambda ^{\left( i\right) }}\left\vert
0\right\rangle +\sum\limits_{j=1}^{N}\beta _{n}^{\left( j\right)
}e^{i\lambda ^{\left( j\right) }}\left\vert 1\right\rangle  \label{109} \\
&\equiv &\tilde{\alpha}_{n}\left\vert 0\right\rangle +\tilde{\beta}%
_{n}\left\vert 1\right\rangle ,  \nonumber
\end{eqnarray}%
where $\tilde{\alpha}_{n}\equiv \sum\limits_{i=1}^{N}\alpha _{n}^{\left(
i\right) }e^{i\lambda ^{\left( i\right) }},\tilde{\beta}_{n}\equiv
\sum\limits_{j=1}^{N}\beta _{n}^{\left( j\right) }e^{i\lambda ^{\left(
j\right) }}$. Then, the formal product state Eq. (\ref{21}) can be written as%
\begin{equation}
\left\vert \Psi \right\rangle =\left( \tilde{\alpha}_{1}\left\vert
0\right\rangle +\tilde{\beta}_{1}\left\vert 1\right\rangle \right) \otimes
\cdots \otimes \left( \tilde{\alpha}_{N}\left\vert 0\right\rangle +\tilde{%
\beta}_{N}\left\vert 1\right\rangle \right) .  \label{110}
\end{equation}

Further, we can obtain each item of the superposition of $\left\vert \Psi
\right\rangle $ as follows%
\begin{equation}
\begin{array}{c}
C_{00\cdots 0}\left\vert 00\cdots 0\right\rangle =\tilde{\alpha}_{1}\tilde{%
\alpha}_{2}\cdots \tilde{\alpha}_{N}\left\vert 00\cdots 0\right\rangle , \\ 
C_{00\cdots 1}\left\vert 00\cdots 1\right\rangle =\tilde{\alpha}_{1}\tilde{%
\alpha}_{2}\cdots \tilde{\beta}_{N}\left\vert 00\cdots 1\right\rangle , \\ 
\vdots \\ 
C_{11\cdots 1}\left\vert 11\cdots 1\right\rangle =\tilde{\beta}_{1}\tilde{%
\beta}_{2}\cdots \tilde{\beta}_{N}\left\vert 11\cdots 1\right\rangle .%
\end{array}
\label{111}
\end{equation}%
The formal product state is expressed as follow%
\begin{equation}
\left\vert \Psi \right\rangle =\sum\limits_{i_{1}=0}^{1}\cdots
\sum\limits_{i_{N}=0}^{1}C_{i_{1}i_{2}\cdots i_{N}}\left\vert
i_{1}i_{2}\cdots i_{N}\right\rangle .  \label{113}
\end{equation}%
After Fourier transform, this state evolves into 
\begin{equation}
\left\vert \Psi \right\rangle _{F}=\hat{F}\left\vert \Psi \right\rangle ,
\label{114}
\end{equation}%
where 
\begin{equation}
\left\vert \Psi \right\rangle _{F}=\sum\limits_{j_{1}=0}^{1}\cdots
\sum\limits_{j_{N}=0}^{1}D_{j_{1}j_{2}\cdots j_{N}}\left\vert
j_{1}j_{2}\cdots j_{N}\right\rangle .  \label{115}
\end{equation}

According\ to the definition Eq. (\ref{99}), the relation between the
coefficients $C_{i_{1}i_{2}\cdots i_{N}}$ and $D_{j_{1}j_{2}\cdots j_{N}}$
of these two states have to be satisfied as Eq. (\ref{103}). To obtain the
relation between these coefficient, we design the following algortithm:

(1) Selected a basis state $\left\vert j_{1}j_{2}\cdots j_{N}\right\rangle $
of $\left\vert \Psi \right\rangle _{F}$;

(2) Applying the following controlled-phase transformation on every field of 
$\left\vert \Psi \right\rangle $ according to the specific value of bits in
the selected basis state, we obtain%
\begin{equation}
\left\{ 
\begin{array}{c}
\left\vert \psi _{1}\right\rangle =\tilde{\alpha}_{1}\left\vert
0\right\rangle +\tilde{\beta}_{1}\left\vert 1\right\rangle \\ 
\left\vert \psi _{2}\right\rangle =\tilde{\alpha}_{2}\left\vert
0\right\rangle +\omega ^{j_{1}\ast 2^{N-2}}\tilde{\beta}_{2}\left\vert
1\right\rangle \\ 
\left\vert \psi _{3}\right\rangle =\tilde{\alpha}_{3}\left\vert
0\right\rangle +\omega ^{j_{2}\ast 2^{N-2}+j_{1}\ast 2^{N-3}}\tilde{\beta}%
_{3}\left\vert 1\right\rangle \\ 
\vdots \\ 
\left\vert \psi _{N}\right\rangle =\tilde{\alpha}_{N}\left\vert
0\right\rangle +\omega ^{j_{N-1}\ast 2^{N-2}+j_{N-2}\ast 2^{N-3}+\cdots
+j_{1}\ast 1}\tilde{\beta}_{N}\left\vert 1\right\rangle%
\end{array}%
\right. ;  \label{116}
\end{equation}

(3) Applying Hadamard gates on these fields, we obtain%
\begin{equation}
\left\{ 
\begin{array}{c}
\left\vert \psi _{1}\right\rangle =\left( \tilde{\alpha}_{1}+\tilde{\beta}%
_{1}\right) \left\vert 0\right\rangle +\left( \tilde{\alpha}_{1}-\tilde{\beta%
}_{1}\right) \left\vert 1\right\rangle \\ 
\left\vert \psi _{2}\right\rangle =\left( \tilde{\alpha}_{2}+\omega
^{j_{1}\ast 2^{N-2}}\tilde{\beta}_{2}\right) \left\vert 0\right\rangle
+\left( \tilde{\alpha}_{2}-\omega ^{j_{1}\ast 2^{N-2}}\tilde{\beta}%
_{2}\right) \left\vert 1\right\rangle \\ 
\left\vert \psi _{3}\right\rangle =\left( \tilde{\alpha}_{3}+\omega
^{j_{2}\ast 2^{N-2}+j_{1}\ast 2^{N-3}}\tilde{\beta}_{3}\right) \left\vert
0\right\rangle +\left( \tilde{\alpha}_{3}-\omega ^{j_{2}\ast
2^{N-2}+j_{1}\ast 2^{N-3}}\tilde{\beta}_{3}\right) \left\vert 1\right\rangle
\\ 
\vdots \\ 
\left\vert \psi _{N}\right\rangle =\left( \tilde{\alpha}_{N}+\omega
^{j_{N-1}\ast 2^{N-2}+j_{N-2}\ast 2^{N-3}+\cdots +j_{1}}\tilde{\beta}%
_{N}\right) \tilde{\alpha}_{N}\left\vert 0\right\rangle +\left( \tilde{\alpha%
}_{N}-\omega ^{j_{N-1}\ast 2^{N-2}+j_{N-2}\ast 2^{N-3}+\cdots +j_{1}}\tilde{%
\beta}_{N}\right) \left\vert 1\right\rangle%
\end{array}%
\right. ;  \label{117}
\end{equation}

(4) Applying the mode control gates on these fields according to the
specific values in $\left\vert j_{1}j_{2}\cdots j_{N}\right\rangle $, the
mode of every field is identical to the corresponding value in $\left\vert
j_{1}j_{2}\cdots j_{N}\right\rangle $, e.g., if $j_{1}=0$, $\left\vert \psi
_{1}\right\rangle $ becomes$\ \left( \tilde{\alpha}_{1}+\tilde{\beta}%
_{1}\right) \left\vert 0\right\rangle $, otherwise if $j_{1}=1$, $\left\vert
\psi _{1}\right\rangle $ becomes$\ \left( \tilde{\alpha}_{1}-\tilde{\beta}%
_{1}\right) \left\vert 1\right\rangle $, and so on;

(5) Applying the coherent demodulation on these fields and obtain the matrix 
$M$, we can obtain the corresponding coefficient $D_{j_{N}j_{N-1}\cdots
j_{1}}$ using the method mentioned in Sec. \ref{Sec III.C}.

\begin{figure}[htbp]
\centering\includegraphics[height=2.3834in, width=5.2736in]{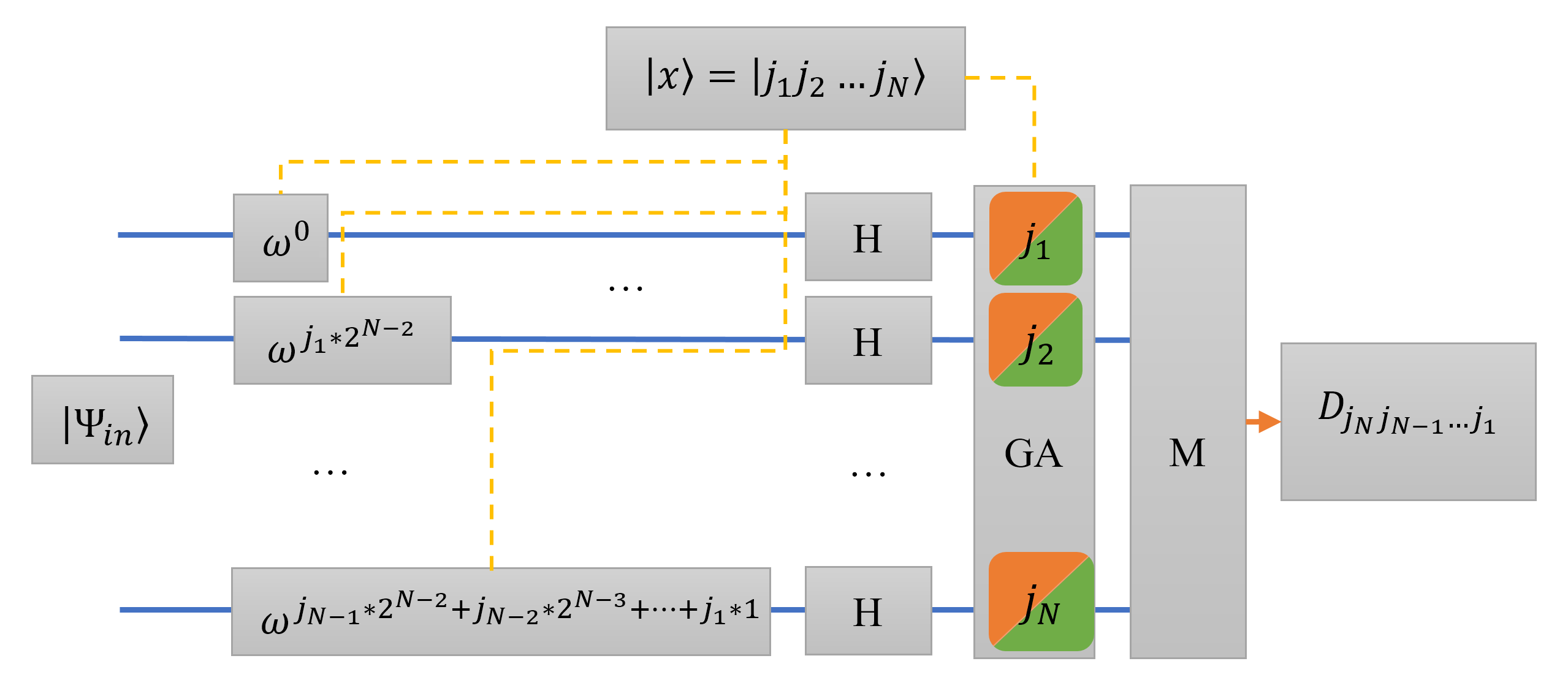}
\caption{The algorithm diagram of optical analogy to quantum Fourier
algorithm is shown.}
\label{fig27}
\end{figure}

The above algorithm can be summarized as the following block diagram in Fig. %
\ref{fig27}. At last, we can analysis the computational complexity: there
are $N$ optical fields in $\left\vert \Psi \right\rangle $ after $N$
controlled-phase gates, $N$ Hadamard gates, $N$ mode selection operations
and finally $O\left( N^{2}\right) $ measurements in the coherent
demodulation. Hence, the total number of operations is in $O(N^{2})$, which
is the same as that in quantum Fourier algorithm. However, the result
obtained in the optical algorithm is with certainty values but not with
probability like quantum Fourier algorithm.

\subsubsection{The equivalence of ensemble-averaged states in optical
algorithm \label{Sec IV.C.4}}

In the phase ensemble, we utilize the characteristic of PPS to define the
EADM $\tilde{\rho}$ as mentioned in Sec. \ref{Sec III.A.2}. As mentioned in
Sec. \ref{Sec III.C}, the imitation states can be constructured by using the
SCPM corresponding to the minimum complete phase ensemble. Actually, the
imitation states can be defined as the following ensemble-averaged state 
\begin{equation}
\left\vert \tilde{\Psi}\right\rangle \equiv
\sum\limits_{k=1}^{N}e^{-i\lambda ^{S}}\left\vert \Psi \right\rangle ,
\label{119}
\end{equation}%
where $\lambda ^{S}=\sum\nolimits_{n=1}^{N}\lambda ^{(n)}$ that is the sum
of all used PPSs of the optical fields. Then, we discuss about Fourier
transform for the ensemble-averaged states. From Eq. (\ref{103}), we obtain
the coefficients of Fourier transform satisfies%
\begin{equation}
\begin{array}{c}
D_{00\cdots 0}=C_{00\cdots 0}+C_{00\cdots 1}+\cdots +C_{11\cdots 1}, \\ 
D_{00\cdots 1}=C_{00\cdots 0}+\omega C_{00\cdots 1}+\cdots +\omega ^{\left(
2^{N}-1\right) }C_{11\cdots 1}, \\ 
\vdots \\ 
D_{11\cdots 0}=C_{00\cdots 0}+\omega ^{\left( 2^{N}-2\right) }C_{00\cdots
1}+\cdots +\omega ^{\left( 2^{N}-2\right) \left( 2^{N}-1\right) }C_{11\cdots
1}, \\ 
D_{11\cdots 1}=C_{00\cdots 0}+\omega ^{\left( 2^{N}-1\right) }C_{00\cdots
1}+\cdots +\omega ^{\left( 2^{N}-1\right) ^{2}}C_{11\cdots 1}.%
\end{array}
\label{120}
\end{equation}%
In these equations, the combinations of $\omega $ and $C_{i_{1}i_{2}\cdots
i_{N}}$ satisfy the following relations%
\begin{equation}
\omega ^{l}C_{i_{1}i_{2}\cdots
i_{N}}=\sum\limits_{j=1}^{N}C_{i_{1}i_{2}\cdots i_{N}}^{\left( j\right)
}e^{i \left[ \lambda ^{\left( j\right) }+\frac{2\pi l}{2^{N}}\right] }.
\label{121}
\end{equation}%
Obviously, these terms also satisfy the balance property of PPS. Hence, the
ensemble-averaged state can also be used in Fourier transform. Then, we can
obtain the Fourier transform%
\begin{equation}
\left\vert \tilde{\Psi}\right\rangle _{F}\equiv
\sum\limits_{k=1}^{N}e^{-i\lambda ^{\left( S\right) }}\left\vert \Psi
\right\rangle _{F}=\hat{F}\sum\limits_{k=1}^{N}e^{-i\lambda ^{\left(
S\right) }}\left\vert \Psi \right\rangle =\hat{F}\left\vert \tilde{\Psi}%
\right\rangle .  \label{122}
\end{equation}%
At last, we show the equivalence of ensemble-averaged states in the Fourier
transform.

\subsubsection{Optical analogies to quantum Fourier transform for three
particles \label{Sec IV.C.5}}

According to Sec. \ref{Sec III.C.1}, three optical fields with PPSs $\lambda
^{\left( i\right) }\left( i=1,2,3\right) $ are required to implement the
imitations of the quantum states consisting of three particles. Modulated
with these PPSs, three optical fields can be expressed as follows%
\begin{equation}
\left\{ 
\begin{array}{c}
\left\vert \psi _{1}\right\rangle =e^{i\lambda ^{\left( 1\right) }}\left(
\left\vert 0\right\rangle +\left\vert 1\right\rangle \right) \\ 
\left\vert \psi _{2}\right\rangle =e^{i\lambda ^{\left( 2\right) }}\left(
\left\vert 0\right\rangle +\left\vert 1\right\rangle \right) \\ 
\left\vert \psi _{3}\right\rangle =e^{i\lambda ^{\left( 3\right) }}\left(
\left\vert 0\right\rangle +\left\vert 1\right\rangle \right)%
\end{array}%
\right. .  \label{123}
\end{equation}%
After gate array models as mentioned in Sec. \ref{Sec IV.A}, we can obtain
arbitrary quantum states which can be expressed as follows%
\begin{equation}
\left\{ 
\begin{array}{c}
\left\vert \psi _{1}\right\rangle =\tilde{\alpha}_{1}\left\vert
0\right\rangle +\tilde{\beta}_{1}\left\vert 1\right\rangle \\ 
\left\vert \psi _{2}\right\rangle =\tilde{\alpha}_{2}\left\vert
0\right\rangle +\tilde{\beta}_{2}\left\vert 1\right\rangle \\ 
\left\vert \psi _{3}\right\rangle =\tilde{\alpha}_{3}\left\vert
0\right\rangle +\tilde{\beta}_{3}\left\vert 1\right\rangle%
\end{array}%
\right. .  \label{124}
\end{equation}

According to the algorithm in Sec. \ref{Sec IV.C.3}, we obtain:

(1) Apply controlled-phase gates on three optical fields respectively

\begin{equation}
\left\{ 
\begin{array}{c}
\left\vert \psi _{1}\right\rangle =\tilde{\alpha}_{1}\left\vert
0\right\rangle +\tilde{\beta}_{1}\left\vert 1\right\rangle \\ 
\left\vert \psi _{2}\right\rangle =\tilde{\alpha}_{2}\left\vert
0\right\rangle +\omega ^{j_{1}\ast 2}\tilde{\beta}_{2}\left\vert
1\right\rangle \\ 
\left\vert \psi _{3}\right\rangle =\tilde{\alpha}_{3}\left\vert
0\right\rangle +\omega ^{j_{2}\ast 2+j_{1}\ast 1}\tilde{\beta}_{3}\left\vert
1\right\rangle%
\end{array}%
\right. .  \label{125}
\end{equation}%
where $\omega =e^{2\pi i/8}$.

(2) Hadamard transformation%
\begin{equation}
\left\{ 
\begin{array}{c}
\left\vert \psi _{1}\right\rangle =\left( \tilde{\alpha}_{1}+\tilde{\beta}%
_{1}\right) \left\vert 0\right\rangle +\left( \tilde{\alpha}_{1}-\tilde{\beta%
}_{1}\right) \left\vert 1\right\rangle \\ 
\left\vert \psi _{2}\right\rangle =\left( \tilde{\alpha}_{2}+\omega
^{j_{1}\ast 2}\tilde{\beta}_{2}\right) \left\vert 0\right\rangle +\left( 
\tilde{\alpha}_{2}-\omega ^{j_{1}\ast 2}\tilde{\beta}_{2}\right) \left\vert
1\right\rangle \\ 
\left\vert \psi _{3}\right\rangle =\left( \tilde{\alpha}_{3}+\omega
^{j_{2}\ast 2+j_{1}\ast 1}\tilde{\beta}_{3}\right) \left\vert 0\right\rangle
+\left( \tilde{\alpha}_{3}-\omega ^{j_{2}\ast 2+j_{1}\ast 1}\tilde{\beta}%
_{3}\right) \left\vert 1\right\rangle%
\end{array}%
\right. .  \label{126}
\end{equation}

(3) Calculate coefficients

(3.1) When $\left\vert j_{1}j_{2}j_{3}\right\rangle =\left\vert
000\right\rangle $ and $\left\vert j_{1}j_{2}j_{3}\right\rangle =\left\vert
001\right\rangle $,%
\begin{equation}
\left\{ 
\begin{array}{c}
\left\vert \psi _{1}\right\rangle =\tilde{\alpha}_{1}\left\vert
0\right\rangle +\tilde{\beta}_{1}\left\vert 1\right\rangle \\ 
\left\vert \psi _{2}\right\rangle =\tilde{\alpha}_{2}\left\vert
0\right\rangle +\omega ^{0\ast 2}\tilde{\beta}_{2}\left\vert 1\right\rangle
\\ 
\left\vert \psi _{3}\right\rangle =\tilde{\alpha}_{3}\left\vert
0\right\rangle +\omega ^{0\ast 2+0\ast 1}\tilde{\beta}_{3}\left\vert
1\right\rangle%
\end{array}%
\right. \rightarrow \left\{ 
\begin{array}{c}
\left\vert \psi _{1}\right\rangle =\left( \tilde{\alpha}_{1}+\tilde{\beta}%
_{1}\right) \left\vert 0\right\rangle +\left( \tilde{\alpha}_{1}-\tilde{\beta%
}_{1}\right) \left\vert 1\right\rangle \\ 
\left\vert \psi _{2}\right\rangle =\left( \tilde{\alpha}_{2}+\tilde{\beta}%
_{2}\right) \left\vert 0\right\rangle +\left( \tilde{\alpha}_{2}-\tilde{\beta%
}_{2}\right) \left\vert 1\right\rangle \\ 
\left\vert \psi _{3}\right\rangle =\left( \tilde{\alpha}_{3}+\tilde{\beta}%
_{3}\right) \left\vert 0\right\rangle +\left( \tilde{\alpha}_{3}-\tilde{\beta%
}_{3}\right) \left\vert 1\right\rangle%
\end{array}%
\right. .  \label{127}
\end{equation}%
Then we obtain the corresponding coefficients $D_{000}$ and $D_{100}$%
\begin{eqnarray}
D_{000} &=&\left( \tilde{\alpha}_{1}+\tilde{\beta}_{1}\right) \left( \tilde{%
\alpha}_{2}+\tilde{\beta}_{2}\right) \left( \tilde{\alpha}_{3}+\tilde{\beta}%
_{3}\right)  \label{128} \\
&=&C_{000}+C_{001}+C_{010}+C_{011}+C_{100}+C_{101}+C_{110}+C_{111}, 
\nonumber
\end{eqnarray}%
\begin{eqnarray}
D_{100} &=&\left( \tilde{\alpha}_{1}+\tilde{\beta}_{1}\right) \left( \tilde{%
\alpha}_{2}+\tilde{\beta}_{2}\right) \left( \tilde{\alpha}_{3}-\tilde{\beta}%
_{3}\right)  \label{129} \\
&=&C_{000}-C_{001}+C_{010}-C_{011}+C_{100}-C_{101}+C_{110}-C_{111}. 
\nonumber
\end{eqnarray}

(3.2) When $\left\vert j_{1}j_{2}j_{3}\right\rangle =\left\vert
010\right\rangle $ and $\left\vert j_{1}j_{2}j_{3}\right\rangle =\left\vert
011\right\rangle $,%
\begin{equation}
\left\{ 
\begin{array}{c}
\left\vert \psi _{1}\right\rangle =\tilde{\alpha}_{1}\left\vert
0\right\rangle +\tilde{\beta}_{1}\left\vert 1\right\rangle \\ 
\left\vert \psi _{2}\right\rangle =\tilde{\alpha}_{2}\left\vert
0\right\rangle +\omega ^{0\ast 2}\tilde{\beta}_{2}\left\vert 1\right\rangle
\\ 
\left\vert \psi _{3}\right\rangle =\tilde{\alpha}_{3}\left\vert
0\right\rangle +\omega ^{1\ast 2+0\ast 1}\tilde{\beta}_{3}\left\vert
1\right\rangle%
\end{array}%
\right. \rightarrow \left\{ 
\begin{array}{c}
\left\vert \psi _{1}\right\rangle =\left( \tilde{\alpha}_{1}+\tilde{\beta}%
_{1}\right) \left\vert 0\right\rangle +\left( \tilde{\alpha}_{1}-\tilde{\beta%
}_{1}\right) \left\vert 1\right\rangle \\ 
\left\vert \psi _{2}\right\rangle =\left( \tilde{\alpha}_{2}+\tilde{\beta}%
_{2}\right) \left\vert 0\right\rangle +\left( \tilde{\alpha}_{2}-\tilde{\beta%
}_{2}\right) \left\vert 1\right\rangle \\ 
\left\vert \psi _{3}\right\rangle =\left( \tilde{\alpha}_{3}+\omega ^{2}%
\tilde{\beta}_{3}\right) \left\vert 0\right\rangle +\left( \tilde{\alpha}%
_{3}-\omega ^{2}\tilde{\beta}_{3}\right) \left\vert 1\right\rangle%
\end{array}%
\right. .  \label{130}
\end{equation}%
Then we obtain the corresponding coefficients $D_{010}$ and $D_{110}$%
\begin{eqnarray}
D_{010} &=&\left( \tilde{\alpha}_{1}+\tilde{\beta}_{1}\right) \left( \tilde{%
\alpha}_{2}-\tilde{\beta}_{2}\right) \left( \tilde{\alpha}_{3}+\omega ^{2}%
\tilde{\beta}_{3}\right)  \label{131} \\
&=&C_{000}+\omega ^{2}C_{001}-C_{010}-\omega ^{2}C_{011}+C_{100}+\omega
^{2}C_{101}-C_{110}-\omega ^{2}C_{111},  \nonumber
\end{eqnarray}%
\begin{eqnarray}
D_{110} &=&\left( \tilde{\alpha}_{1}+\tilde{\beta}_{1}\right) \left( \tilde{%
\alpha}_{2}-\tilde{\beta}_{2}\right) \left( \tilde{\alpha}_{3}-\omega ^{2}%
\tilde{\beta}_{3}\right)  \label{132} \\
&=&C_{000}-\omega ^{2}C_{001}-C_{010}+\omega ^{2}C_{011}+C_{100}-\omega
^{2}C_{101}-C_{110}+\omega ^{2}C_{111}.  \nonumber
\end{eqnarray}

(3.3) When $\left\vert j_{1}j_{2}j_{3}\right\rangle =\left\vert
100\right\rangle $ and $\left\vert j_{1}j_{2}j_{3}\right\rangle =\left\vert
101\right\rangle $,%
\begin{equation}
\left\{ 
\begin{array}{c}
\left\vert \psi _{1}\right\rangle =\tilde{\alpha}_{1}\left\vert
0\right\rangle +\tilde{\beta}_{1}\left\vert 1\right\rangle \\ 
\left\vert \psi _{2}\right\rangle =\tilde{\alpha}_{2}\left\vert
0\right\rangle +\omega ^{1\ast 2}\tilde{\beta}_{2}\left\vert 1\right\rangle
\\ 
\left\vert \psi _{3}\right\rangle =\tilde{\alpha}_{3}\left\vert
0\right\rangle +\omega ^{0\ast 2+1\ast 1}\tilde{\beta}_{3}\left\vert
1\right\rangle%
\end{array}%
\right. \rightarrow \left\{ 
\begin{array}{c}
\left\vert \psi _{1}\right\rangle =\left( \tilde{\alpha}_{1}+\tilde{\beta}%
_{1}\right) \left\vert 0\right\rangle +\left( \tilde{\alpha}_{1}-\tilde{\beta%
}_{1}\right) \left\vert 1\right\rangle \\ 
\left\vert \psi _{2}\right\rangle =\left( \tilde{\alpha}_{2}+\omega ^{2}%
\tilde{\beta}_{2}\right) \left\vert 0\right\rangle +\left( \tilde{\alpha}%
_{2}-\omega ^{2}\tilde{\beta}_{2}\right) \left\vert 1\right\rangle \\ 
\left\vert \psi _{3}\right\rangle =\left( \tilde{\alpha}_{3}+\omega \tilde{%
\beta}_{3}\right) \left\vert 0\right\rangle +\left( \tilde{\alpha}%
_{3}-\omega \tilde{\beta}_{3}\right) \left\vert 1\right\rangle%
\end{array}%
\right. .  \label{133}
\end{equation}%
Then we obtain the corresponding coefficients $D_{001}$ and $D_{101}$%
\begin{eqnarray}
D_{001} &=&\left( \tilde{\alpha}_{1}-\tilde{\beta}_{1}\right) \left( \tilde{%
\alpha}_{2}+\omega ^{2}\tilde{\beta}_{2}\right) \left( \tilde{\alpha}%
_{3}+\omega \tilde{\beta}_{3}\right)  \label{134} \\
&=&C_{000}+\omega C_{001}+\omega ^{2}C_{010}+\omega
^{3}C_{011}-C_{100}-\omega C_{101}-\omega ^{2}C_{110}-\omega ^{3}C_{111}, 
\nonumber
\end{eqnarray}%
\begin{eqnarray}
D_{101} &=&\left( \tilde{\alpha}_{1}-\tilde{\beta}_{1}\right) \left( \tilde{%
\alpha}_{2}+\omega ^{2}\tilde{\beta}_{2}\right) \left( \tilde{\alpha}%
_{3}-\omega \tilde{\beta}_{3}\right)  \label{135} \\
&=&C_{000}-\omega C_{001}+\omega ^{2}C_{010}-\omega
^{3}C_{011}-C_{100}+\omega C_{101}-\omega ^{2}C_{110}+\omega ^{3}C_{111}. 
\nonumber
\end{eqnarray}

(3.4) When $\left\vert j_{1}j_{2}j_{3}\right\rangle =\left\vert
110\right\rangle $ and $\left\vert j_{1}j_{2}j_{3}\right\rangle =\left\vert
111\right\rangle $,%
\begin{equation}
\left\{ 
\begin{array}{c}
\left\vert \psi _{1}\right\rangle =\tilde{\alpha}_{1}\left\vert
0\right\rangle +\tilde{\beta}_{1}\left\vert 1\right\rangle \\ 
\left\vert \psi _{2}\right\rangle =\tilde{\alpha}_{2}\left\vert
0\right\rangle +\omega ^{1\ast 2}\tilde{\beta}_{2}\left\vert 1\right\rangle
\\ 
\left\vert \psi _{3}\right\rangle =\tilde{\alpha}_{3}\left\vert
0\right\rangle +\omega ^{1\ast 2+1\ast 1}\tilde{\beta}_{3}\left\vert
1\right\rangle%
\end{array}%
\right. \rightarrow \left\{ 
\begin{array}{c}
\left\vert \psi _{1}\right\rangle =\left( \tilde{\alpha}_{1}+\tilde{\beta}%
_{1}\right) \left\vert 0\right\rangle +\left( \tilde{\alpha}_{1}-\tilde{\beta%
}_{1}\right) \left\vert 1\right\rangle \\ 
\left\vert \psi _{2}\right\rangle =\left( \tilde{\alpha}_{2}+\omega ^{2}%
\tilde{\beta}_{2}\right) \left\vert 0\right\rangle +\left( \tilde{\alpha}%
_{2}-\omega ^{2}\tilde{\beta}_{2}\right) \left\vert 1\right\rangle \\ 
\left\vert \psi _{3}\right\rangle =\left( \tilde{\alpha}_{3}+\omega ^{3}%
\tilde{\beta}_{3}\right) \left\vert 0\right\rangle +\left( \tilde{\alpha}%
_{3}-\omega ^{3}\tilde{\beta}_{3}\right) \left\vert 1\right\rangle%
\end{array}%
\right. .  \label{136}
\end{equation}%
Then we obtain the corresponding coefficients $D_{011}$ and $D_{111}$%
\begin{eqnarray}
D_{011} &=&\left( \tilde{\alpha}_{1}-\tilde{\beta}_{1}\right) \left( \tilde{%
\alpha}_{2}-\omega ^{2}\tilde{\beta}_{2}\right) \left( \tilde{\alpha}%
_{3}+\omega ^{3}\tilde{\beta}_{3}\right)  \label{137} \\
&=&C_{000}+\omega ^{3}C_{001}-\omega ^{2}C_{010}-\omega
^{5}C_{011}-C_{100}-\omega ^{3}C_{101}+\omega ^{2}C_{110}+\omega ^{5}C_{111},
\nonumber
\end{eqnarray}%
\begin{eqnarray}
D_{111} &=&\left( \tilde{\alpha}_{1}-\tilde{\beta}_{1}\right) \left( \tilde{%
\alpha}_{2}-\tilde{\beta}_{2}\right) \left( \tilde{\alpha}_{3}-\omega ^{2}%
\tilde{\beta}_{3}\right)  \label{138} \\
&=&C_{000}-\omega ^{3}C_{001}-\omega ^{2}C_{010}+\omega
^{5}C_{011}-C_{100}+\omega ^{3}C_{101}+\omega ^{2}C_{110}-\omega ^{5}C_{111}.
\nonumber
\end{eqnarray}%
At last, we obtain the transform matrix of all coefficients as follow%
\begin{equation}
\left( 
\begin{array}{c}
D_{000} \\ 
D_{001} \\ 
D_{010} \\ 
D_{011} \\ 
D_{100} \\ 
D_{101} \\ 
D_{110} \\ 
D_{111}%
\end{array}%
\right) =\left( 
\begin{array}{cccccccc}
1 & 1 & 1 & 1 & 1 & 1 & 1 & 1 \\ 
1 & \omega & \omega ^{2} & \omega ^{3} & -1 & -\omega & -\omega ^{2} & 
-\omega ^{3} \\ 
1 & \omega ^{2} & -1 & -\omega ^{2} & 1 & \omega ^{2} & -1 & \omega ^{2} \\ 
1 & \omega ^{3} & -\omega ^{2} & \omega & -1 & -\omega ^{3} & \omega ^{2} & 
-\omega \\ 
1 & -1 & 1 & -1 & 1 & -1 & 1 & -1 \\ 
1 & -\omega & \omega ^{2} & -\omega ^{3} & -1 & \omega & -\omega ^{2} & 
\omega ^{3} \\ 
1 & -\omega ^{2} & -1 & \omega ^{2} & 1 & -\omega ^{2} & -1 & \omega ^{2} \\ 
1 & -\omega ^{3} & -\omega ^{2} & -\omega & -1 & \omega ^{3} & \omega ^{2} & 
\omega%
\end{array}%
\right) \left( 
\begin{array}{c}
C_{000} \\ 
C_{001} \\ 
C_{010} \\ 
C_{011} \\ 
C_{100} \\ 
C_{101} \\ 
C_{110} \\ 
C_{111}%
\end{array}%
\right) .  \label{139}
\end{equation}%
Obviously, the result is completely similar to quantum Fourier algorithm. We
will utilize the above algorithm applying to some imitation states of three
optical fields as following examples:

(1) the product state

In quantum mechanics, the product state of three particles is $\left\vert
\Psi \right\rangle =\frac{1}{\sqrt{8}}\left( \left\vert 000\right\rangle
+\left\vert 001\right\rangle +\cdots +\left\vert 111\right\rangle \right) $.
We can expressed three fields as Eq. (\ref{123}), except for normalization
constant. According to the definition of the ensemble-averaged state Eq. (%
\ref{119}), we obtain the imitation state%
\begin{equation}
\left\vert \tilde{\Psi}\right\rangle =\left\vert 000\right\rangle
+\left\vert 001\right\rangle +\cdots +\left\vert 111\right\rangle .
\label{141}
\end{equation}%
Using the above algorithm, we can easily obtain the Fourier transform
coefficients $D_{000}=C_{000}+C_{001}+\cdots +C_{111}=8e^{i\left( \lambda
^{\left( 1\right) }+\lambda ^{\left( 2\right) }+\lambda ^{\left( 3\right)
}\right) }$, while the other terms is $0$. Then we obtain the
ensemble-averaged state 
\begin{equation}
\left\vert \tilde{\Psi}\right\rangle _{F}=8\left\vert 000\right\rangle ,
\label{142}
\end{equation}%
which is identical to quantum Fourier transform, except for the
normalization constant.

(2) GHZ state

In quantum mechanics, GHZ state is the biggest entanglement state of three
particles. According to Sec. \ref{Sec III.C.1}, we can obtain the following
form of three optical fields 
\begin{equation}
\left\{ 
\begin{array}{c}
\left\vert \psi _{1}\right\rangle =\tilde{\alpha}_{1}\left\vert
0\right\rangle +\tilde{\beta}_{1}\left\vert 1\right\rangle =e^{i\lambda
^{\left( 1\right) }}\left\vert 0\right\rangle +e^{i\lambda ^{\left( 2\right)
}}\left\vert 1\right\rangle \\ 
\left\vert \psi _{2}\right\rangle =\tilde{\alpha}_{2}\left\vert
0\right\rangle +\tilde{\beta}_{2}\left\vert 1\right\rangle =e^{i\lambda
^{\left( 2\right) }}\left\vert 0\right\rangle +e^{i\lambda ^{\left( 3\right)
}}\left\vert 1\right\rangle \\ 
\left\vert \psi _{3}\right\rangle =\tilde{\alpha}_{3}\left\vert
0\right\rangle +\tilde{\beta}_{3}\left\vert 1\right\rangle =e^{i\lambda
^{\left( 3\right) }}\left\vert 0\right\rangle +e^{i\lambda ^{\left( 1\right)
}}\left\vert 1\right\rangle%
\end{array}%
\right. .  \label{143}
\end{equation}%
The formal product state can be expressed as 
\begin{eqnarray}
\left\vert \Psi \right\rangle &=&\left\vert \psi _{1}\right\rangle \otimes
\left\vert \psi _{2}\right\rangle \otimes \left\vert \psi _{3}\right\rangle
=e^{i\left( \lambda ^{\left( 1\right) }+\lambda ^{\left( 2\right) }+\lambda
^{\left( 3\right) }\right) }\left[ \left\vert 000\right\rangle +\left\vert
111\right\rangle +e^{i\left( \lambda ^{\left( 1\right) }-\lambda ^{\left(
3\right) }\right) }\left\vert 001\right\rangle \right.  \label{144} \\
&&+e^{i\left( \lambda ^{\left( 3\right) }-\lambda ^{\left( 2\right) }\right)
}\left\vert 010\right\rangle +e^{i\left( \lambda ^{\left( 1\right) }-\lambda
^{\left( 2\right) }\right) }\left\vert 011\right\rangle +e^{i\left( \lambda
^{\left( 2\right) }-\lambda ^{\left( 1\right) }\right) }\left\vert
100\right\rangle  \nonumber \\
&&\left. +e^{i\left( \lambda ^{\left( 2\right) }-\lambda ^{\left( 3\right)
}\right) }\left\vert 101\right\rangle +e^{i\left( \lambda ^{\left( 3\right)
}-\lambda ^{\left( 1\right) }\right) }\left\vert 110\right\rangle \right] . 
\nonumber
\end{eqnarray}%
According to the definition Eq. (\ref{119}), we obtain the ensemble-averaged
state%
\begin{equation}
\left\vert \tilde{\Psi}\right\rangle =\left\vert 000\right\rangle
+\left\vert 111\right\rangle .  \label{145}
\end{equation}%
Similarly, except for normalization constant and overall phase factor, the
state is identical to GHZ state. Using the above algorithm, we can easily
obtain the Fourier transform coefficients as follows%
\begin{eqnarray}
D_{000} &=&\left( \tilde{\alpha}_{1}+\tilde{\beta}_{1}\right) \left( \tilde{%
\alpha}_{2}+\tilde{\beta}_{2}\right) \left( \tilde{\alpha}_{3}+\tilde{\beta}%
_{3}\right) =2e^{i\left( \lambda ^{\left( 1\right) }+\lambda ^{\left(
2\right) }+\lambda ^{\left( 3\right) }\right) }+e^{i\left( 2\lambda ^{\left(
1\right) }+\lambda ^{\left( 2\right) }\right) }+e^{i\left( 2\lambda ^{\left(
1\right) }+\lambda ^{\left( 3\right) }\right) }  \label{146} \\
&&+e^{i\left( 2\lambda ^{\left( 2\right) }+\lambda ^{\left( 1\right)
}\right) }+e^{i\left( 2\lambda ^{\left( 2\right) }+\lambda ^{\left( 3\right)
}\right) }+e^{i\left( 2\lambda ^{\left( 3\right) }+\lambda ^{\left( 1\right)
}\right) }+e^{i\left( 2\lambda ^{\left( 3\right) }+\lambda ^{\left( 2\right)
}\right) },  \nonumber
\end{eqnarray}%
\begin{eqnarray}
D_{001} &=&\left( \tilde{\alpha}_{1}-\tilde{\beta}_{1}\right) \left( \tilde{%
\alpha}_{2}+\omega ^{2}\tilde{\beta}_{2}\right) \left( \tilde{\alpha}%
_{3}+\omega \tilde{\beta}_{3}\right) =\left( 1-\omega ^{3}\right) e^{i\left(
\lambda ^{\left( 1\right) }+\lambda ^{\left( 2\right) }+\lambda ^{\left(
3\right) }\right) }+\omega e^{i\left( 2\lambda ^{\left( 1\right) }+\lambda
^{\left( 2\right) }\right) }  \label{147} \\
&&+\omega ^{3}e^{i\left( 2\lambda ^{\left( 1\right) }+\lambda ^{\left(
3\right) }\right) }-\omega e^{i\left( 2\lambda ^{\left( 2\right) }+\lambda
^{\left( 1\right) }\right) }-e^{i\left( 2\lambda ^{\left( 2\right) }+\lambda
^{\left( 3\right) }\right) }+\omega ^{2}e^{i\left( 2\lambda ^{\left(
3\right) }+\lambda ^{\left( 1\right) }\right) }  \nonumber \\
&&-\omega ^{2}e^{i\left( 2\lambda ^{\left( 3\right) }+\lambda ^{\left(
2\right) }\right) },  \nonumber
\end{eqnarray}%
\begin{eqnarray}
D_{010} &=&\left( \tilde{\alpha}_{1}+\tilde{\beta}_{1}\right) \left( \tilde{%
\alpha}_{2}-\tilde{\beta}_{2}\right) \left( \tilde{\alpha}_{3}+\omega ^{2}%
\tilde{\beta}_{3}\right) =\left( 1-\omega ^{2}\right) e^{i\left( \lambda
^{\left( 1\right) }+\lambda ^{\left( 2\right) }+\lambda ^{\left( 3\right)
}\right) }+\omega ^{2}e^{i\left( 2\lambda ^{\left( 1\right) }+\lambda
^{\left( 2\right) }\right) }  \label{148} \\
&&-\omega ^{2}e^{i\left( 2\lambda ^{\left( 1\right) }+\lambda ^{\left(
3\right) }\right) }+\omega ^{2}e^{i\left( 2\lambda ^{\left( 2\right)
}+\lambda ^{\left( 1\right) }\right) }+e^{i\left( 2\lambda ^{\left( 2\right)
}+\lambda ^{\left( 3\right) }\right) }-e^{i\left( 2\lambda ^{\left( 3\right)
}+\lambda ^{\left( 1\right) }\right) }  \nonumber \\
&&-e^{i\left( 2\lambda ^{\left( 3\right) }+\lambda ^{\left( 2\right)
}\right) },  \nonumber
\end{eqnarray}%
\begin{eqnarray}
D_{011} &=&\left( \tilde{\alpha}_{1}-\tilde{\beta}_{1}\right) \left( \tilde{%
\alpha}_{2}-\omega ^{2}\tilde{\beta}_{2}\right) \left( \tilde{\alpha}%
_{3}+\omega ^{3}\tilde{\beta}_{3}\right) =\left( 1+\omega ^{5}\right)
e^{i\left( \lambda ^{\left( 1\right) }+\lambda ^{\left( 2\right) }+\lambda
^{\left( 3\right) }\right) }+\omega ^{3}e^{i\left( 2\lambda ^{\left(
1\right) }+\lambda ^{\left( 2\right) }\right) }  \label{149} \\
&&-\omega ^{5}e^{i\left( 2\lambda ^{\left( 1\right) }+\lambda ^{\left(
3\right) }\right) }-\omega ^{3}e^{i\left( 2\lambda ^{\left( 2\right)
}+\lambda ^{\left( 1\right) }\right) }-e^{i\left( 2\lambda ^{\left( 2\right)
}+\lambda ^{\left( 3\right) }\right) }-\omega ^{2}e^{i\left( 2\lambda
^{\left( 3\right) }+\lambda ^{\left( 1\right) }\right) }  \nonumber \\
&&+\omega ^{2}e^{i\left( 2\lambda ^{\left( 3\right) }+\lambda ^{\left(
2\right) }\right) },  \nonumber
\end{eqnarray}%
\begin{eqnarray}
D_{100} &=&\left( \tilde{\alpha}_{1}+\tilde{\beta}_{1}\right) \left( \tilde{%
\alpha}_{2}+\tilde{\beta}_{2}\right) \left( \tilde{\alpha}_{3}-\tilde{\beta}%
_{3}\right) =-e^{i\left( 2\lambda ^{\left( 1\right) }+\lambda ^{\left(
2\right) }\right) }-e^{i\left( 2\lambda ^{\left( 1\right) }+\lambda ^{\left(
3\right) }\right) }  \label{150} \\
&&-e^{i\left( 2\lambda ^{\left( 2\right) }+\lambda ^{\left( 1\right)
}\right) }+e^{i\left( 2\lambda ^{\left( 2\right) }+\lambda ^{\left( 3\right)
}\right) }+e^{i\left( 2\lambda ^{\left( 3\right) }+\lambda ^{\left( 1\right)
}\right) }+e^{i\left( 2\lambda ^{\left( 3\right) }+\lambda ^{\left( 2\right)
}\right) },  \nonumber
\end{eqnarray}%
\begin{eqnarray}
D_{101} &=&\left( \tilde{\alpha}_{1}-\tilde{\beta}_{1}\right) \left( \tilde{%
\alpha}_{2}+\omega ^{2}\tilde{\beta}_{2}\right) \left( \tilde{\alpha}%
_{3}-\omega \tilde{\beta}_{3}\right) =\left( 1+\omega ^{3}\right) e^{i\left(
\lambda ^{\left( 1\right) }+\lambda ^{\left( 2\right) }+\lambda ^{\left(
3\right) }\right) }-\omega e^{i\left( 2\lambda ^{\left( 1\right) }+\lambda
^{\left( 2\right) }\right) }  \label{151} \\
&&-\omega ^{3}e^{i\left( 2\lambda ^{\left( 1\right) }+\lambda ^{\left(
3\right) }\right) }+\omega e^{i\left( 2\lambda ^{\left( 2\right) }+\lambda
^{\left( 1\right) }\right) }-e^{i\left( 2\lambda ^{\left( 2\right) }+\lambda
^{\left( 3\right) }\right) }+\omega ^{2}e^{i\left( 2\lambda ^{\left(
3\right) }+\lambda ^{\left( 1\right) }\right) }  \nonumber \\
&&-\omega ^{2}e^{i\left( 2\lambda ^{\left( 3\right) }+\lambda ^{\left(
2\right) }\right) },  \nonumber
\end{eqnarray}%
\begin{eqnarray}
D_{110} &=&\left( \tilde{\alpha}_{1}+\tilde{\beta}_{1}\right) \left( \tilde{%
\alpha}_{2}-\tilde{\beta}_{2}\right) \left( \tilde{\alpha}_{3}-\omega ^{2}%
\tilde{\beta}_{3}\right) =\left( 1+\omega ^{2}\right) e^{i\left( \lambda
^{\left( 1\right) }+\lambda ^{\left( 2\right) }+\lambda ^{\left( 3\right)
}\right) }-\omega ^{2}e^{i\left( 2\lambda ^{\left( 1\right) }+\lambda
^{\left( 2\right) }\right) }  \label{152} \\
&&+\omega ^{2}e^{i\left( 2\lambda ^{\left( 1\right) }+\lambda ^{\left(
3\right) }\right) }-\omega ^{2}e^{i\left( 2\lambda ^{\left( 2\right)
}+\lambda ^{\left( 1\right) }\right) }+e^{i\left( 2\lambda ^{\left( 2\right)
}+\lambda ^{\left( 3\right) }\right) }-e^{i\left( 2\lambda ^{\left( 3\right)
}+\lambda ^{\left( 1\right) }\right) }  \nonumber \\
&&-e^{i\left( 2\lambda ^{\left( 3\right) }+\lambda ^{\left( 2\right)
}\right) },  \nonumber
\end{eqnarray}%
\begin{eqnarray}
D_{111} &=&\left( \tilde{\alpha}_{1}-\tilde{\beta}_{1}\right) \left( \tilde{%
\alpha}_{2}-\omega ^{2}\tilde{\beta}_{2}\right) \left( \tilde{\alpha}%
_{3}-\omega ^{3}\tilde{\beta}_{3}\right) =\left( 1-\omega ^{5}\right)
e^{i\left( \lambda ^{\left( 1\right) }+\lambda ^{\left( 2\right) }+\lambda
^{\left( 3\right) }\right) }-\omega ^{3}e^{i\left( 2\lambda ^{\left(
1\right) }+\lambda ^{\left( 2\right) }\right) }  \label{153} \\
&&+\omega ^{5}e^{i\left( 2\lambda ^{\left( 1\right) }+\lambda ^{\left(
3\right) }\right) }+\omega ^{3}e^{i\left( 2\lambda ^{\left( 2\right)
}+\lambda ^{\left( 1\right) }\right) }-e^{i\left( 2\lambda ^{\left( 2\right)
}+\lambda ^{\left( 3\right) }\right) }-\omega ^{2}e^{i\left( 2\lambda
^{\left( 3\right) }+\lambda ^{\left( 1\right) }\right) }  \nonumber \\
&&+\omega ^{2}e^{i\left( 2\lambda ^{\left( 3\right) }+\lambda ^{\left(
2\right) }\right) }.  \nonumber
\end{eqnarray}%
According to the definition Eq. (\ref{119}), we obtain the ensemble-averaged
state 
\begin{eqnarray}
\left\vert \tilde{\Psi}\right\rangle _{F}
&=&\sum\limits_{k=1}^{N}e^{-i\lambda ^{\left( S\right) }}\left\vert \Psi
\right\rangle _{F}=2\left\vert 000\right\rangle +\left( 1-\omega ^{3}\right)
\left\vert 001\right\rangle +\left( 1-\omega ^{2}\right) \left\vert
010\right\rangle  \label{154} \\
&&+\left( 1-\omega \right) \left\vert 011\right\rangle +\left( 1+\omega
^{3}\right) \left\vert 101\right\rangle +\left( 1+\omega ^{2}\right)
\left\vert 110\right\rangle +\left( 1+\omega \right) \left\vert
111\right\rangle .  \nonumber
\end{eqnarray}%
In conclusion, $\left\vert \tilde{\Psi}\right\rangle _{F}$ is the Fourier
transform of $\left\vert \tilde{\Psi}\right\rangle $ for the imitaion of GHZ
states.

(3) W state

In quantum mechanics, W state is the most robust entanglement state $%
\left\vert \Psi \right\rangle =\frac{1}{\sqrt{3}}\left( \left\vert
100\right\rangle +\left\vert 010\right\rangle +\left\vert 001\right\rangle
\right) $. According to Sec. \ref{Sec III.C.1}, we can obtain the expression
of three fields as follows%
\begin{equation}
\left\{ 
\begin{array}{c}
\left\vert \psi _{1}\right\rangle =\tilde{\alpha}_{1}\left\vert
0\right\rangle +\tilde{\beta}_{1}\left\vert 1\right\rangle =e^{i\lambda
^{\left( 1\right) }}\left\vert 1\right\rangle +e^{i\lambda ^{\left( 2\right)
}}\left\vert 0\right\rangle +e^{i\lambda ^{\left( 3\right) }}\left\vert
0\right\rangle \\ 
\left\vert \psi _{2}\right\rangle =\tilde{\alpha}_{2}\left\vert
0\right\rangle +\tilde{\beta}_{2}\left\vert 1\right\rangle =e^{i\lambda
^{\left( 1\right) }}\left\vert 1\right\rangle +e^{i\lambda ^{\left( 2\right)
}}\left\vert 0\right\rangle +e^{i\lambda ^{\left( 3\right) }}\left\vert
0\right\rangle \\ 
\left\vert \psi _{3}\right\rangle =\tilde{\alpha}_{3}\left\vert
0\right\rangle +\tilde{\beta}_{3}\left\vert 1\right\rangle =e^{i\lambda
^{\left( 1\right) }}\left\vert 1\right\rangle +e^{i\lambda ^{\left( 2\right)
}}\left\vert 0\right\rangle +e^{i\lambda ^{\left( 3\right) }}\left\vert
0\right\rangle%
\end{array}%
\right. .  \label{155}
\end{equation}%
The formal product state can be expressed as%
\begin{eqnarray}
\left\vert \Psi \right\rangle &=&\left\vert \psi _{1}\right\rangle \otimes
\left\vert \psi _{2}\right\rangle \otimes \left\vert \psi _{3}\right\rangle
=e^{i\left( \lambda ^{\left( 1\right) }+\lambda ^{\left( 2\right) }+\lambda
^{\left( 3\right) }\right) }\left\{ \left[ 2+e^{i\left( \lambda ^{\left(
2\right) }-\lambda ^{\left( 3\right) }\right) }+e^{i\left( \lambda ^{\left(
3\right) }-\lambda ^{\left( 2\right) }\right) }\right] \times \left(
\left\vert 100\right\rangle +\left\vert 010\right\rangle \right. \right.
\label{156} \\
&&\left. +\left\vert 001\right\rangle \right) +\left[ e^{i\left( \lambda
^{\left( 1\right) }-\lambda ^{\left( 3\right) }\right) }+e^{i\left( \lambda
^{\left( 1\right) }-\lambda ^{\left( 2\right) }\right) }\right] \left(
\left\vert 011\right\rangle +\left\vert 110\right\rangle +\left\vert
101\right\rangle \right) +e^{i\left( 2\lambda ^{\left( 1\right) }-\lambda
^{\left( 2\right) }-\lambda ^{\left( 3\right) }\right) }\left\vert
111\right\rangle  \nonumber \\
&&\left. +\left[ e^{i\left( 2\lambda ^{\left( 2\right) }-\lambda ^{\left(
1\right) }-\lambda ^{\left( 3\right) }\right) }+e^{i\left( 2\lambda ^{\left(
3\right) }-\lambda ^{\left( 2\right) }-\lambda ^{\left( 1\right) }\right)
}+3e^{i\left( \lambda ^{\left( 2\right) }-\lambda ^{\left( 1\right) }\right)
}+3e^{i\left( \lambda ^{\left( 3\right) }-\lambda ^{\left( 1\right) }\right)
}\right] \left\vert 000\right\rangle \right\} .  \nonumber
\end{eqnarray}%
According to the definition Eq. (\ref{119}), we obtain the ensemble-averaged
state%
\begin{equation}
\left\vert \tilde{\Psi}\right\rangle =\left\vert 100\right\rangle
+\left\vert 010\right\rangle +\left\vert 001\right\rangle .  \label{157}
\end{equation}%
Similarly, except for normalization constant and overall phase factor, the
state is identical to W state. Using the above algorithm, we can easily
obtain the Fourier transform coefficients as follows%
\begin{eqnarray}
D_{000} &=&\left( \tilde{\alpha}_{1}+\tilde{\beta}_{1}\right) \left( \tilde{%
\alpha}_{2}+\tilde{\beta}_{2}\right) \left( \tilde{\alpha}_{3}+\tilde{\beta}%
_{3}\right) =6e^{i\left( \lambda ^{\left( 1\right) }+\lambda ^{\left(
2\right) }+\lambda ^{\left( 3\right) }\right) }+3\left[ e^{i\left( 2\lambda
^{\left( 1\right) }+\lambda ^{\left( 2\right) }\right) }\right.  \label{158}
\\
&&\left. +e^{i\left( 2\lambda ^{\left( 1\right) }+\lambda ^{\left( 3\right)
}\right) }+e^{i\left( 2\lambda ^{\left( 2\right) }+\lambda ^{\left( 1\right)
}\right) }+e^{i\left( 2\lambda ^{\left( 2\right) }+\lambda ^{\left( 3\right)
}\right) }+e^{i\left( 2\lambda ^{\left( 3\right) }+\lambda ^{\left( 1\right)
}\right) }+e^{i\left( 2\lambda ^{\left( 3\right) }+\lambda ^{\left( 2\right)
}\right) }\right]  \nonumber \\
&&+e^{3i\lambda ^{\left( 1\right) }}+e^{3i\lambda ^{\left( 2\right)
}}+e^{3i\lambda ^{\left( 3\right) }},  \nonumber
\end{eqnarray}%
\begin{eqnarray}
D_{001} &=&\left( \tilde{\alpha}_{1}-\tilde{\beta}_{1}\right) \left( \tilde{%
\alpha}_{2}+\omega ^{2}\tilde{\beta}_{2}\right) \left( \tilde{\alpha}%
_{3}+\omega \tilde{\beta}_{3}\right) =2\left( -1+\omega +\omega ^{2}\right)
e^{i\left( \lambda ^{\left( 1\right) }+\lambda ^{\left( 2\right) }+\lambda
^{\left( 3\right) }\right) }  \label{159} \\
&&-\left( \omega +\omega ^{2}-\omega ^{3}\right) \left[ e^{i\left( 2\lambda
^{\left( 1\right) }+\lambda ^{\left( 2\right) }\right) }+e^{i\left( 2\lambda
^{\left( 1\right) }+\lambda ^{\left( 3\right) }\right) }\right] -\left(
1-\omega -\omega ^{2}\right)  \nonumber \\
&&\times \left[ e^{i\left( 2\lambda ^{\left( 2\right) }+\lambda ^{\left(
1\right) }\right) }+e^{i\left( 2\lambda ^{\left( 3\right) }+\lambda ^{\left(
1\right) }\right) }\right] +e^{i\left( 2\lambda ^{\left( 3\right) }+\lambda
^{\left( 2\right) }\right) }+e^{i\left( 2\lambda ^{\left( 2\right) }+\lambda
^{\left( 3\right) }\right) }  \nonumber \\
&&-\omega ^{3}e^{3i\lambda ^{\left( 1\right) }}+e^{3i\lambda ^{\left(
2\right) }}+e^{3i\lambda ^{\left( 3\right) }},  \nonumber
\end{eqnarray}%
\begin{eqnarray}
D_{010} &=&\left( \tilde{\alpha}_{1}+\tilde{\beta}_{1}\right) \left( \tilde{%
\alpha}_{2}-\tilde{\beta}_{2}\right) \left( \tilde{\alpha}_{3}+\omega ^{2}%
\tilde{\beta}_{3}\right) =2\omega ^{2}e^{i\left( \lambda ^{\left( 1\right)
}+\lambda ^{\left( 2\right) }+\lambda ^{\left( 3\right) }\right) }-\left[
e^{i\left( 2\lambda ^{\left( 1\right) }+\lambda ^{\left( 2\right) }\right)
}\right.  \label{160} \\
&&\left. +e^{i\left( 2\lambda ^{\left( 1\right) }+\lambda ^{\left( 3\right)
}\right) }\right] +\omega ^{2}\left[ e^{i\left( 2\lambda ^{\left( 2\right)
}+\lambda ^{\left( 1\right) }\right) }+e^{i\left( 2\lambda ^{\left( 3\right)
}+\lambda ^{\left( 1\right) }\right) }\right] +e^{i\left( 2\lambda ^{\left(
3\right) }+\lambda ^{\left( 2\right) }\right) }  \nonumber \\
&&+e^{i\left( 2\lambda ^{\left( 2\right) }+\lambda ^{\left( 3\right)
}\right) }-\omega ^{2}e^{3i\lambda ^{\left( 1\right) }}+e^{3i\lambda
^{\left( 2\right) }}+e^{3i\lambda ^{\left( 3\right) }},  \nonumber
\end{eqnarray}%
\begin{eqnarray}
D_{011} &=&\left( \tilde{\alpha}_{1}-\tilde{\beta}_{1}\right) \left( \tilde{%
\alpha}_{2}-\omega ^{2}\tilde{\beta}_{2}\right) \left( \tilde{\alpha}%
_{3}+\omega ^{3}\tilde{\beta}_{3}\right) =2\left( -1-\omega ^{2}+\omega
^{3}\right) e^{i\left( \lambda ^{\left( 1\right) }+\lambda ^{\left( 2\right)
}+\lambda ^{\left( 3\right) }\right) }  \label{161} \\
&&+\left( \omega ^{2}-\omega ^{3}-\omega ^{5}\right) \left[ e^{i\left(
2\lambda ^{\left( 1\right) }+\lambda ^{\left( 2\right) }\right) }+e^{i\left(
2\lambda ^{\left( 1\right) }+\lambda ^{\left( 3\right) }\right) }\right]
-\left( 1+\omega ^{2}-\omega ^{3}\right)  \nonumber \\
&&\times \left[ e^{i\left( 2\lambda ^{\left( 2\right) }+\lambda ^{\left(
1\right) }\right) }+e^{i\left( 2\lambda ^{\left( 3\right) }+\lambda ^{\left(
1\right) }\right) }\right] +e^{i\left( 2\lambda ^{\left( 3\right) }+\lambda
^{\left( 2\right) }\right) }+e^{i\left( 2\lambda ^{\left( 2\right) }+\lambda
^{\left( 3\right) }\right) }  \nonumber \\
&&+\omega ^{5}e^{3i\lambda ^{\left( 1\right) }}+e^{3i\lambda ^{\left(
2\right) }}+e^{3i\lambda ^{\left( 3\right) }},  \nonumber
\end{eqnarray}%
\begin{eqnarray}
D_{100} &=&\left( \tilde{\alpha}_{1}+\tilde{\beta}_{1}\right) \left( \tilde{%
\alpha}_{2}+\tilde{\beta}_{2}\right) \left( \tilde{\alpha}_{3}-\tilde{\beta}%
_{3}\right) =2e^{i\left( \lambda ^{\left( 1\right) }+\lambda ^{\left(
2\right) }+\lambda ^{\left( 3\right) }\right) }-2\left[ e^{i\left( 2\lambda
^{\left( 1\right) }+\lambda ^{\left( 2\right) }\right) }\right.  \label{162}
\\
&&\left. +e^{i\left( 2\lambda ^{\left( 1\right) }+\lambda ^{\left( 3\right)
}\right) }\right] +2\left[ e^{i\left( 2\lambda ^{\left( 2\right) }+\lambda
^{\left( 1\right) }\right) }+e^{i\left( 2\lambda ^{\left( 3\right) }+\lambda
^{\left( 1\right) }\right) }\right] +e^{i\left( 2\lambda ^{\left( 3\right)
}+\lambda ^{\left( 2\right) }\right) }  \nonumber \\
&&+e^{i\left( 2\lambda ^{\left( 2\right) }+\lambda ^{\left( 3\right)
}\right) }-e^{3i\lambda ^{\left( 1\right) }}+e^{3i\lambda ^{\left( 2\right)
}}+e^{3i\lambda ^{\left( 3\right) }},  \nonumber
\end{eqnarray}%
\begin{eqnarray}
D_{101} &=&\left( \tilde{\alpha}_{1}-\tilde{\beta}_{1}\right) \left( \tilde{%
\alpha}_{2}+\omega ^{2}\tilde{\beta}_{2}\right) \left( \tilde{\alpha}%
_{3}-\omega \tilde{\beta}_{3}\right) =-2\left( 1+\omega -\omega ^{2}\right)
e^{i\left( \lambda ^{\left( 1\right) }+\lambda ^{\left( 2\right) }+\lambda
^{\left( 3\right) }\right) }  \label{163} \\
&&+\left( \omega -\omega ^{2}-\omega ^{3}\right) \left[ e^{i\left( 2\lambda
^{\left( 1\right) }+\lambda ^{\left( 2\right) }\right) }+e^{i\left( 2\lambda
^{\left( 1\right) }+\lambda ^{\left( 3\right) }\right) }\right] -\left(
1+\omega -\omega ^{2}\right)  \nonumber \\
&&\times \left[ e^{i\left( 2\lambda ^{\left( 2\right) }+\lambda ^{\left(
1\right) }\right) }+e^{i\left( 2\lambda ^{\left( 3\right) }+\lambda ^{\left(
1\right) }\right) }\right] +e^{i\left( 2\lambda ^{\left( 3\right) }+\lambda
^{\left( 2\right) }\right) }+e^{i\left( 2\lambda ^{\left( 2\right) }+\lambda
^{\left( 3\right) }\right) }  \nonumber \\
&&+\omega ^{3}e^{3i\lambda ^{\left( 1\right) }}+e^{3i\lambda ^{\left(
2\right) }}+e^{3i\lambda ^{\left( 3\right) }},  \nonumber
\end{eqnarray}%
\begin{eqnarray}
D_{110} &=&\left( \tilde{\alpha}_{1}+\tilde{\beta}_{1}\right) \left( \tilde{%
\alpha}_{2}-\tilde{\beta}_{2}\right) \left( \tilde{\alpha}_{3}-\omega ^{2}%
\tilde{\beta}_{3}\right) =-2\omega ^{2}e^{i\left( \lambda ^{\left( 1\right)
}+\lambda ^{\left( 2\right) }+\lambda ^{\left( 3\right) }\right) }-\left[
e^{i\left( 2\lambda ^{\left( 1\right) }+\lambda ^{\left( 2\right) }\right)
}\right.  \label{164} \\
&&\left. +e^{i\left( 2\lambda ^{\left( 1\right) }+\lambda ^{\left( 3\right)
}\right) }\right] -\omega ^{2}\left[ e^{i\left( 2\lambda ^{\left( 2\right)
}+\lambda ^{\left( 1\right) }\right) }+e^{i\left( 2\lambda ^{\left( 3\right)
}+\lambda ^{\left( 1\right) }\right) }\right] +e^{i\left( 2\lambda ^{\left(
3\right) }+\lambda ^{\left( 2\right) }\right) }  \nonumber \\
&&+e^{i\left( 2\lambda ^{\left( 2\right) }+\lambda ^{\left( 3\right)
}\right) }+\omega ^{2}e^{3i\lambda ^{\left( 1\right) }}+e^{3i\lambda
^{\left( 2\right) }}+e^{3i\lambda ^{\left( 3\right) }},  \nonumber
\end{eqnarray}%
\begin{eqnarray}
D_{111} &=&\left( \tilde{\alpha}_{1}-\tilde{\beta}_{1}\right) \left( \tilde{%
\alpha}_{2}-\omega ^{2}\tilde{\beta}_{2}\right) \left( \tilde{\alpha}%
_{3}-\omega ^{3}\tilde{\beta}_{3}\right) =-2\left( 1+\omega ^{2}+\omega
^{3}\right) e^{i\left( \lambda ^{\left( 1\right) }+\lambda ^{\left( 2\right)
}+\lambda ^{\left( 3\right) }\right) }  \label{165} \\
&&+\left( \omega ^{2}+\omega ^{3}+\omega ^{5}\right) \left[ e^{i\left(
2\lambda ^{\left( 1\right) }+\lambda ^{\left( 2\right) }\right) }+e^{i\left(
2\lambda ^{\left( 1\right) }+\lambda ^{\left( 3\right) }\right) }\right]
-\left( 1+\omega ^{2}+\omega ^{3}\right)  \nonumber \\
&&\times \left[ e^{i\left( 2\lambda ^{\left( 2\right) }+\lambda ^{\left(
1\right) }\right) }+e^{i\left( 2\lambda ^{\left( 3\right) }+\lambda ^{\left(
1\right) }\right) }\right] +e^{i\left( 2\lambda ^{\left( 3\right) }+\lambda
^{\left( 2\right) }\right) }+e^{i\left( 2\lambda ^{\left( 2\right) }+\lambda
^{\left( 3\right) }\right) }  \nonumber \\
&&-\omega ^{5}e^{3i\lambda ^{\left( 1\right) }}+e^{3i\lambda ^{\left(
2\right) }}+e^{3i\lambda ^{\left( 3\right) }}.  \nonumber
\end{eqnarray}%
According to the definition Eq. (\ref{119}), we obtain the ensemble-averaged
state 
\begin{eqnarray}
\left\vert \tilde{\Psi}\right\rangle _{F}
&=&\sum\limits_{k=1}^{N}e^{-i\lambda ^{\left( S\right) }}\left\vert \Psi
\right\rangle _{F}=6\left\vert 000\right\rangle -2\left( 1-\omega -\omega
^{2}\right) \left\vert 001\right\rangle +2\omega ^{2}\left\vert
010\right\rangle -2\left( 1+\omega ^{2}-\omega ^{3}\right) \left\vert
011\right\rangle  \label{166} \\
&&+2\left\vert 100\right\rangle -2\left( 1+\omega -\omega ^{2}\right)
\left\vert 101\right\rangle +2\omega ^{2}\left\vert 110\right\rangle
-2\left( 1+\omega ^{2}+\omega ^{3}\right) \left\vert 111\right\rangle . 
\nonumber
\end{eqnarray}%
In conclusion, $\left\vert \tilde{\Psi}\right\rangle _{F}$ is also the
Fourier transform of $\left\vert \tilde{\Psi}\right\rangle $ for the
imitaion of W states.

Based on the phase ensemble model, we propose an optical Fourier algorithm
similar to quantum Fourier algorithm. The computational resources required
for this algorithm is in $O(N^{2})$ also similar to quantum Fourier
algorithm, which means an exponential speedup compared with classical
Fourier algorithm.

\section{Conclusion \label{Sec V}}

In this paper, we have discussed a new approach to imitate quantum states
using the optical fields modulated with PPSs. We demonstrated that $N$
optical fields modulated with $N$\ different PPSs can span a $N2^{N}$
dimensional Hilbert space that contains tensor product structure similar to
quantum systems. It is noteworthy that a classical optical field\ is the
most similar to a quantum state, especially for coherent superposition
state. This is why the space spanned by optical fields can imitate a quantum
system yet not the space of probability distributions of classical coins
that also contains a tensor product.

In this paper, we only build a simple framework for this approach. However,
there are still many problems that need to be studied continuously, such as
the imitation forms of all quantum states, more general algorithms, unitary
universal gate like CNOT, etc. It is particularly interesting to simulate
higher-dimensional real quantum systems applying this approach, such as
qutrits, higher-dimensional Hilbert spaces, even quantum fields. The
greatest benefits of this approach is an arbitrary dimensional Hilbert space
can be provided by using linear growth resources. Finally, we look forward
to verifying the feasibility of the approach through the relevant
experiments. We believe the experiments are not difficult to achieve,
because all the technologies have been applied in the mature optical
communication system.

\begin{acknowledgments}
I would like to thank those who have supported me for ten years, including
Dr. Shuo Sun who took part in the discussion of Shor's algorithm, Dr. Xutai
Ma who took part in the discussion of the gate array model, Prof. Xunkun Wu
who helped me to revise some subjects of English, Prof. Wei Fang who took
part in the discussion of quantum Fourier algorithm, and Mr. Yongzheng Ye
who took part in the discussion of the software OPTISYSTEM.
\end{acknowledgments}

\end{document}